\documentclass[11pt]{article}
\usepackage{color}
\usepackage{epic}
\usepackage{amsthm,amsmath,amsfonts,mathtools,mathrsfs,amssymb,latexsym,bm,soul}
\usepackage{graphicx}
\usepackage{setspace}
\usepackage{comment,  verbatim} 
\usepackage{enumerate,enumitem}
\usepackage{graphicx}
\usepackage{float}
\usepackage{setspace}
\usepackage{hyperref}
\usepackage{subcaption}
\usepackage[table]{xcolor}
\usepackage{setspace}
\usepackage[margin=1in]{geometry}
\usepackage[affil-it]{authblk}
\usepackage{color}
\usepackage{amsmath}
\usepackage{graphicx,psfrag,epsf}
\usepackage{enumerate}
\usepackage[comma,numbers]{natbib}
\usepackage{algorithm,algpseudocode}

\usepackage[normalem]{ulem}

\RequirePackage{latexsym}
\RequirePackage{amsmath}
\RequirePackage{amssymb}
\RequirePackage{bm}             %
\RequirePackage{amsbsy}             %
\RequirePackage{soul}

\DeclareMathOperator{\Dcal}{\mathcal{D}}

\DeclareMathOperator{\Pcal}{\mathcal{P}}

\DeclareMathOperator{\MM}{\mathbb{M}} 
\DeclareMathOperator{\NN}{\mathbb{N}} 
\DeclareMathOperator{\RR}{\mathbb{R}} 

\DeclareMathOperator{\diag}{diag}

\ifx\BlackBox\undefined
\newcommand{\BlackBox}{\rule{1.5ex}{1.5ex}}  
\fi

\newtheorem{theorem}{Theorem}[section]

\newtheorem{lemma}{Lemma}[section]

\ifx\proof\undefined
\newenvironment{proof}{\par\noindent{\bf Proof\ }}{\hfill\BlackBox\\[2mm]}
\fi

\ifx\axiom\undefined
\newtheorem{axiom}[theorem]{Axiom}
\fi

\newcommand{\T}{T^{*}}

\allowdisplaybreaks

\def\T{{ \mathrm{\scriptscriptstyle T} }}

\theoremstyle{plain}
\newtheorem{thm}{Theorem}[section]
\newtheorem{assump}{Assumption}[section]
\newtheorem{lem}{Lemma}[]

\title{Asynchronous and Distributed Data Augmentation for Massive Data Settings}

\author[1]{Jiayuan Zhou \thanks{\url{zhou.j@ufl.edu}}}
\author[1]{Kshitij Khare \thanks{\url{kdkhare@stat.ufl.edu}}}
\author[2]{Sanvesh Srivastava \thanks{\url{sanvesh-srivastava@uiowa.edu}}}

\affil[1]{Department of Statistics, University of Florida}
\affil[2]{Department of Statistics and Actuarial Science, The University of Iowa}

\date{\today}

\begin{document}

\maketitle

\begin{abstract}  
  Data augmentation (DA) algorithms are widely used for Bayesian inference due to their simplicity. In massive data settings, however, DA algorithms are prohibitively slow because they pass through the full data in any iteration, imposing serious restrictions on their usage despite the advantages. Addressing this problem, we develop a framework for extending any DA that exploits asynchronous and distributed computing. The extended DA algorithm is indexed by a parameter $r \in (0, 1)$ and is called Asynchronous and Distributed (AD) DA with the original DA as its parent. Any ADDA starts by dividing the full data into $k$ smaller disjoint subsets and storing them on $k$ processes, which could be machines or processors. Every iteration of ADDA augments only an $r$-fraction of the $k$ data subsets with some positive probability and leaves the remaining $(1-r)$-fraction of the augmented data unchanged. The parameter draws are obtained using the $r$-fraction of new and $(1-r)$-fraction of old augmented data. For many choices of $k$ and $r$, the fractional updates of ADDA lead to a significant speed-up over the parent DA in massive data settings, and it reduces to the distributed version of its parent DA when $r=1$. {We show that the ADDA Markov chain is Harris ergodic with the desired stationary distribution under mild conditions on the parent DA algorithm.} We demonstrate the numerical advantages of the ADDA in three representative examples {corresponding to different kinds of massive data settings encountered in applications. In all these examples, our DA generalization is significantly faster than its parent DA algorithm for all the choices of $k$ and $r$. We also establish geometric ergodicity of the ADDA Markov chain for all three examples, which in turn yields asymptotically valid standard errors for estimates of desired posterior quantities.}
\end{abstract}

\section{Introduction}

DA algorithms are a popular choice for Bayesian inference using Markov chain Monte Carlo. A variety of DA algorithms have been developed over the years for sampling-based posterior inference in a large class of hierarchical Bayesian models. The development of DA algorithms still remains popular because they are simple, easily implemented, and numerically stable; however, DA algorithms are very slow in massive data applications because they pass through the full data in every iteration. Taking advantage of asynchronous and distributed computations, we propose the ADDA as a scalable  generalization of any DA-type algorithm. ADDA is useful for fitting flexible Bayesian models to arbitrarily large data sets, retains the convergence properties of its parent DA, and reduces to the parent DA under certain assumptions. 

Consider the setup of DA algorithms. In a typical DA application, we augment ``missing data'' to the observed data and obtain the ``augmented data'' model, where the term missing data is interpreted broadly to include any additional parameters or latent variables. Under the augmented data model, the imputation (I) step draws the missing data from their conditional distribution given the observed data and the current parameter value. The I step is followed by the prediction (P) step that draws the ``original" parameter from its conditional distribution given the augmented data. This completes one cycle of a DA algorithm. Starting from some initial value of the parameter, the I and P steps are repeated sequentially to obtain a Markov chain for the missing data and parameter. {The marginal parameter samples from the DA algorithm form a Markov chain with the posterior distribution of parameters given the observed data as its stationary distribution \citep{Hob11,RobCas13}.}

The convergence of the DA Markov chain to its stationary distribution can be extremely slow. This problem has been addressed by developing generalizations of many DA algorithms based on missing data models indexed by a ``working parameter'' that is identified under the augmented data model but not under the observed data model. Parameter expanded DA chooses any working parameter, conditional DA chooses the working parameter that leads to maximum acceleration in the convergence to the stationary distribution, and marginal DA averages over values of the working parameter drawn from its prior distribution \citep{DykMen01, Hobert:Marchev:2008}. All these generalizations accelerate convergence to the stationary distribution by relaxing many restrictions in the observed data model. {Modern massive data settings, however, pose severe computational challenges for DA and its parameter expanded versions in that the size of the augmented missing data can be really large, resulting in a prohibitively slow I step in every iteration of the DA or its extension.} 

The inefficiency of DA-type algorithms in massive data settings has received significant attention. Expectation-Maximization (EM) algorithm is the deterministic version of DA and its inefficiency is due to the slow E step \citep{DemLaiRub77}. \citet{CapMou09} have addressed this issue by developing the family of \emph{online} EMs that modify the E step using stochastic approximation. They compute the conditional expectation of complete data log likelihood obtained using a small fraction of the full data and the online E step increases in accuracy as greater fraction of the full data are processed. The stochastic counterparts of online EMs are the suite of Monte Carlo algorithms based on subsampling \citep{KorCheWel13,Baretal17,Dasetal18,Deetal18,ZhaDe19,Quietal19} and stochastic gradient descent \citep{WelTeh11,AhnKorWel12,Bouetal18,Feaetal18,Bieetal19,plassier2021dg}; see \citet{NemFea19} for a recent overview. All these algorithms are extremely scalable, have convergence guarantees, and are broadly applicable; however, the performance of these algorithms is sensitive to the specification of tuning parameters in hierarchical Bayesian models, including subsample size, gradient step-size, and the \emph{learning rate}. Their optimal performance requires proposal tuning, which is a major limitation in their use as off-the-shelf algorithms for Bayesian inference. 

Distributed Bayesian inference approaches based on the divide-and-conquer technique also rely on DA and exploit its ease of implementation. A typical application involves dividing the full data into smaller subsets, obtaining parameter draws in parallel on all the subsets using a DA-type algorithm, and combining parameter draws from all the subsets. The combined parameter draws are used as a computationally efficient alternative to the draws from the true posterior distribution. A variety of algorithms have been developed over the years at an increasing level of generality and applicability \citep{Scoetal16,Lietal16,Minetal17,Srietal18,XueLia17,Xuetal18,Joretal19,WuRob19,WanSri21,Ouetal21}. There are two major limitations in all these methods. {First, the combined parameter draws do not provide a Markov chain with the target posterior based on the entire data as the stationary distribution, which implies that quantification of the Monte Carlo error is challenging.} Second, the theoretical guarantees of these methods are based on a normal approximation of the posterior distribution as the subset sample size and the number of subsets tend to infinity. No guarantees are available about the distance between the distribution of the combined parameter draws and the target posterior distribution. 

Addressing both these problems, the proposed ADDA offers an simple approach for bypassing problems due to the slow I step using {\it asynchronous} and {\it distributed} computations, but at the same time producing a {\it single Markov 
chain with the target posterior as a marginal of its stationary distribution}. The key observation which enables such a construction is that for many DA settings, the missing data can be partitioned into several sub-blocks such that two conditions are satisfied. First, all these sub-blocks are mutually independent given the original parameter and the observed data. Second, the conditional posterior distribution (given the original parameter) of each sub-block depends only on an exclusive subset of the observed data; see the representative examples in Sections \ref{sec:adda-logist-reg}-\ref{sec:adda-Mixed-Effects} for greater details.

An ADDA algorithm starts by reserving $(k + l)$ computing units called \emph{processes}, where $l \ll k$, $k$ and $l$ are the number of workers and managers, respectively. The processes could be cores in a CPU, threads in a GPU, or machines in a cluster. Every implementation of ADDA divides the full data into smaller $k$ disjoint subsets, stores them on the $k$ workers, and performs the I step in parallel on the $k$ workers. The workers communicate the results of their I steps to the manager processes that are responsible for performing the P step. The managers also track progress of the algorithm by maintaining the latest copies of I step results for every worker. Let an $r \in (0, 1)$ and an $\epsilon \in (0, 1)$ be given. Then, the managers wait for all workers to return their results with probability $\epsilon$. With the probability $1-\epsilon$, the managers wait to receive the I step results from an $r$-fraction of the workers and perform the P step using an augmented data model that depends on the $r$-fraction of new I step results and the ($1-r$)-fraction of old I step results.  This sequence of I and P steps is repeated until convergence, which generalizes the classical DA-type algorithms by using $r$ as an additional working parameter. If $r=1$, then we recover a distributed generalization of the parent DA algorithm.

{The parameter $\epsilon$ is introduced as a theoretical device for ensuring that any ADDA algorithm produces a Markov chain. Any positive $\epsilon$ ensures that the missing data sub-block corresponding to each worker has a positive probability to be updated at every iteration. This is crucial for establishing theoretical results about the ADDA chain, including Harris ergodicity and geometric ergodicity. In practice, $\epsilon$ can be set to be a small number for faster computations.}

It is important to compare and contrast the proposed ADDA algorithm with Asynchronous Gibbs sampling algorithms that have been developed in machine learning motivated by topic models. \citet{Newetal09} developed the first such algorithm, which has been extended by exploiting various synchronization schemes and computer architecture \citep{Ahmetal11,Leeetal14}. The only similarity between this class of algorithms and ADDA is that the data are partitioned and stored on the workers, which run the sampling algorithm locally. The manager process is absent in asynchronous Gibbs sampling because every worker draws its local set of parameters and latent variables from the local full conditional without waiting for the full Gibbs cycle to finish. This also implies that the parameter and latent variable updates do not form a Markov chain. In simple Gaussian models, this strategy produces draws from a distribution that fails to converge to the target \citep{Johetal13}. 

\citet{Teretal20} fix this issue by introducing a correction step that allows a worker to accept or reject a draw with a probability based on the corresponding acceptance ratio. This additional Metropolis step increases the computational burden and the amount of information that needs to be communicated among the workers, but the authors are able to establish convergence to the desired stationary distribution under additional regularity conditions. The sequence of iterates generated by this exact algorithm still do not form a Markov chain in general, and hence one does not have access to the standard Markov chain central limit theorem (CLT) based approaches to quantify the MCMC standard error (see the discussion below). More recently, these ideas have been extended to Bayesian variable selection using Gaussian spike-and-slab priors \citep{AtcWan21}. Developing a general theoretical understanding of these asynchronous Gibbs algorithms is still an active area of research. 

On the other hand, any ADDA algorithm inherits several attractive properties of its parent DA. First, being a generalization of the parent DA for a valid choice of $(k, r)$, any ADDA is easy to implement and numerically stable. Second, ADDA is the stochastic counterpart of the distributed EM, is independent of the computer architecture, and has broad applicability \citep{Srietal19}. Finally, while the marginal sequence of original parameter values generated by the ADDA algorithm is not a Markov chain (unlike the DA algorithm), the joint sequence of original and augmented parameter values is a Markov chain whose stationary distribution has the targeted posterior as a marginal. This facilitates a simpler theoretical analysis of the Monte Carlo error based on the familiar tools for analyzing Markov chains in sampling-based posterior inference; see Sections \ref{sec:adda-logist-reg}-\ref{sec:adda-Mixed-Effects} below.

We now describe the structure and key contributions of this paper. The general framework for the ADDA algorithm is developed in Section \ref{sec:adda:-gener-fram}, and Harris ergodicity of the resulting Markov chain is established in Theorem \ref{Harris}. We then develop ADDA extensions of three representative DA algorithms in Sections \ref{sec:adda-logist-reg}-\ref{sec:adda-Mixed-Effects}. There are two kinds of massive data settings that are encountered in modern applications. The first is the ``massive $n$ setting", where the number of observations or samples $n$ is extremely large, typically in the order of millions or larger. For this setting, we consider two illustrative examples: the Polya-Gamma DA algorithm \citep{Poletal13} for Bayesian logistic regression and the marginal DA algorithm \citep{DykMen01} for linear mixed-effects modeling. For both algorithms, the number of augmented parameters is equal to the number of observations, which makes the I step computationally very expensive, and provides motivation for applying the ADDA methodology. The corresponding ADDA extensions are specified in Algorithms \ref{ADPG} and \ref{ADGLM}. The second big data setting is the ``massive $p$ setting", where the number of variables $p$ in the dataset is extremely large. For this setting, we consider the Bayesian lasso DA algorithm \citep{rajaratnam2015scalable} for high-dimensional regression, where the number of augmented parameters equals the (large) number of regression parameters $p$. This provides another motivation for applying the ADDA methodology and the corresponding ADDA for the Bayesian lasso   is presented in Algorithm \ref{ADBL}.

\begin{figure}[H]
\begin{subfigure}{.5\textwidth}
  \centering
  \includegraphics[scale=0.21]{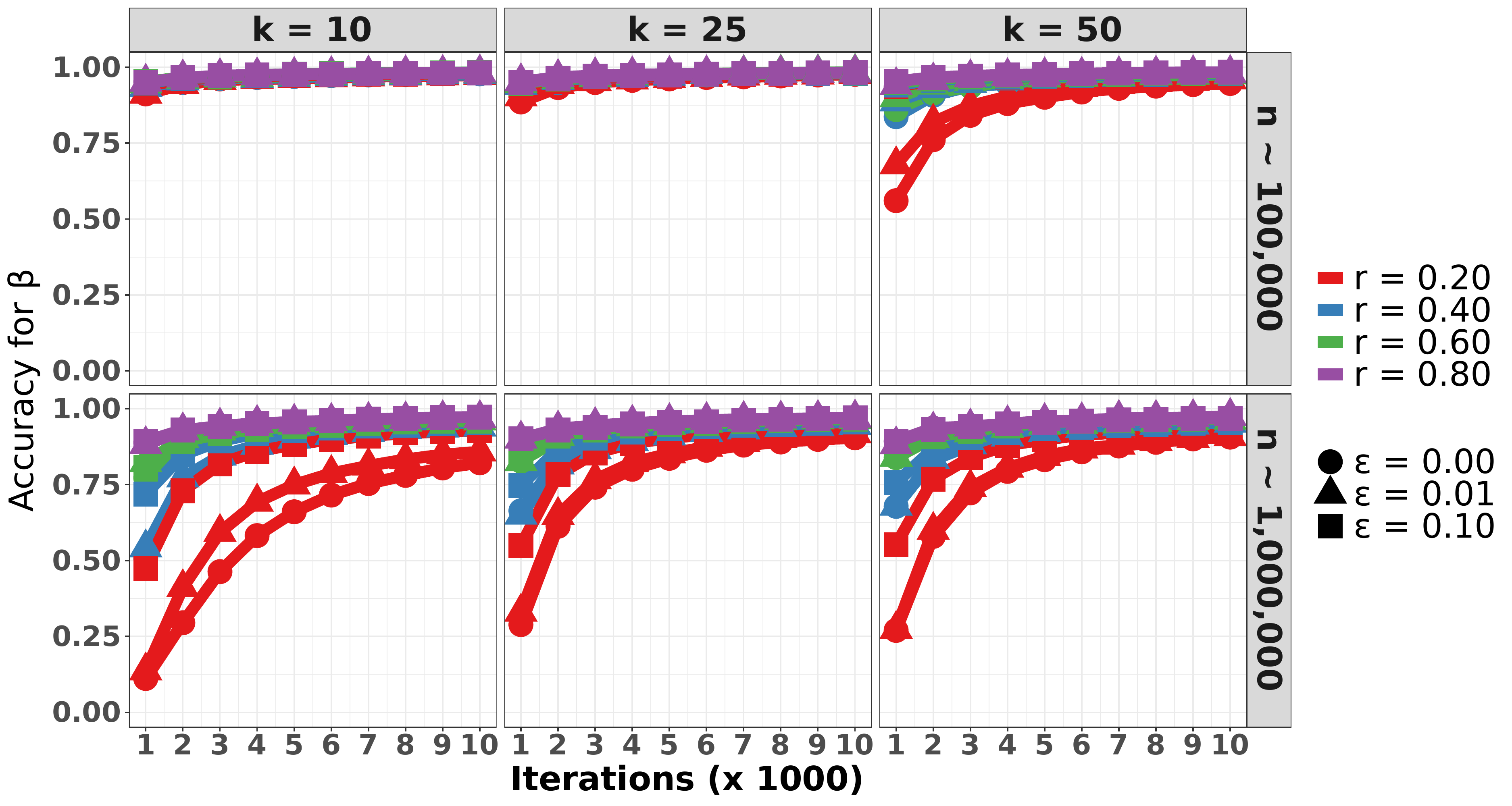}
  \caption{$\beta$}
  \label{fig:mov_beta}
\end{subfigure}%
\begin{subfigure}{.5\textwidth}
  \centering
  \includegraphics[scale=0.21]{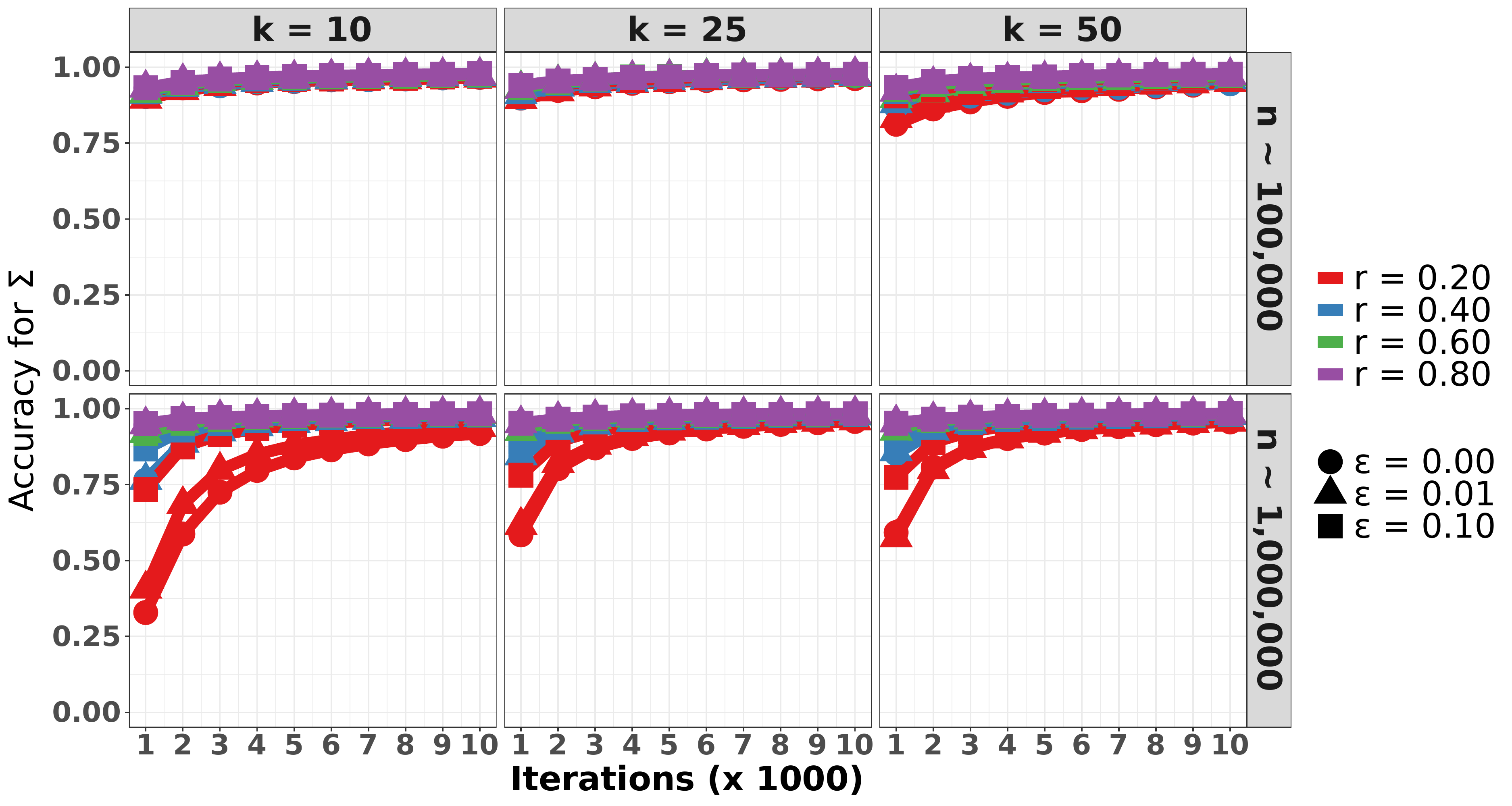}
  \caption{ $\Sigma$}
  \label{fig:mov_dmat}
\end{subfigure}
\caption{$\text{Acc}_{rk}(t)$ metric for inference on $\beta$ and $\Sigma$ in the mixed-effects model for MovieLens data with $r = 0.20, 0.40, 0.60, 0.80$ and $k=10, 25, 50$.}
\label{fig:mov_beta_dmat}
\end{figure}

\begin{figure}[H]
\begin{subfigure}{.5\textwidth}
  \centering
  \includegraphics[scale=0.21]{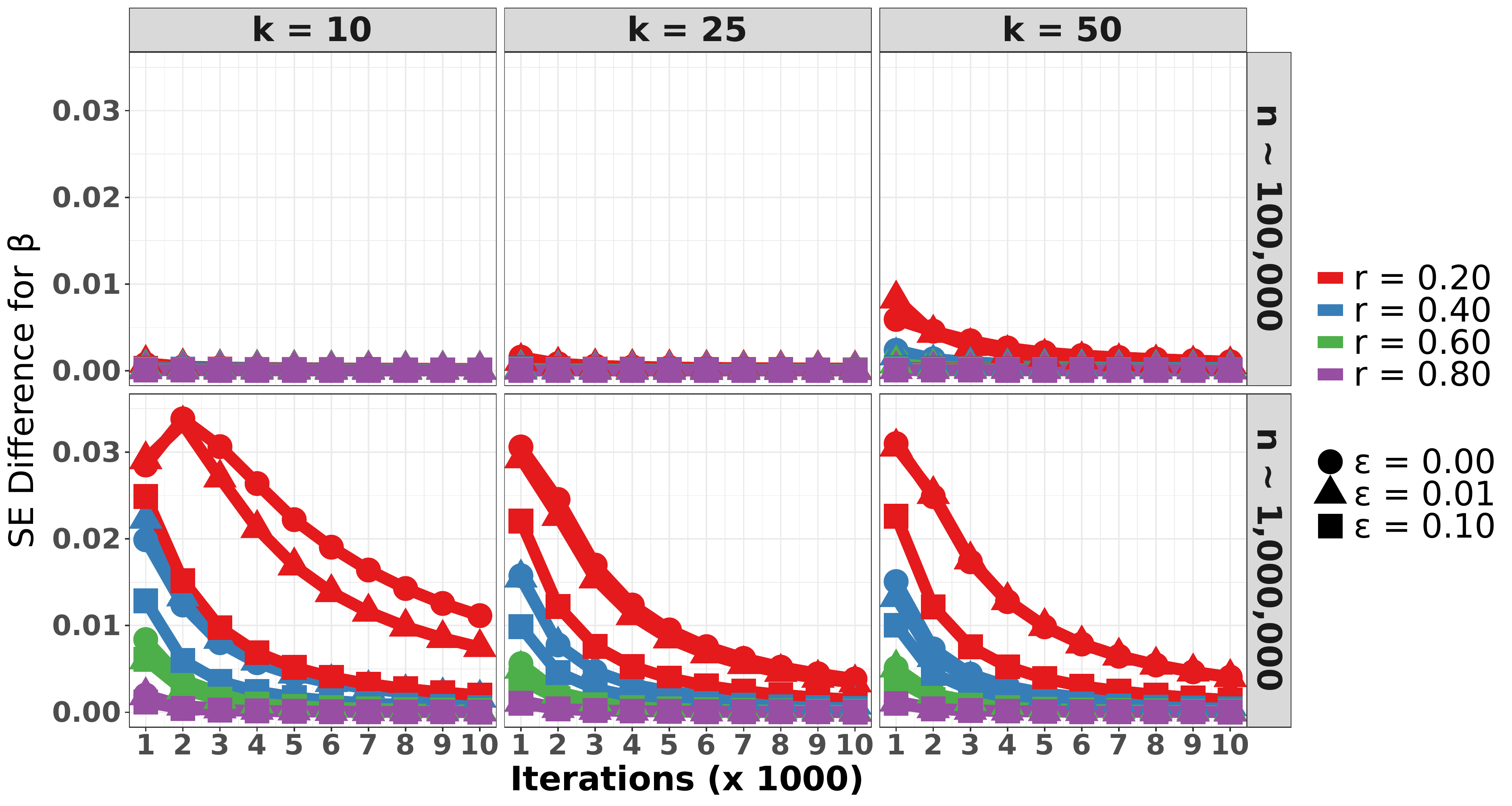}
  \caption{$\beta$}
  \label{fig:mcse_mov_beta}
\end{subfigure}%
\begin{subfigure}{.5\textwidth}
  \centering
  \includegraphics[scale=0.21]{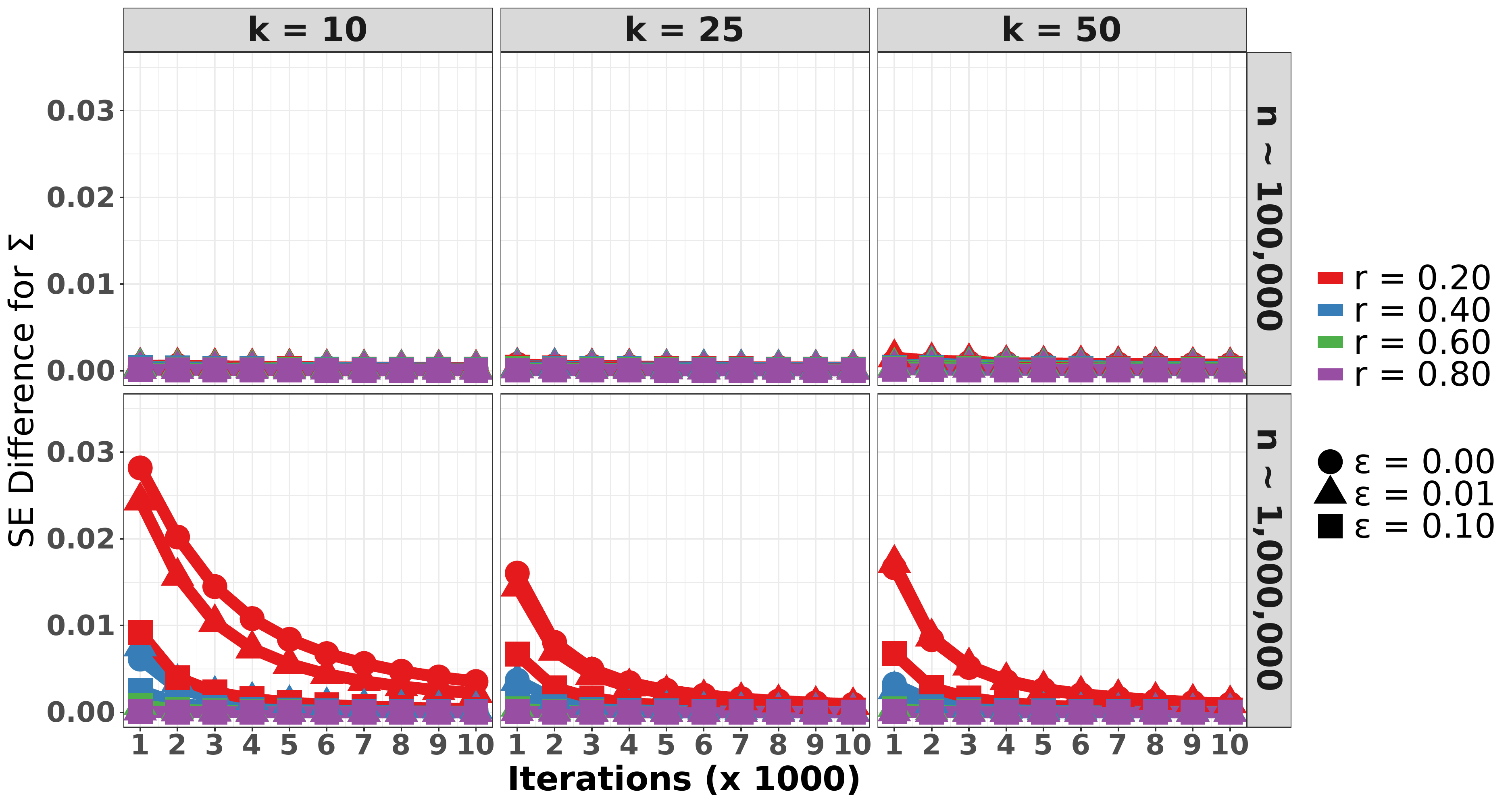}
  \caption{ $\Sigma$}
  \label{fig:mcse_mov_dmat}
\end{subfigure}
\caption{$\text{SE}_{rk}(t)$ metric for inference on $\beta$ and $\Sigma$ in the mixed-effects model for MovieLens data  with $r = 0.20, 0.40, 0.60, 0.80$ and $k=10, 25, 50$.}
\label{fig:mcse_mov_beta_dmat}
\end{figure}

\begin{figure}[H]
\begin{subfigure}{.5\textwidth}
  \centering
  \includegraphics[scale=0.25]{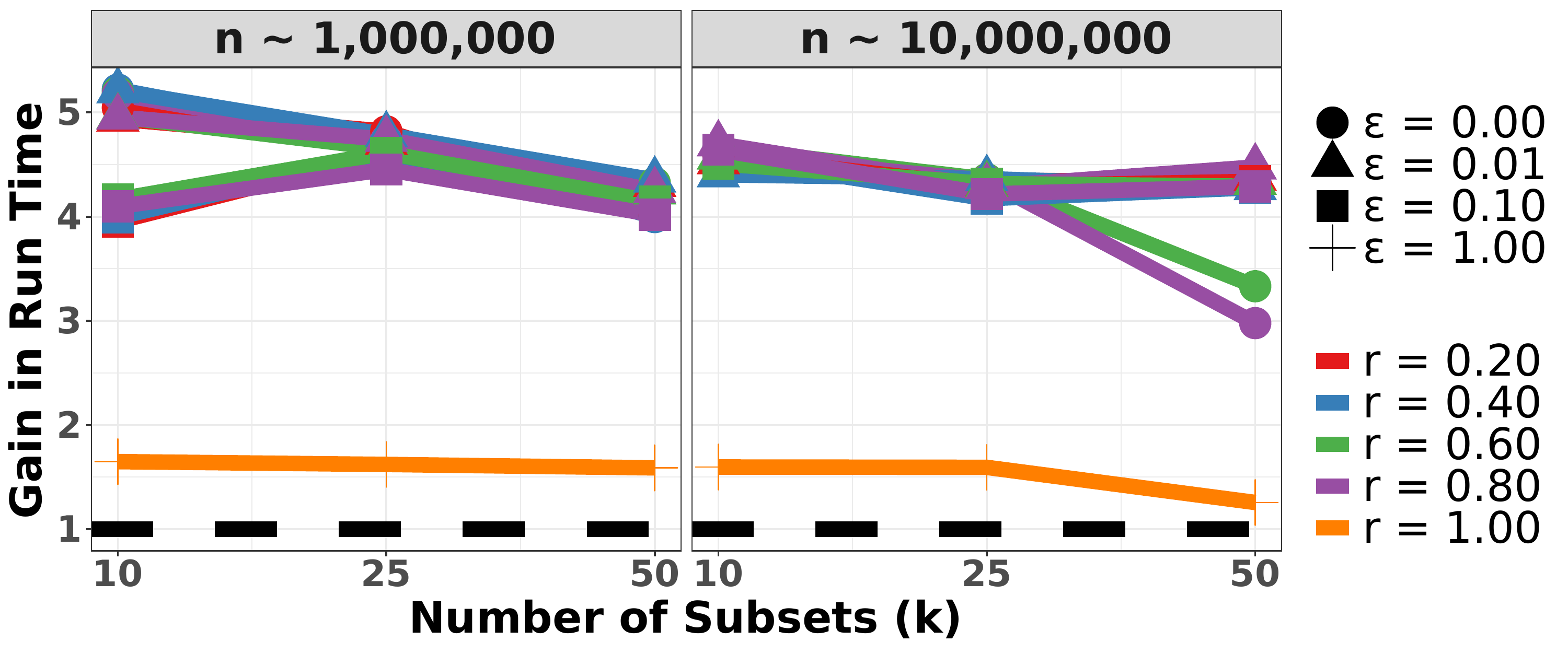}
  \caption{Logistic regression}
  \label{fig:log_mov_time}
\end{subfigure}%
\begin{subfigure}{.5\textwidth}
  \centering
  \includegraphics[scale=0.25]{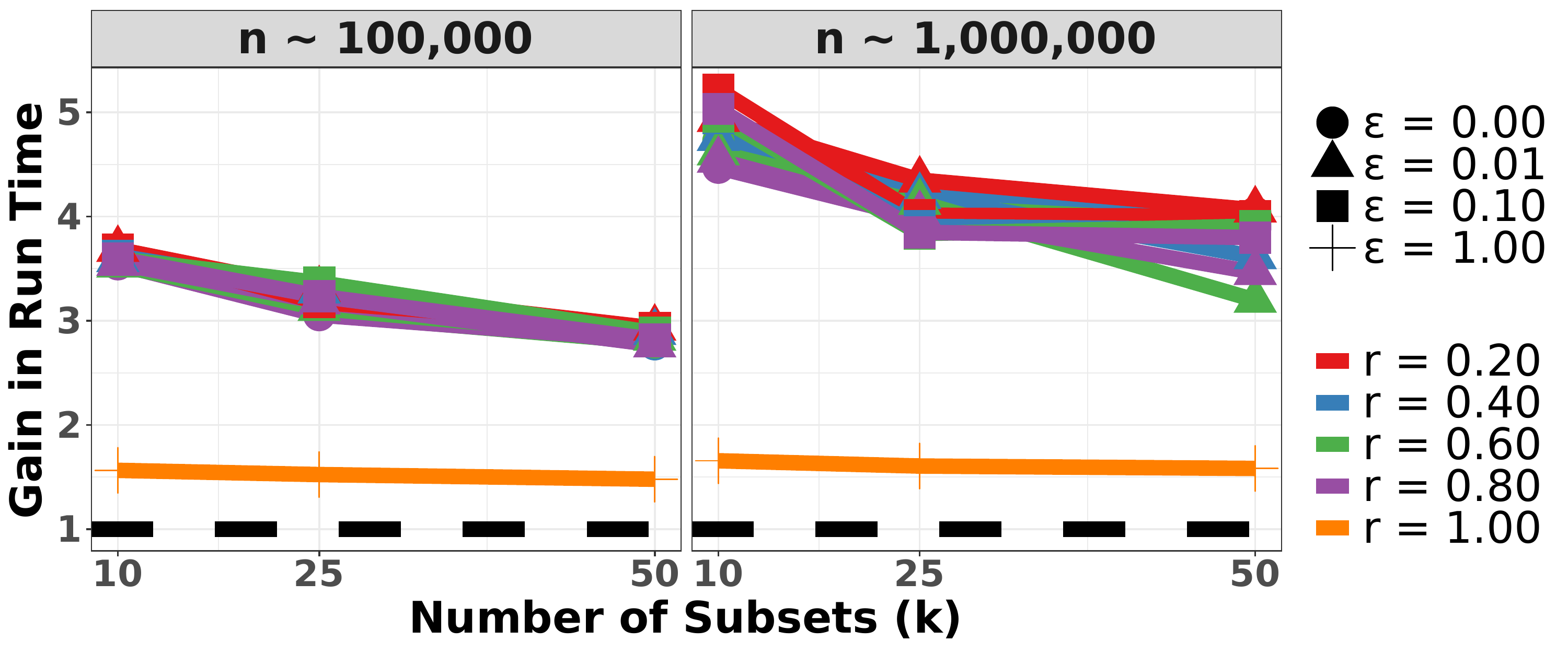}
 \caption{Linear mixed-effects model}
 \label{fig:mov_time}
\end{subfigure}
\caption{Gain in run times for the logistic regression and linear mixed-effects models used in MovieLens data analysis for $r = 0.20, 0.40, 0.60, 0.80, 1.00$ and $k=10, 25, 50$. The distributed (or parallel) implementation of the parent DA corresponds to $r=1.00$ (or $\epsilon = 1.00$). The gain is defined as the ratio of run times of the parent DA and its ADDA-based extension. The dashed black line indicates equal run times for an ADDA algorithm and its parent.}
\label{fig:mov-run-time}
\end{figure}

We then undertake extensive simulation studies for each of the three representative examples to understand and evaluate the performance of the respective ADDA extensions depending on the number of workers $k$ and the update fraction $r$. The two hallmarks of ADDA are distributed and asynchronous processing. Parallel processing by distributing the independent draws of the appropriate sub-blocks of the missing data to the $k$ workers obviously leads to faster computations with no adverse impact on mixing of the Markov chain \citep{Wil18}. The asynchronous updates, however, come with a tradeoff. Smaller values of the fraction $r$ lead to faster computations per iteration, but also slow down the mixing of the Markov chain; therefore, compared to the DA algorithm, the same number of iterations of the ADDA algorithm take much less time but provide less accurate estimates of desired posterior quantities. The idea, as demonstrated by our extensive simulations in Section \ref{sec:experiments}, is that typically the computational gain is quite significant with a comparatively small loss of accuracy. For example, MovieLens ratings data analysis in Section \ref{sec:movi-data-analys} using logistic regression and linear mixed-effects models shows that the ADDA algorithm is anywhere between three to five times faster and only around $2\%$ less accurate than its parent after $10,000$ iterations; see Figures \ref{fig:mov_beta_dmat}--\ref{fig:mov-run-time}.

As mentioned above, the iterates generated by the ADDA algorithm form a Markov chain whose stationary distribution has the target posterior as a marginal. This potentially allows us to directly leverage standard MCMC approaches to quantify the Monte Carlo standard error of the resulting estimators of posterior quantities. All asymptotically consistent estimators of the MCMC standard error, however, rely on the existence of a Markov chain CLT. A standard method for establishing the existence of such a CLT is to show geometric ergodicity; that is, the distribution of the Markov chain converges geometrically fast to the stationary distribution as the number of iterations increase. It is well-known that establishing geometric ergodicity of statistical continuous state space Markov chains is in general very challenging \citep{jones2001honest}. The drift and minorization analysis typically needed for this purpose can be quite involved and ``a matter of art'', requiring the analysis to be heavily adapted to the structure of individual Markov chains.

While geometric ergodicity has been established for the Markov chains corresponding to all the three DA algorithms considered in the paper, adapting these analyses to establish a similar result for the ADDA extensions is not feasible given the complications introduced by the asynchronous updates. The main reason is that the subset of missing data that is updated in any given iteration depends on the current parameter value, which implies that the marginal parameter process of the ADDA chain is not Markov in general. Also, establishing minorization conditions, which are integral parts of the original DA proofs, becomes very challenging. Through a fresh analysis which only uses appropriate drift conditions that are unbounded off compact sets, we establish geometric ergodicity for all the three ADDA extensions in Theorems \ref{pg_thm}, \ref{hdim_varsel}, \ref{thm_GLM}. These results not only provide asymptotically valid standard errors but also assure the user that the ADDA Markov chain converges geometrically fast to its stationary distribution despite the asynchronous updates. Based on our analysis for the three representative examples, we believe that similar results should be true for other ADDA-based generalizations of geometrically ergodic DA algorithms, but such a general treatment is restricted by the aforementioned challenges in theoretical techniques for establishing geometric ergodicity. The proofs of the geometric ergodicity results are provided in a Supplementary Document. We conclude the paper with a discussion in Section \ref{discussion}.

\section{Asynchronous and Distributed Data Augmentation (ADDA)}

\subsection{The General Framework}
\label{sec:adda:-gener-fram}

Consider the setup of a general ADDA algorithm. Assume that we have chosen an $r \in (0, 1)$ and reserved $k$ worker and $1$ manager processes. Let $\theta$ be the model parameter, $\Dcal_{\text{obs}}$ and $\Dcal$ be the observed and missing data, $\Dcal_{\text{aug}}=\{\Dcal_{\text{obs}}, \Dcal\}$ be the augmented data. {Let $\MM$ and $\Theta$ denote the missing data and the parameter spaces, respectively. We assume that both sets are equipped with countably generated $\sigma$-algebras, and respective $\sigma$-finite measures $\mu$ and $\nu$. Let $f(\Dcal, \theta \mid \Dcal_{\text{obs}})$ denote the joint posterior density (with respect to $\mu \times \nu$) of $\Dcal$ and $\theta$, and $f(\Dcal \mid \theta, \Dcal_{\text{obs}})$ and $f( \theta \mid \Dcal, \Dcal_{\text{obs}} )$ denote the conditional posterior densities of the missing data given the parameter, and the parameter given the augmented data, respectively.} Let $\Dcal_{i}$ denote the missing data subset distributed to worker $i$ so that $\Dcal = (\Dcal_1, \ldots, \Dcal_k)$. {We assume that $\Dcal_1, \Dcal_2, \cdots, \Dcal_k$ are conditionally independent given $\Dcal_{\text{obs}}, \theta$ so that  $f(\Dcal \mid \theta, \Dcal_{\text{obs}}) = \prod_{i=1}^k f_i (\Dcal_i \mid \theta, \Dcal_{\text{obs}})$. If the parameter $\theta$ can also be divided into conditionally independent blocks, one could introduce more than one manager process, but for ease of exposition we restrict ourselves to a setting with one manager process.}

For an $r\in (0, 1)$ and an $\epsilon \in (0,1)$, a general implementation of an ADDA-type algorithm is described as follows:
\begin{enumerate}[label={(\arabic*)},ref=(\arabic*)]
\item The manager starts with some initial values ($\Dcal^{(0)}$, $\theta^{(0)}$) at $t=0$ and sends $\theta^{(0)}$ to the workers. For $t=0, 1, \ldots, \infty$, the manager 
  \begin{enumerate}[label={(M-\alph*)},ref=(\alph*)]
  \item waits to receive only an $r$-fraction of  updated 
      $\Dcal_{i}^{(t+1)}$s (see below) from the workers with probability $1-\epsilon$, and with probability $\epsilon$, waits to 
      receive all the updated $\Dcal_i^{(t+1)}$s from the workers;
  \item creates $\Dcal^{(t+1)}$ by replacing {the relevant} $\Dcal_{i}^{(t)}$s with the newly received 
    $\Dcal_{i}^{(t+1)}$;
  \item draws $\theta^{(t+1)}$ from $p( \theta \mid \Dcal^{(t+1)}, \Dcal_{\text{obs}})$; and 
  \item sends $\theta^{(t+1)}$ to all the worker processes and resets $t = t+1$.  
  \end{enumerate}
\item For $t=0, \ldots,\infty$, the worker $i$ ($i=1, \ldots, k$) 
  \begin{enumerate}[label={(W-\alph*)},ref=(\alph*)]
  \item waits to receive $\theta^{(t)}$ from the manager process;
  \item draws $\Dcal_{i}^{(t+1)}$ from $p(\Dcal_{i} \mid \Dcal_{\text{obs}}, \theta^{(t)})$;   and
  \item sends $\Dcal_{i}^{(t+1)}$ to the manager process, resets $t = t+1$, and goes to (W-a) if $\theta^{(t+1)}$ is not received from the manager before the draw is complete; otherwise, it truncates the sampling process, resets $t = t+1$, and goes to (W-b).
  \end{enumerate}
\end{enumerate}

The general steps for an ADDA algorithm are easily summarized into Asynchronous and Distributed (AD) I and P steps. The ADDA algorithm initializes ($\Dcal$, $\theta$) at ($\Dcal^{(0)}$, $\theta^{(0)}$) and the AD-I and AD-P steps in 
the $(t+1)$th iteration for $t=0, 1, \ldots , \infty$ are as follows:
\begin{description}
\item[AD-I step:] Each worker draws $\Dcal_{i}^{(t+1)}$ from $f(\Dcal_{i} \mid \Dcal_{\text{obs}}, \theta^{(t)})$ for $i=1, 
  \ldots, k$ in parallel and sends $\Dcal_{i}^{(t+1)}$ to the manager (if the draw is finished before receiving 
    $\theta^{(t+1)}$).  
\item[AD-P step:] {As soon as the manager receives the required fraction ($r$ with probability $1-\epsilon$ 
    and $1$ with probability $\epsilon$) of updated $\Dcal_{i}^{(t+1)}$ values from the workers, $\theta^{(t+1)}$ is 
    drawn from $f( \theta \mid \Dcal_{1}^{(t+1)}, \ldots,  \Dcal_{k}^{(t+1)}, \Dcal_{\text{obs}})$, and sent to the 
    workers. }
\end{description}
Note that if the $r$-fraction update regime is chosen, then the manager sets $\Dcal_{i}^{t+1} = \Dcal_{i}^t$ for the remaining $(1-r)$-fraction. As soon as the workers receive $\theta^{(t+1)}$, they stop any ongoing activity and proceed with the AD-I step for the $(t+2)$th iteration.

One cycle of ADDA consists of AD-I step followed by AD-P step, and they are repeated sequentially to obtain the $\{(\Dcal^{(t)}, \theta^{(t)}): t = 0, 1, \ldots, \infty\}$ chain for the missing data and parameter. We emphasize again that at the end of iteration $t+1$, with probability $1-\epsilon$, only an $r$-fraction of $\Dcal_{1}^{(t)}, \ldots, \Dcal_{k}^{(t)}$ are replaced by the  manager with new draws received from the workers, whereas the remaining ($1-r$)-fraction of them haven't changed. The AD-I and AD-P in ADDA are distributed generalizations of the I and P steps in its parent DA. If $r=1$, then AD-I and AD-P steps reduce to the I and P steps of their parent DA. 

ADDA inherits the simplicity and stability of its parent DA. Its implementation requires communication between the manager and worker processes that is easily implemented using software packages such as the parallel package in R. The most important advantage of ADDA is that its cycle is faster than the parent DA cycle by a factor of $O(k)$ for every $r$. The main reason is that the $k$ workers perform the AD-I steps in parallel on an $(1/k)$-fraction of the full data, resulting in a $O(k)$-fold speed-up over the I step of parent DA. The P steps of ADDA and its parent DA are identical given the missing data, so their timings are also similar. If $r$ is sufficiently small, then ADDA offers a simple and general strategy for scaling any existing DA-type algorithm to massive data settings by choosing a sufficiently large $k$. The next section carefully investigates crucial theoretical properties of the ADDA algorithm which guarantee that draws from the ADDA algorithm can be used to effectively approximate relevant posterior quantities.

\subsection{Markov Property and Harris Ergodicity} \label{secMH}

The classical (parent) DA algorithm (that is, the ADDA with $r=1$) corresponds to a systematic scan two-block Gibbs sampler from the joint density $f(\Dcal, \theta \mid \Dcal_{\text{obs}})$. Furthermore, in the classical DA setting, $\theta^{(t+1)}$ solely depends on $\Dcal^{(t+1)}$ given $\Dcal^{(t+1)}, \{(\Dcal^{(j)}, \theta^{(j)})\}_{j=0}^{t}$ and 
$\Dcal^{(t+1)}$ depends solely on $\theta^{(t)}$ given $\{(\Dcal^{(j)}, \theta^{(j)})\}_{j=0}^{t}$. This implies that $\theta^{(t+1)}$ solely depends on  $\theta^{(t)}$ given $\{\theta^{(j)}\}_{j=0}^{t}$ so that the marginal $\{\theta^{(t)}\}$ process is Markov. This fact is often useful when establishing theoretical properties such as geometric ergodicity of the DA Markov chain (see below).

The ADDA can be interpreted as a {\it hybrid Gibbs sampler}. The AD-P step is a systematic scan step for sampling the parameter $\theta$ from its conditional posterior distribution given $\Dcal$, but the AD-I step is a ``random subset scan" step which updates a random subset of the missing data $\Dcal = (\Dcal_1, \cdots, \Dcal_k)$. Other hybrid versions of systematic and random scan Gibbs steps have been considered in the literature. \citet{backlund2018hybrid} consider a DA setup where the {\it parameter} $\theta$ is partitioned into two blocks. They construct and study a hybrid sampler that performs a systematic scan step for the entire missing data $\Dcal$
at each iteration and updates exactly one of the two parameter blocks with fixed probabilities $s$ and $1-s$, respectively. A general (non-DA) setting with multiple blocks is considered in \cite{Levine:2005}, where a sampler updates a single coordinate in each iteration. The choice of the block to update is made randomly based on a vector of fixed probabilities which depend on the block index that was updated in the previous iteration; therefore, this strategy subsumes the traditional systematic and random scan samplers as special cases. The ADDA is significantly different from the above strategies and updates a randomly chosen {\it subset} of the conditionally independent blocks of the missing data $\Dcal$, while performing a systematic scan update of the entire parameter $\theta$. Note also that the probabilities of choosing various relevant subsets at a given iteration can in general depend on the {\it current value} of the parameter $\theta$.

A simple observation shows that $\left\{ \Dcal^{(t)}, \theta^{(t)}  \right\}_{t=0}^\infty$ process in ADDA is Markov. For an ADDA algorithm with $r \in (0, 1)$, the knowledge of $(\Dcal^{(t)}, \theta^{(t)})$ makes the entire past irrelevant for generating $(\Dcal^{(t+1)}, \theta^{(t+1)})$. Specifically, $(\Dcal^{(t+1)}, \theta^{(t+1)})$ is conditionally independent of $\left\{ (\Dcal^{(j)}, \theta^{(j)}) \right\}_{j=0}^{t-1}$ given $(\Dcal^{(t)}, \theta^{(t)})$; therefore, the process $\left\{ \Dcal^{(t)}, \theta^{(t)} \right\}_{t=0}^\infty$ is still Markov. It is also easily seen that both the AD-P step and the AD-I step leave $f(\Dcal, \theta \mid  \Dcal_{\text{obs}})$ invariant. This is promising but to use the ADDA iterates to approximate relevant posterior expectations  and quantiles, we need to establish Harris ergodicity. The following theorem shows that the ADDA Markov chain is Harris ergodic as long as $\epsilon$ is positive, the original DA chain is Harris ergodic, and its transition kernel satisfies a mild absolute continuity condition.
\begin{thm} \label{Harris}
  Let $K_{DA}$ denote the transition kernel of the parent DA chain which is equivalent to the ADDA chain with 
  $r = 1$, and $\Pi \left( \cdot \mid \Dcal_{obs} \right)$ denote the joint posterior distribution of $\Dcal$ and 
  $\theta$. Assume that $K_{DA} ((\Dcal, \theta), \cdot)$ is absolutely continuous with respect to $\Pi$ for 
  every $(\Dcal, \theta)$. Then, if $\epsilon>0$ and the parent DA chain is Harris ergodic, the ADDA chain is 
  Harris ergodic.
\end{thm}

The proof of the theorem is given in Supplementary Section \ref{prove_harris}. The assumption $\epsilon > 0$ is needed because it ensures that there is a positive probability that each worker returns an updated value at each iteration. This allows us to leverage the Harris ergodicity of the parent DA chain to establish relevant properties of the ADDA chain. If $\epsilon = 0$, then assumptions regarding the computer architecture and the processing power of the workers will need to be made to guarantee Harris ergodicity; for example, see \citet{Teretal20}. This can get quite messy and tricky to ensure in real-life computing. Using $\epsilon > 0$ is an elegant and clean way of side-stepping these issues. Of course, smaller values of epsilon are preferred for faster computations.

Theorem \ref{Harris} is a fundamental step towards understanding theoretical properties of the $\{ (\Dcal^{(t)}, \theta^{(t)})\}$ process in ADDA but is still far from the end. An important next step is to establish a CLT for the  $\{ (\Dcal^{(t)}, \theta^{(t)})\}$ process in order to provide asymptotically valid standard error bounds for MCMC based approximations. As discussed before, the standard method of establishing a CLT is to establish geometric ergodicity; however, proving geometric ergodicity can be quite challenging in general. The proof of geometric ergodicity for a large majority of statistical Markov chains is not available. For Markov chains where a proof is available, it is typically based on drift and minorization conditions, which are specifically tailored for the particular chain at hand.

In the next section, we illustrate the application of ADDA on logistic regression, high-dimensional variable selection, and linear mixed-effects modeling. For the three examples, the $\{(\Dcal^{(t)},\theta^{(t)})\}$ process in the parent DA is known to be geometrically ergodic. Extension of each of these proofs to ADDA is difficult for at least one of two main reasons. First, each cycle of ADDA updates (with $1-\epsilon$ probability) a random $r$-fraction of the $\Dcal_1, \ldots, \Dcal_k$. This random subset update significantly complicates the transition density in terms of establishing a minorization condition involved in the proof of geometric ergodicity. Second, some of the parent DA geometric ergodicity proofs focus on the marginal process $\{\theta^{(t)}\}$, which is a Markov chain in the parent DA setting, and extend the result to the full process  $\{ (\Dcal^{(t)}, \theta^{(t)})\}$. This strategy fails in the ADDA setting because the $\{\theta^{(t)}\}$ process is not in general  Markov. We provide results establishing geometric ergodicity for the ADDA chains in the three above-mentioned models in 
the next section. A key idea in the three proofs is to identify and establish geometric drift conditions, where the drift functions are unbounded off compact sets. Some of these proofs are established under weaker conditions than those required for the parent DA chains.

\section{Applications}
\subsection{ADDA for Bayesian Logistic Regression}
\label{sec:adda-logist-reg}

Consider the implementation of P\'olya-Gamma DA for Bayesian logistic regression \citep{Poletal13}. Let $n$ be the sample 
size, $\beta = (\beta_1, \ldots, \beta_p)^\T \in \RR^p$ be the regression coefficients, and $y_i, s_i, x_i = (x_{i1}, \ldots, 
x_{ip})^\T$ be the response, number of trials, covariates for sample $i$, respectively, where $y_i \in \{0, 1, \ldots, s_i\}$, $s_i 
\in \NN$, $x_i \in \RR^p$. The hierarchical model for logistic regression is
\begin{align}
\label{eq:l1}
y_i \mid \beta \overset{\text{ind.}}{\sim} \text{Binomial}\{s_i, 1 / (1 + e^{-\psi_i})\}, \quad \psi_i = x_i^\T \beta,  \quad i = 1, 
\ldots, n, \quad \beta \sim N(\mu_{\beta}, \Sigma_{\beta}).
\end{align}
The DA algorithm augments the model in \eqref{eq:l1} with $n$ P\'olya-Gamma random variables $\omega_1, \ldots, \omega_n$ specific to the $n$ samples, and cycles between the I and P steps for a large number of iterations starting from a given value of $\beta$:
\begin{enumerate}[label={PG.\arabic*},ref=PG.\arabic*]
	\item \label{logs1} (I step) Draw $\omega_i$ given $\beta$ from $\text{PG}$($s_i$, $|x_i^\T \beta|$) for $i = 1, \ldots, n$, where PG is the P\'olya-Gamma distribution.   
	\item \label{logs2} (P step) Draw $\beta$ given $\omega_1, \ldots, \omega_n$ and $y_1, \ldots, y_n$ from $N$($m_{\omega}$, $V_{\omega}$), where $V_{\omega} = (X^\T \Omega X + \Sigma_{\beta}^{-1})^{-1}$, $m_{\omega} = V_{\omega} (X^\T \kappa + \Sigma_{\beta}^{-1} \mu_{\beta})$, $\kappa = (y_1 - s_1 / 2, \ldots, y_n - s_n / 2)^\T$, and $\Omega$ is the diagonal matrix of $\omega_i$s. 
\end{enumerate}
The marginal Markov chain $\{\beta^{(t)}\}$ of the $\beta$ draws collected in step \ref{logs2} has the posterior distribution of $\beta$ in \eqref{eq:l1} as its invariant distribution \citep{choi2013polya,WanRoy18,WanRoy18b}.

The ADDA algorithm with P\'olya-Gamma DA algorithm as its parent DA modifies step \ref{logs1}. Let $s= (s_1, \ldots, s_n)$ and $\omega = \{\omega_1, \ldots, \omega_n\}$. Then, following the notation of Section \ref{sec:adda:-gener-fram}, the P\'olya-Gamma DA has 
\begin{align*}
\theta = \{\beta\}, \quad  \Dcal_{\text{obs}} = \{y, s, X\}, \quad \Dcal = \{\omega\}, \quad \Dcal_{\text{aug}} = \{(y_i, s_i, x_i, \omega_i): i = 1, \ldots, n\}. 
\end{align*}
{For many datasets, the sample size $n$ can be really large. For example, the MovieLens data analyzed in Section \ref{sec:movi-data-analys} has 10 million samples.} Instead of updating $n$ $\omega_i$s in every iteration, ADDA updates only an $r$ fraction of them (with probability $1-\epsilon$) and reduces to its parent DA when $r=1$. The ADDA algorithm runs after we randomly split the $n$ samples into $k$ disjoint subsets and store them on $k$ worker processes. Let $n_i$ be the number of samples assigned to worker $i$, $(y_{j(i)}, s_{j(i)}, x_{j(i)}, \omega_{j(i)})$ be the $j$th augmented sample on worker $i$ ($j=1, \ldots, n_i$) and $\Dcal_{i} = \{\omega_{1(i)}, \ldots, \omega_{n_i(i)}\}$. Note that $\omega_{j(i)}$s in 
$\Dcal_{i}$ are rearranged to match the ordering of samples in $\Dcal_{\text{obs}}$ so that $\Dcal = \omega = \{\omega_1, 
\ldots, \omega_n\}$. 

{The ADDA extension of P\'olya-Gamma DA for a given $k$ and $r$ initializes ($\omega$, $\beta$) at  ($\omega^{(0)}$, $\beta^{(0)}$) and the AD-I and AD-P steps in the $(t+1)$th iteration are as follows:}
\begin{algorithm}[H] 
	\caption{AD P\'olya-Gamma DA} 
	\label{ADPG}
	\begin{algorithmic}[1]	
		\Statex \label{logas1} \textbf{(AD-I step)} Draw $\omega^{(t+1)}_{j(i)}$ given $\beta^{(t)}$ from 
		$\text{PG}$($s_{j(i)}$, $|x_{j(i)}^T \beta^{(t)}|$) on worker $i$ for $j = 1, \ldots, n_i$ and send $\Dcal_{i}^{(t+1)} = 
		\left\{ \omega_{1(i)}^{(t+1)}, \ldots, \omega_{n_i(i)}^{(t+1)} \right\}$ to the 
		manager process if these draws are finished before receiving $\beta^{(t+1)}$ ($i=1, \ldots, k$).  	
		\Statex \label{logas2} \textbf{(AD-P step)} {{With probability $1-\epsilon$, wait to receive updated 
		$\Dcal_{i}^{(t+1)}$ values from an $r$-fraction of workers, set $\Dcal_{i}^{(t+1)} = \Dcal_{i}^{(t)}$ for the remaining 
		$(1-r)$-fraction of workers. Then, draw $\beta^{(t+1)}$ given $\omega^{(t+1)}$ and $\Dcal_{\text{obs}}$ from $N$
		($m_{\omega^{(t+1)}}$,$V_{\omega^{(t+1)}}$), where $\Omega^{(t+1)}=\diag(\omega^{(t+1)})$, 
		$V_{\omega^{(t+1)}} = (X^\top\Omega^{(t+1)} X + \Sigma_{\beta}^{-1})^{-1}$, and $m_{\omega^{(t+1)}} = 
		V_{\omega^{(t+1)}} (X^T \kappa + \Sigma_{\beta}^{-1} \mu_{\beta})$. With probability $\epsilon$, wait to receive 
		updated $\Dcal_{i}^{(t+1)}$ values from all the workers, and then proceed to sampling $\beta^{(t+1)}$ given 
		$\omega^{(t+1)}$ and $\Dcal_{\text{obs}}$.}} 
	\end{algorithmic} 
\end{algorithm}

\noindent The ADDA cycles generate the Markov chain  $\Phi_{\text{ADPG}} = \{ \omega^{(t)}, \beta^{(t)}\}_{t=0}^{\infty}$. The following Theorem establishes the geometric ergodicity of $\Phi_{\text{ADPG}}$ when $\epsilon>0$ and its proof is  in Supplementary Materials Section \ref{prove_pg}.

\begin{thm}\label{pg_thm}
	If $\epsilon>0$, the Markov chain $\Phi_{\text{ADPG}}$ as described in Algorithm \ref{ADPG} is geometrically ergodic. 
\end{thm}

The convergence analysis of the parent PG DA chain has been performed in the literature. In particular, \citet{choi2013polya} show that the parent PG DA chain is uniformly ergodic (using the marginal $\beta^{(t)}$ chain) if proper priors are used. Following this work, \citet{WanRoy18} show the geometric ergodicity of the PG DA chain with flat priors. The uniform ergodicity proof for the parent DA chain in \citet{choi2013polya} for the binary and proper prior setting is based on a minorization argument on the marginal $\{\beta^{(t)}\}$ chain; however, this proof strategy fails for ADDA because the $\{\beta^{(t)}\}$ process for $\Phi_{ADPG}$ is not Markov. As a result, we take a different approach where we establish a geometric drift condition using a function of $(\beta, \omega)$, which is unbounded off compact sets. \citet{WanRoy18} follow a similar strategy for the  proving geometric ergodicity of the parent DA chain with improper priors, but they focus on the marginal $\omega$ chain instead of the $(\beta, \omega)$ chain. Furthermore, the previous two works consider the parent DA chain when the response is binary, whereas Theorem \ref{pg_thm} deals with the ADDA chain in a more general binomial setting.

\subsection{ADDA for Bayesian High Dimensional Variable Selection}
\label{sec:adda-Lasso}

Consider Bayesian high dimensional variable selection using a Bayesian lasso shrinkage prior \cite{rajaratnam2015scalable}. Let $n$ be the sample size, $\beta = (\beta_1, \ldots, \beta_p)$ be the regression coefficients, $y=(y_1, \ldots, y_n) \in \mathbb{R}^n$ be the response vector, and $X \in \RR^{n\times p}$ be the design matrix. The Bayesian lasso DA algorithm augments dimension $j$ with local shrinkage parameter $\tau_j$ ($j=1, \ldots, p$), and the hierarchical model for variable selection is
\begin{align}\label{eq:v1}
y &\mid \beta,\sigma^2 \sim N( X\beta, \sigma^2 I_n), \quad \beta \mid \sigma^2, \tau \sim N(0, \sigma^2 D_{\tau}), \quad \tau = (\tau_1, \ldots, \tau_p), \; D_{\tau} = \diag(\tau), \nonumber\\
\sigma^2 &\sim \text{Inverse-Gamma}(\alpha, b), \quad \tau_j \overset{\text{ind.}}{\sim} \text{Exponential}(\lambda^2/2), \quad j = 1, \ldots, p,
\end{align}
where $I_n$ is an $n \times n$ identity matrix, $\sigma^2>0$ is the variance of idiosyncratic error term, $\tau = (\tau_1, \ldots, \tau_p)$, and $\alpha, b, \lambda$ are hyper-parameters. The DA algorithm for variable selection starts with some initial values of $\beta$ and $\sigma^2$ and cycles between the I and P steps for a large number of iterations:
\begin{enumerate}[label={BL.\arabic*},ref=BL.\arabic*]
	\item \label{bvss1} (I step) Draw $\tau_j$ given $\beta, \sigma^2$ from $\text{Inverse-Gaussian}$($ \tfrac{|\lambda| |\sigma|} {|\beta_j|}, \lambda^2$)
	for $j = 1, \ldots, p$.   
	\item \label{bvss2} (P step) Draw $(\beta, \sigma^2)$ given $\tau$ and $y$ as
	\begin{align}    \label{2block}
	\sigma^2 &\mid \tau, y \sim \text{Inverse-Gamma} \left( \frac{n}{2} + \alpha, \frac{y^\top (I - X A_\tau^{-1} X^\top) y + 2b}{2} \right), \quad A_{\tau} = X^\top X+D_{\tau}^{-1}, \nonumber\\ 
	\beta &\mid \sigma^2, \tau, y \sim N \left( A_{\tau}^{-1}X^\top y, \sigma^ 2A_\tau^{-1} \right).
	\end{align}
\end{enumerate}
The Markov chain $\{\beta^{(t)}, \sigma^{2(t)}\}$ of the $(\beta, \sigma^2)$ draws collected in step \ref{bvss2} has the posterior distribution of $(\beta, \sigma^2)$ in \eqref{eq:v1} as its invariant distribution \citep[Lemma 1]{rajaratnam2015scalable}.

This DA is different from P\'olya-Gamma DA in that the I step in \ref{bvss1} draws $p$ latent variables instead of $n$. In high-dimensional problems, $n \ll p$ and updating $\tau_1, \ldots, \tau_p$ in every iteration is time consuming even when $n$ is small. Following Section \ref{sec:adda:-gener-fram}, the Bayesian lasso DA has 
\begin{align*}
\theta = \{\beta, \sigma^2\}, \quad  \Dcal_{\text{obs}} = \{y, X\}, \quad \Dcal = \{\tau \}, \quad 
\Dcal_{\text{aug}} = \{(y_i, x_{ij}, \tau_j): i = 1, \ldots, n; j=1, \ldots p\}.   
\end{align*}
Unlike P\'olya-Gamma DA, the computational bottleneck in the I step happens when $p$ is large. The ADDA algorithm starts by partitioning $\{1, \ldots, p\}$  into $k$ disjoint subsets denoted as $\Pcal_1, \ldots, \Pcal_k$. The elements of $\tau$ and $\beta$ are stored on $k$ worker processes. Let $p_i = |\Pcal_i|$,  $((\beta_{j(i)}, \tau_{j(i)}): j \in \Pcal_i)$ be the
$\beta$ and $\tau$ elements on worker $i$, and $\Dcal_i = \{\tau_{1(i)}, \ldots, \tau_{p_i(i)}\}$ ($i=1, \ldots, k$). The disjoint partitioning ensures that $p_1 + \cdots + p_k = p$. Note that the $\tau_{j(i)}$'s are rearranged so that $\Dcal = \tau = 
\{\tau_1, \ldots, \tau_p\}$

The ADDA extension of Bayesian lasso DA, at any iteration, updates (with probability $1-\epsilon$) only an $r$ fraction of 
$\Dcal_{1}, \ldots, \Dcal_{k}$, and with probability $\epsilon$ updates all of $\Dcal_{1}, \ldots, \Dcal_{k}$. Clearly, the ADDA 
extension reduces to the parent DA when $r=1$. For a given $k$ and $r$, the ADDA extension of the Bayesian Lasso DA 
initializes $(\tau, \beta, \sigma^{2})$ at $(\tau^{(0)}, \beta^{(0)}, \sigma^{2(0)})$ and the AD-I and AD-P steps in the $(t+1)$th 
iteration are as follows: 


\begin{algorithm}[H] 
	\caption{AD Bayesian Lasso DA} 
	\label{ADBL}
	\begin{algorithmic}[1]
		\Statex \label{bvsas1} \textbf{(AD-I step)} Draw $\tau_{j(i)}^{(t+1)}$ given $\beta^{(t)}, \sigma^{2(t)}$ from 
		$\text{Inverse-Gaussian}$($ \tfrac{|\lambda| |\sigma^{(t)}|} {|\beta_{j(i)}^{(t)}|}, \lambda^2$)
		on worker $i$ for $j = 1, \ldots, p_i$ and send $\Dcal_{i}^{(t+1)} = \left\{ \tau_{1(i)}^{(t+1)}, \ldots, 
		\tau_{p_i(i)}^{(t+1)} \right\}$ to the manager process if these draws are finished before receiving 
		$\beta^{(t+1)}, \sigma^{2(t+1)}$ ($i=1, \ldots, k$).  	
		\Statex \label{bvsas2} \textbf{(AD-P step)} {With probability $1-\epsilon$, wait to receive updated 
		$\Dcal_{i}^{(t+1)}$ values from an $r$-fraction of workers, set $\Dcal_{i}^{(t+1)} = \Dcal_{i}^{(t)}$ for the remaining 
		$(1-r)$-fraction of workers}. Then draw $(\beta^{(t+1)}, \sigma^{2(t+1)})$ given $\Dcal_{1}^{(t+1)}, \ldots, 
		\Dcal_{k}^{(t+1)}$ and $\Dcal_{\text{obs}}$ as
		\begin{itemize}
			\item $\sigma^{2(t+1)}$ given $\tau^{(t+1)}, y$ from Inverse-Gamma $\left( \tfrac{n}{2} + \alpha, \frac{y^\top 
			(I - X A_{\tau^{(t+1)}}^{-1} X^\top) y + 2b}{2} \right)$, where $D_{\tau^{(t+1)}} = \diag(\tau^{(t+1)})$, and 
			$A_{\tau^{(t+1)}} = X^\top X+D_{\tau^{(t+1)}}^{-1}$; and 
			\item $\beta^{(t+1)}$ given $\sigma^{2(t+1)}, \tau^{(t+1)}, y$ from N$\left( A_{\tau^{(t+1)}}^{-1}X^\top y, 
			\sigma^{2(t+1)} A_{\tau^{(t+1)}}^{-1} \right)$.           
		\end{itemize}
		
		\noindent
		With probability $\epsilon$, wait to receive updated $\Dcal_{i}^{(t+1)}$ values from all the workers, and then 
		proceed to sampling $\beta^{(t+1)}, \sigma^{2(t+1)}$ given $\Dcal_{1}^{(t+1)}, \ldots, 
		\Dcal_{k}^{(t+1)}$ and $\Dcal_{\text{obs}}$. 
	\end{algorithmic} 
\end{algorithm}

\noindent Similar to the P\'olya-Gamma DA extension, this ADDA cycles generate the Markov chain $\Phi_{\text{ADBL}} = 
\{\tau^{(t)}, \beta^{(t)}, \sigma^{2(t)} \}_{t=0}^{\infty}$. The following theorem establishes the geometric ergodicity of 
$\Phi_{\text{ADBL}}$ when $\epsilon>0$, and the proof is given in Supplementary Materials Section \ref{prove_BL}.


\begin{thm}\label{hdim_varsel}
	If $\epsilon>0$ and $n \geq 3$, the Markov chain $\Phi_{ADBL}$  described in Algorithm \ref{ADBL} is geometrically ergodic.
\end{thm}
\noindent Geometric ergodicity of a three block version of the parent DA chain was established in \cite{khare2013geometric}, and geometric ergodicity for the parent DA chain described in BL.1 and BL.2 above was established in \cite{rajaratnam2019uncertainty}. Both these proofs make use of drift {\it and} minorization. In particular, \cite{rajaratnam2019uncertainty} prove results on the marginal $(\beta, \sigma^2)$ chain, and leverage them to establish geometric ergodicity of the parent DA chain. As mentioned previously, this route is not available to us, as the marginal $(\beta, \sigma^2)$ is not Markov, and establishing minorization with the asynchronous updates is nontrivial. As such, similar to the Polya-Gamma ADDA setting, our proof of Theorem \ref{hdim_varsel} is based on establishing a geometric drift condition using a drift function that is unbounded off compact sets.

\subsection{ADDA for Linear Mixed-Effects Modeling}
\label{sec:adda-Mixed-Effects}

Consider the setup for linear mixed-effects models \citep{DykMen01}. Let $n$ be the number of observations, $m$ be the number of subjects, $(y_i, X_i, Z_i)$ be the response, fixed effects covariates, and random effects covariates, respectively, for subject $i$, $p$ and $q$ be the number of fixed and random effects, $\beta \in \RR^p$ be the fixed effect, and $\Sigma$ be a $q \times q$ positive definite covariance matrix. Then, the linear 
mixed-effects model assumes that 
\begin{align}
\label{eq:mix1}
y_i = X_i \beta + Z_i b_i + e_i, \quad b_i \sim N(0, \Sigma), \quad e_i \sim N(0, \sigma^2 I), \quad b_i \perp e_i, \quad i=1, \ldots, m,
\end{align}
where $b_i \in \RR^q$ and $e_i \in \RR^{n_i}$ are the random effect and idiosyncratic error for subject $i$, $b_i$ and $e_i$ are mutually independent, $\Sigma$ is the covariance matrix of random effects, $\sigma^2$ is the error variance, $I$ is an identity matrix, $y_i \in \RR^{n_i}$, $n = \sum_{i=1}^m n_i$, and the dimensions of $X_i$, $Z_i$, $e_i$, $I$ are chosen appropriately. The $b_i$s are unobserved and the interest lies in inference on $\theta = \{\beta, \Sigma, \sigma^2\}$.

The DA algorithm for posterior inference on $\theta$ is based on modified random effects. Let $\Gamma$ be a non-singular 
matrix, $\gamma = \text{vec}(\Gamma)$ stacks the columns of $\Gamma$ into a $q^2$-dimensional vector, $\tilde p = p + 
q^2$, $y^\T = (y_1^\T, \ldots, y_m^\T) \in \RR^n$, $\tilde \Sigma = \Gamma^{-1} \Sigma \Gamma^{-1}$, $\alpha = (\beta, 
\gamma)$, and $\tilde X \in \RR^{n \times \tilde p}$  be the matrix with $[X_i \; (d_i^\T \otimes Z_i)]$ as its $i$th row block,  where $d_i = \Gamma^{-1} b_i$ for $i=1, \ldots, m$. The prior distributions on $(\sigma^2, \alpha)$ and $\tilde{\Sigma}$ are assigned as 
\begin{align}\label{ab-prior}
  \sigma^2 \sim \frac{M}{\chi^{2}_{a}}, \quad \alpha \mid \sigma^{2} \sim N(0,\sigma^{2}V_{\alpha}^{-1}), \quad  \tilde{\Sigma} \sim \text{IW}(W,s), \quad M> 0, \; a>0, \; s>q+1,
\end{align}
where $V_{\alpha}\succ 0, W\succ 0$, $\succ 0$ denotes a symmetric positive definite matrix, and  IW stands for inverse Wishart distribution.
Then, the DA algorithm for posterior inference in linear mixed-effects modeling starts with some given value of $\theta$ and repeats the following I and P steps for a large number of iterations: 
\begin{enumerate}[label={LM.\arabic*},ref=LM.\arabic*]
	\item \label{ms1} (I step) Draw $d_i$ given $\theta$ and $y_i$ from Normal$(m_{d_i}, V_{d_i})$ for $i=1, \ldots, m$, where
	\begin{align}
	\label{eq:mixp}
	m_{d_i} &= \tilde \Sigma \Gamma^\T Z_{i}^\T \left(  Z_{i} \Gamma \tilde \Sigma \Gamma^\T Z_{i}^\T  + \sigma^{2} I \right)^{-1} (y_{i} - X_{i} \beta), \nonumber\\ 
	V_{d_i} &=  \tilde \Sigma  - \tilde \Sigma \Gamma^{\T} Z_{i}^\T \left(  Z_{i} \Gamma \tilde \Sigma \Gamma^\T Z_{i}^\T  + \sigma^{2} I \right)^{-1} Z_{i} \Gamma \tilde \Sigma.                                   
	\end{align}
	\item \label{ms2} (P step) Draw $\Sigma, \sigma^2$ and $\alpha = (\beta, \gamma)$ given $d_1, \ldots, d_m, y_1, 
	\ldots, y_m$ in a sequence of three steps:  
	\begin{eqnarray}
	& & \tilde \Sigma^{-1} \sim \text{Wishart}_{m{+s} - q}\left\{\left( \sum_{i=1}^m d_i d_i^\T  {+W} \right)^{-1} \right\}, \quad 
	\sigma^2 \sim \frac{\| y - \hat y \|^2_2 {+M}}{\chi^2_{n +a- p - q^2}}, \nonumber\\
	& & \alpha \mid \sigma^2 \sim N \{ \hat \alpha, \sigma^2 (\tilde X^\T \tilde X { + V_{\alpha}})^{-1}\}, \label{eq:mixi}
	\end{eqnarray}
	where $\hat \alpha =(\tilde X^\T \tilde X { + V_{\alpha}})^{-1} \tilde X^\T y$, $\hat y = \tilde X \hat \alpha$, $\Gamma=\text{unvec}(\gamma)$, and $\Sigma = \Gamma \tilde \Sigma^{-1} \Gamma^\T$.
\end{enumerate}

This DA is different from the previous two DA algorithms in Sections \ref{sec:adda-logist-reg} and \ref{sec:adda-Lasso} due 
to the presence of random effects. Following Section \ref{sec:adda:-gener-fram}, we have
\begin{align*} 
\theta = \{\beta, \Gamma, \Sigma, \sigma^2\}, \quad  \Dcal_{\text{obs}} = \{y, \tilde X\}, \quad \Dcal = \{d_1, \ldots, d_m\}, \quad 
\Dcal_{\text{aug}} = \{(y_i, X_i, Z_i, d_i): i = 1, \ldots, m\},   
\end{align*}
The main computational bottleneck here is the repeated sampling of $d_1, \ldots, d_m$ in the I step \ref{ms1} when $m$ is large. For example, the MovieLens data analyzed 
in Section \ref{sec:movi-data-analys}  has $m \approx 72,000$. The ADDA-based extension of this algorithm starts by partitioning $\{1, \ldots, m\}$  into $k$ disjoint subsets denoted as $\Pcal_1, \ldots, \Pcal_k$. Let $m_i = |\Pcal_i|$, $\{(y_{j(i)}, X_{j(i)}, Z_{j(i)}, d_{j(i)}): j \in \Pcal_i\}$ be the augmented sample on worker $i$, and $\tilde X_{j(i)} = [X_{j(i)} \; (d_{j(i)}^\T \otimes Z_{j(i)})]$  ($j=1, \ldots, m_i$). Note that $m_1 + \cdots + m_k = m$ due to the the disjoint partitioning.  Define the statistics $S_{dd (i)}, S_{xx(i)}, S_{xy(i)}$ for worker $i$ as \begin{align}\label{lme-suff-stat}
S_{dd (i)} = \sum_{j=1}^{m_i} d_{j(i)} d_{j(i)}^\T , \quad S_{xx (i)} = \sum_{j=1}^{m_i} \tilde X_{j(i)}^\T \tilde X_{j(i)}, \quad S_{xy(i)} =\sum_{j=1}^{m_i} \tilde X_{j(i)}^\T y_{j(i)}.
\end{align}
Then, the augmented data sufficient statistics $\{\tilde X_{1(i)}, \ldots, \tilde X_{m_i(i)}, S_{dd (i)}, S_{xx(i)}, S_{xy(i)}\}$ for worker $i$ play an important role in both the DA and ADDA algorithms. In particular, $\hat \alpha$  and $\hat y$ in P step of the parent 
DA in \eqref{eq:mixi} become 
\begin{align}
\hat \alpha = (S_{XX}{+V_{\alpha}})^{-1} S_{XY}, \quad \hat y = \tilde X \hat \alpha, \label{eq:upd-lme2}
\end{align}
\noindent
where 
\begin{align}
\quad S_{XX} = \sum_{i=1}^k S_{xx}^{i}, \quad S_{XY} = \sum_{i=1}^k S_{xy}^{i}, \quad S_{DD} = \sum_{i=1}^k S_{dd (i)} 
{+W}. \label{eq:upd-lme2.1}
\end{align}
\noindent
Here, 
$S_{DD}^{-1} = \left( \sum_{i=1}^k S_{dd (i)} {+W} \right)^{-1}$ is the scale parameter of the Wishart density in 
\eqref{eq:mixi}. Note that $\tilde X$ is obtained by arranging $\tilde X_{1(i)}, \ldots, \tilde X_{m_i(i)}$ ($i=1, \ldots, k$) in the 
order of samples in $\tilde X$ and binding them along the rows. 

The ADDA extension of the linear mixed effects DA, at any iteration, updates (with probability $1-\epsilon$) only an $r$ 
fraction of $\Dcal_{1}, \ldots, \Dcal_{k}$, and with probability $\epsilon$ updates all of $\Dcal_{1}, \ldots, \Dcal_{k}$. Clearly, 
the ADDA  extension reduces to the parent DA in steps \ref{ms1}--\ref{ms2} when $r=1$. For a given $k$ and $r$, the ADDA 
extension of the marginal DA initializes $(d_1, d_2, \cdots, d_m, \alpha, \Sigma, \sigma^{2})$ at $(d_1^{(0)}, \ldots, 
d^{(0)}_m, \alpha^{(0)}, \Sigma^{(0)}, \sigma^{2(0)})$ and the AD-I and AD-P steps in the $(t+1)$th iteration are as follows: \
\begin{algorithm}[H]
	\caption{AD Linear Mixed-Effects Model DA} 
	\label{ADGLM}
	\begin{algorithmic}[1]    
		\State \label{mixas1} (AD-I step) Draw $d_{j(i)}^{(t+1)}$ given $\theta^{(t)} = (\alpha^{(t)}, \Sigma^{(t)}, \sigma^{2(t)})$ and $y_{j(i)}$ from $N(m_{d_{j(i)}}, V_{d_{j(i)}})$ on worker $i$ for $j = 1, \ldots, m_i$, where $m_{d_{j(i)}}, V_{d_{j(i)}}$ are obtained by replacing $(y_{i}, X_{i}, Z_{i})$  in \eqref{eq:mixp} with $(y_{j(i)}, X_{j(i)}, Z_{j(i)})$. Compute $\Dcal_i^{(t+1)}$ from the sampled $d_{j(i)}^{(t+1)}$ values and send $\Dcal_i^{(t+1)}$ to the manager process if these draws are finished before receiving $\theta^{(t+1)} = (\alpha^{(t+1)}, \Sigma^{(t+1)}, \sigma^{2(t+1)})$ ($i=1, \ldots, k$).  
		
		\State \label{mixas2} (AD-P step) {With probability $1-\epsilon$, wait to receive updated $\Dcal_{i}^{(t+1)}$ values 
		from an $r$-fraction of workers, set $\Dcal_{i}^{(t+1)} = \Dcal_{i}^{(t)}$ for the remaining $(1-r)$-fraction of 
		workers}. Then, draw $(\alpha^{(t+1)}, \Sigma^{(t+1)}, \sigma^{2(t+1)})$ given $\Dcal_{1}^{(t+1)}, \ldots, 
		\Dcal_{k}^{(t+1)}$ and $\Dcal_{\text{obs}}$ using a sequence of three steps:
		\begin{eqnarray}
		& & \left( \tilde{\Sigma}^{(t+1)} \right)^{-1}  \sim \text{Wishart}_{m - q}\left\{ \left( S_{DD}^{(t+1)} \right)^{-1} \right\}, 
		\quad \sigma^{2(t+1)} \sim \frac{\| y - {\hat y}^{(t+1)} \|^2_2 + M }{\chi^2_{n +a - p - q^2}}, \nonumber\\
		& & \alpha^{(t+1)} \mid \sigma^{2(t+1)} \sim N \left( {\hat \alpha}^{(t+1)}, \sigma^{2(t+1)} \left( S_{XX}^{(t+1)} + V_{\alpha}
		\right)^{-1} \right), \label{eq:upd-lme1}
		\end{eqnarray}
		where $\hat \alpha^{(t+1)}$, $\hat y^{(t+1)}$, $S_{DD}^{(t+1)}$, and $S_{XX}^{(t+1)}$ are computed based on 
		\eqref{eq:upd-lme2} and \eqref{eq:upd-lme2.1} using the latest $d_1^{(t+1)}, \ldots, d_m^{(t+1)}$ values. 
	\end{algorithmic} 
	
\end{algorithm}
\noindent
Denote the Markov chain obtained from this ADDA algorithm as $\Phi_{\text{ADLME}} = \{d_1^{(t)}, \ldots, d^{(t)}_m, \alpha^{(t)}, \Sigma^{(t)}, \sigma^{2(t)}\}_{t=0}^{\infty}$. 

The convergence analysis of the linear mixed-effect model parent DA chain is extremely challenging. In fact, 
the literature on geometric ergodicity of the parent DA chain described in \ref{ms1} and \ref{ms2}, to the best of our 
knowledge, focuses only on the special case when the $\Sigma^{(t)}$ is diagonal and $\Gamma$ is fixed to be the identity matrix; for example, see \cite{roman2015geometric} and \cite{abrahamsen2019fast}. Following
\cite{roman2015geometric, abrahamsen2019fast}, we assume that $\Gamma=I_q$ for the convergence analysis; however, unlike these analyses of the parent DA chain, we do {\it not} restrict $\Sigma$ to be a diagonal matrix. Algorithm \ref{ADGLM}, without the $\Gamma$ updates (i.e., $\Gamma$ is fixed at $I_p$), provides a Markov chain $\{d_1^{(t)}, \ldots, d^{(t)}_m, \beta^{(t)}, \Sigma^{(t)}, \sigma^{2(t)}\}_{t=0}^{\infty}$. In this setting, from the priors used in \eqref{ab-prior}, only the prior for $\alpha$ given $\sigma^2$ is modified to the prior for $\beta$ given $\sigma^2$ as $N(0,\sigma^{2}V_{\beta}^{-1})$ for some $V_{\beta} \succ 0$. We need the following assumptions for our geometric ergodicity result. 
\begin{assump}\label{GLM_As1}
\begin{enumerate}[label=(\alph*)]
\item $Z_i^\T Z_i \succ 0$ ($i=1,\dots,m$) and $X^\T X \succ 0$, where $X=\left[ X_{1}^\T \; X_{2}^\T \; 
\cdots \; X_{m}^\T \right]^\T$. 
\item The prior shape parameter $s$ for $\Sigma$ satisfies $s-q-1>\frac{(1-\epsilon)m}{\epsilon}$. 
\item The matrix $V_{\beta}$ from the prior distribution on $\beta$, and the degrees of freedom $a$ from the prior distribution  on $\sigma^2$ satisfy $\frac{(p+mq)}{n+a-p-2}(I_{p}-H) \prec \epsilon I_{p} - H$, where $H = 
X(V_{\beta}+X^{\T}X)^{-1}X^{\T}$ and $A_1 \prec A_2$ denotes that $A_2 - A_1 \succ 0$.  
\end{enumerate}
\end{assump}

\noindent
Assumption (a) is quite reasonable, especially since the ADDA algorithm is geared towards applications with large $m, n$ and typically small values of $p$ and $q$ (such as the MovieLens data analyzed in Section \ref{mov-lme}). Assumptions (b) 
and (c) on the prior hyper-parameters $s, a, V_{\beta}$ might seem restrictive, but it is important to keep in mind that we  are in a challenging setting with a general non-diagonal $\Sigma$ for which convergence results are  unavailable even in the parent DA setting.  As discussed previously, the asynchronous updates in the ADDA setting present additional challenges in the analysis; however, we are still able to establish geometric ergodicity in this more general ADDA setting, as described in the result 
below. 
\begin{thm} \label{thm_GLM}
If $\epsilon>0$ and Assumption \ref{GLM_As1} holds, the Markov chain $\Phi_{\text{ADLME}}$ described in 
Algorithm \ref{ADGLM} (adjusted for $\Gamma = I_q$) is geometrically ergodic. 
\end{thm}

\noindent
The proof of Theorem \ref{thm_GLM} is provided in Supplementary Materials Section \ref{prove_LME}.

\section{Experiments}
\label{sec:experiments}

\subsection{Setup}
\label{sec:setup}

We evaluate the numerical accuracy and computational efficiency of ADDA algorithms using their parent DA algorithms as the benchmarks. In all our simulated and real data analyses, $k \in \{10, 25, 50\}$ and $r \in \{0.20$, $0.40$, $0.60$, $0.80\}$, where the first two and last two values of $r$ represent small and large choices of $r$, respectively. Every experiment with a given choice of $(n,k,r)$ is replicated ten times. Before running an ADDA algorithm, the full data are split (based on the samples or parameter dimensions) into $k$ disjoint subsets that are stored on $k$ worker machines. The parent DA algorithm uses the full data. We use the choice $\epsilon=0$ in our simulations and $\epsilon=0, 0.01, 0.10$ in the real data analyses. The choice $\epsilon = 0$ is best for fast computations that are required for multiple simulation replications; however,  it is not clear if the ADDA chain is Harris ergodic with this choice because Theorem \ref{Harris} requires that $\epsilon > 0$. In our simulations with 20,000 iterations for ADDA algorithms with $\epsilon = 0$, we have rarely encountered a situation  where a worker never sends its I step results to the manager. Our observations from the real data analyses suggests that one could set $\epsilon$ to be $10^{-4}$ or $10^{-5}$ to maintain theoretical convergence guarantees while compromising very little on the computational gains. 

The (empirical) accuracy of an ADDA algorithm relative to its parent DA algorithm is defined using the total variation 
distance. Let $\eta = h(\theta) = (\eta_1, \ldots, \eta_c)$ be a function of the underlying parameter $\theta$ with $c$ components that is of interest to the practitioner. Denote the posterior density estimates of $\eta_j$ using $t$ iterations of the ADDA algorithm and its parent DA as  $q^{t}_{rk}(\eta_j)$ and $p^{t}(\eta_j)$, respectively. Then, the accuracy metric based on $\eta_j$ for the ADDA algorithm at the end of $t$ iterations 
 is defined as
\begin{align} 
  \text{Acc}_{rk,j}(t) = 1 - \text{TV}(p^{t}, q^{t}_{rk}) = 1 - \tfrac{1}{2} \int_{\mathbb{R}} \left| p^{t}(\eta_j) -  q^{t}_{rk}(\eta_j) \right| \, d 
\eta_j, \label{eq:acc:component}
\end{align}
where $r$ and $k$ are given, $\text{TV}(p^{t}, q^{t}_{rk})$ is the total variation distance between $p^{t}$ and $q^{t}_{rk}$, and $\text{Acc}_{rk,j}(t) \in [0, 1]$ because $\text{TV}(p^{t}, q^{t}_{rk}) \in [0, 1]$ 
for every $r, k, t$. We estimate $\text{Acc}_{rk,j}(t)$ using kernel density estimation in two steps. First, we select the first $t$ draws of $\eta_j$ from the ADDA algorithm and its parent DA, respectively, to estimate $q^{t}_{rk}$ and $p^t$ using the {\it bkde} function in the KernSmooth R package. Second, these estimates are used to numerically approximate the integral in \eqref{eq:acc:component} and obtain the $\text{Acc}_{rk,j}(t)$ ($j=1, \ldots, c$) estimates. The overall accuracy metric estimate based on $\eta$ for the ADDA algorithm is computed by averaging the estimates of $\text{Acc}_{rk,1}(t), \ldots, \text{Acc}_{rk,c}(t)$ as
\begin{align}
  \label{eq:acc}
  \text{Acc}_{rk}(t) =  c^{-1} \sum_{j=1}^c \text{Acc}_{rk,j}(t). 
\end{align}
\noindent
The larger the $\text{Acc}_{rk}(t)$ estimate, the better an ADDA algorithm is at approximating its parent DA algorithm. 

The Monte Carlo standard errors in an application of ADDA algorithm and its parent DA algorithm are computed using the 
overlapping batch means method. This method is implemented in the \emph{mcse} function of mcmcse R package 
\citep{Jonetal06,Vatetal19}. Following the notation from the previous paragraph, let $\text{SE}_{rk, j}^{\text{A}}(t)$ and 
$\text{SE}_{j}^{\text{P}}(t)$ respectively be outputs of the mcse function using the first $t$ draws of $\eta_j$ from the ADDA algorithm for a given $(r, k)$ and its parent DA algorithm ($j=1, \ldots, c$). Then, a metric for comparing their 
overall Monte Carlo standard errors is 
\begin{align}
  \label{eq:mcse}
  \text{SE}_{rk}(t) = c^{-1} \sum_{j=1}^c \Delta \text{SE}_{rk, j}(t), \quad \Delta \text{SE}_{rk, j}(t) = |\text{SE}_{rk, j}^{\text{A}}(t) - \text{SE}_{j}^{\text{P}}(t)|. 
\end{align}
The smaller the value of $\text{SE}_{rk}(t)$, better is the agreement between Monte Carlo standard errors of the ADDA algorithm and its parent DA.

We now undertake an extensive experimental evaluation of the proposed ADDA framework using both simulated datasets 
(Section \ref{sec:simulation}) and the MovieLens data (Section \ref{sec:movi-data-analys}). The simulation study focuses 
on using the metrics in (\ref{eq:acc}) and (\ref{eq:mcse}) to evaluate the accuracy of the ADDA algorithm in approximating its 
parent DA algorithm. Due to the relatively small sample/missing data size to facilitate multiple replications for various 
$(k,r)$ choices, a runtime comparison is not undertaken in Section \ref{sec:simulation}. For the MovieLens data in 
Section \ref{sec:movi-data-analys} however (with up to $10^7$ samples), we present both an accuracy and runtime 
comparison for both the logistic regression and the linear mixed model settings. 
All ADDA algorithms are implemented in R using the parallel package. The ease of implementation and reproducibility  drive this choice.
For this data, our ADDA implementations 
using the parallel package show three to five times gains in run-times for all the choices of $(k, r)$ (see 
Figure \ref{fig:mov-run-time}). We expect the run-time gains to be even better if the asynchronous updates for the ADDA 
algorithms are implemented carefully using  MPI \citep{Nie16}. However, this requires significant additional programming 
efforts and will be addressed in future research. 

\subsection{Simulated Data Analyses} \label{sec:simulation}

This section focuses on two broad classes of applications. The first class consists of logistic regression and linear mixed-effects models, where the computational bottlenecks arise due to massive sample size; see Sections \ref{sec:sim-adda-logist-reg} and \ref{sec:sim-adda-Mixed-Effects}. In these models, the samples are partitioned into $k$ subsets and stored on the worker machines. Section \ref{sec:sim-adda-Lasso} focuses on high-dimensional variable selection using the Bayesian lasso, where the parent DA is inefficient due to the augmentation of $p$ local shrinkage parameters specific to the regression coefficients. Unlike the former two models, here we partition the augmented parameters into $k$ subsets and store them on worker machines. 

\subsubsection{Logistic Regression}
\label{sec:sim-adda-logist-reg}

We evaluate the empirical performance of the ADDA Algorithm \ref{ADPG} for logistic regression in \eqref{eq:l1} with $p=10$, and $s_i=10$ for every $i$. The entries of $\beta$ are set to $-2$ and $2$ alternatively starting from $-2$ and the entries of $X$ are independent standard normal random variables. We use them to simulate $y_1, \ldots, y_n$ using the hierarchical model in \eqref{eq:l1}. The rows of $X$ and $y$ are randomly split into $k$ subsets that are stored on the worker machines. The ADDA and its parent run for 20000 iterations and $\beta$ draws are collected at the end of every iteration.

\begin{figure}[H]
\begin{subfigure}{.5\textwidth}
  \centering
  \includegraphics[scale=0.27]{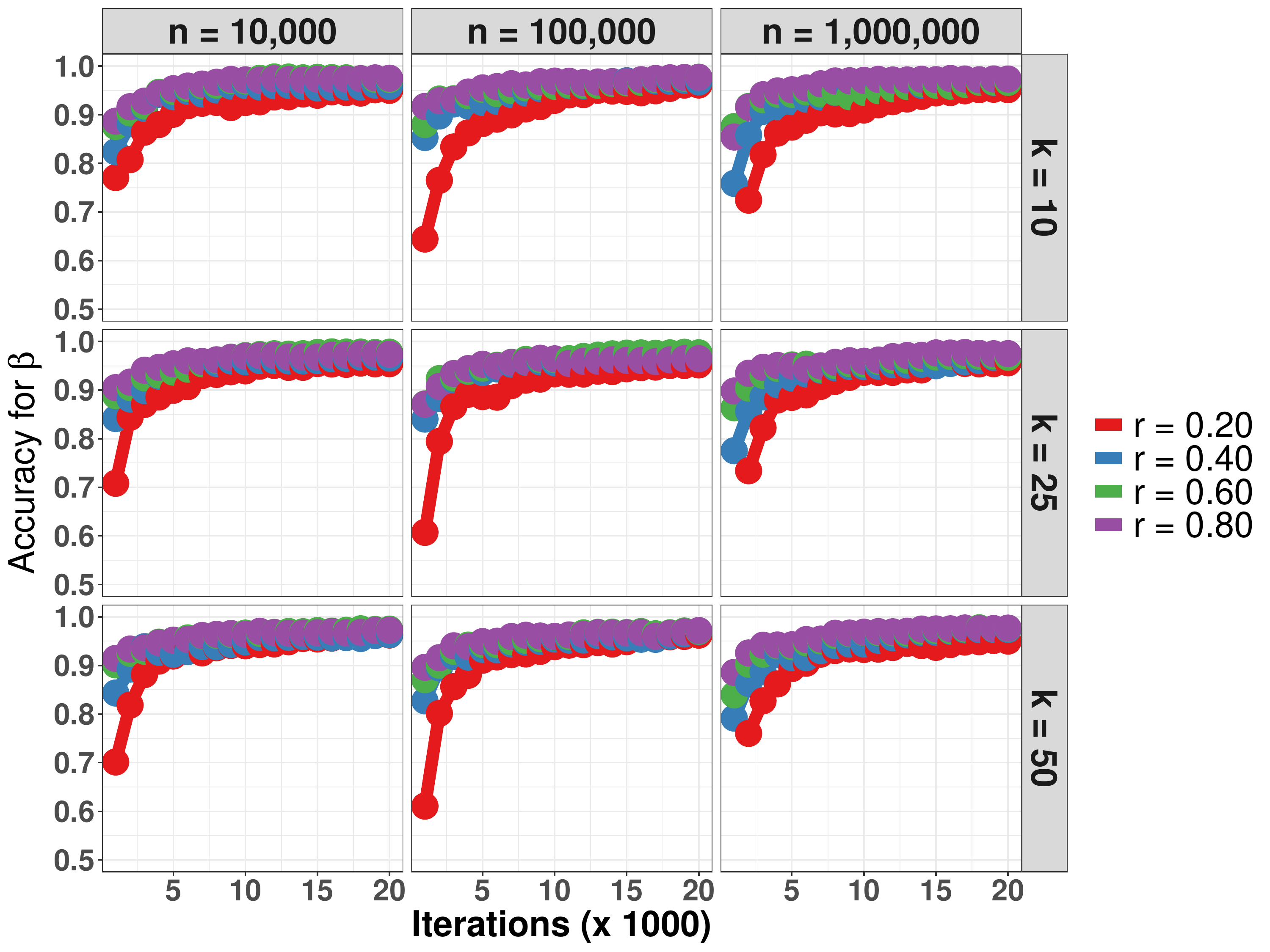}
  \caption{$\beta$}
  \label{fig:log1}
\end{subfigure}%
\begin{subfigure}{.5\textwidth}
  \centering
  \includegraphics[scale=0.27]{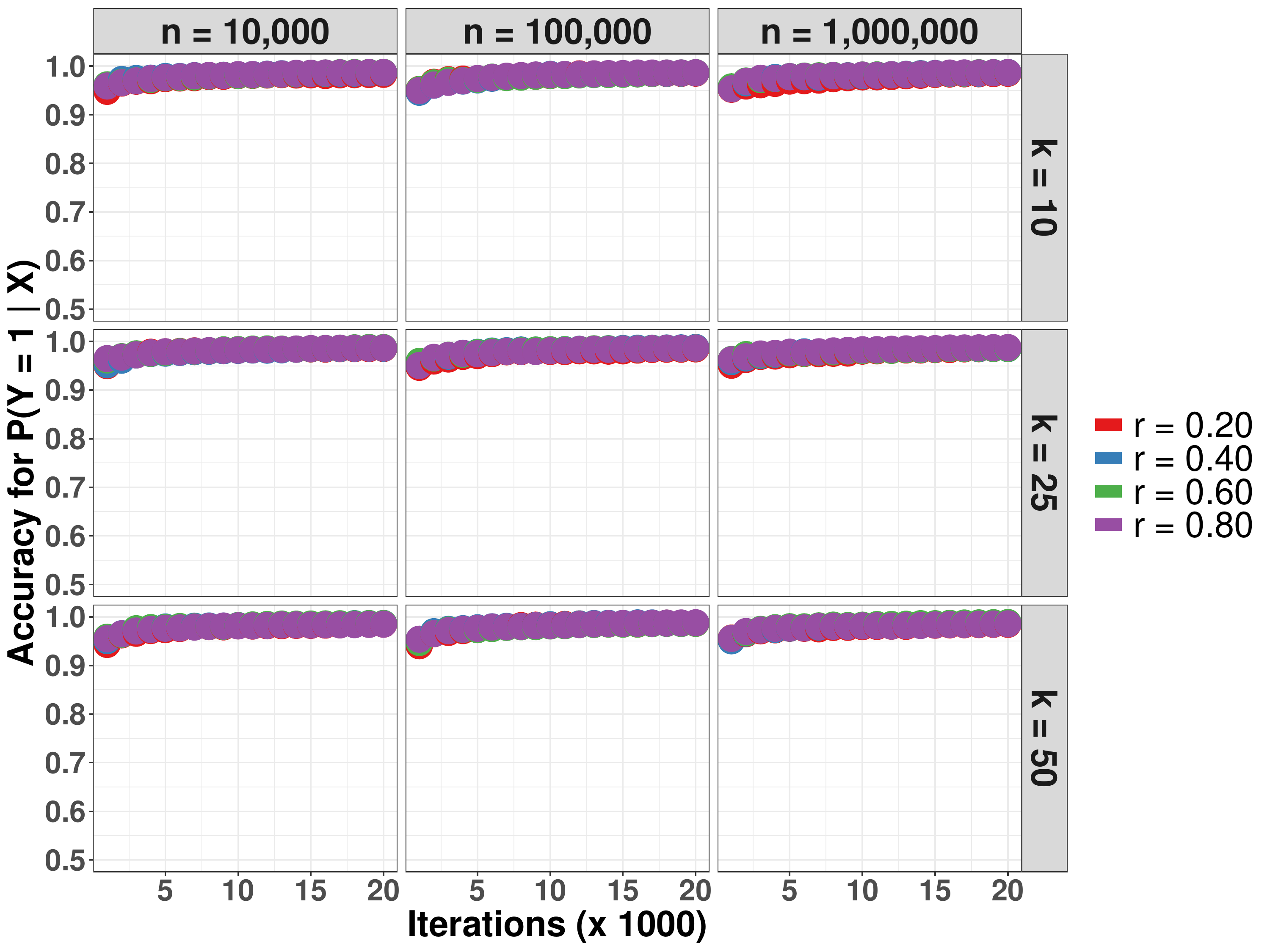}
  \caption{ $\text{P}(Y = 1 \mid x_1 = 1, \ldots, x_{p}=1)$}
  \label{fig:log2}
\end{subfigure}
\caption{$\text{Acc}_{rk}(t)$ metric for inference on $\beta$ and $\text{P}(Y = 1 \mid x_1 = 1, \ldots, x_{p}=1)$ using the ADDA Algorithm \ref{ADPG} based on P\'olya-Gamma DA in logistic regression with $r = 0.20, 0.40, 0.60, 0.80$ and $k=10, 25, 50$.}
\label{fig:log}
\end{figure}

The $\text{Acc}_{rk}(t)$ metric is evaluated for $\eta = \beta$ and $\eta = \text{P}(Y = 1 \mid x_1 = 1, \ldots, x_p = 1)$, respectively  (Figure \ref{fig:log}). The $\text{Acc}_{rk}(t)$ values for $\beta$ depend on $n, r$  values but are fairly insensitive to the choice of $k$. For every $n,k, t$,  the $\text{Acc}_{rk}(t)$ values for $\beta$ are (initially) the largest when $r=0.80$ and decrease with $r$. However, The differences between $\text{Acc}_{rk}(t)$ values for different choices of $r,k,n$ are negligible after $t=5000$ for $\beta$. On the other hand, $\text{Acc}_{rk}(t)$ values for $\text{P}(Y = 1 \mid x_1 = 1, \ldots, x_p = 1)$ are comparatively insensitive to the choice of $n, r, k$, and after the differences for different choices of 
$r,k,n$ are negligible after $t=2000$.

\begin{figure}[H]
\begin{subfigure}{.5\textwidth}
  \centering
  \includegraphics[scale=0.27]{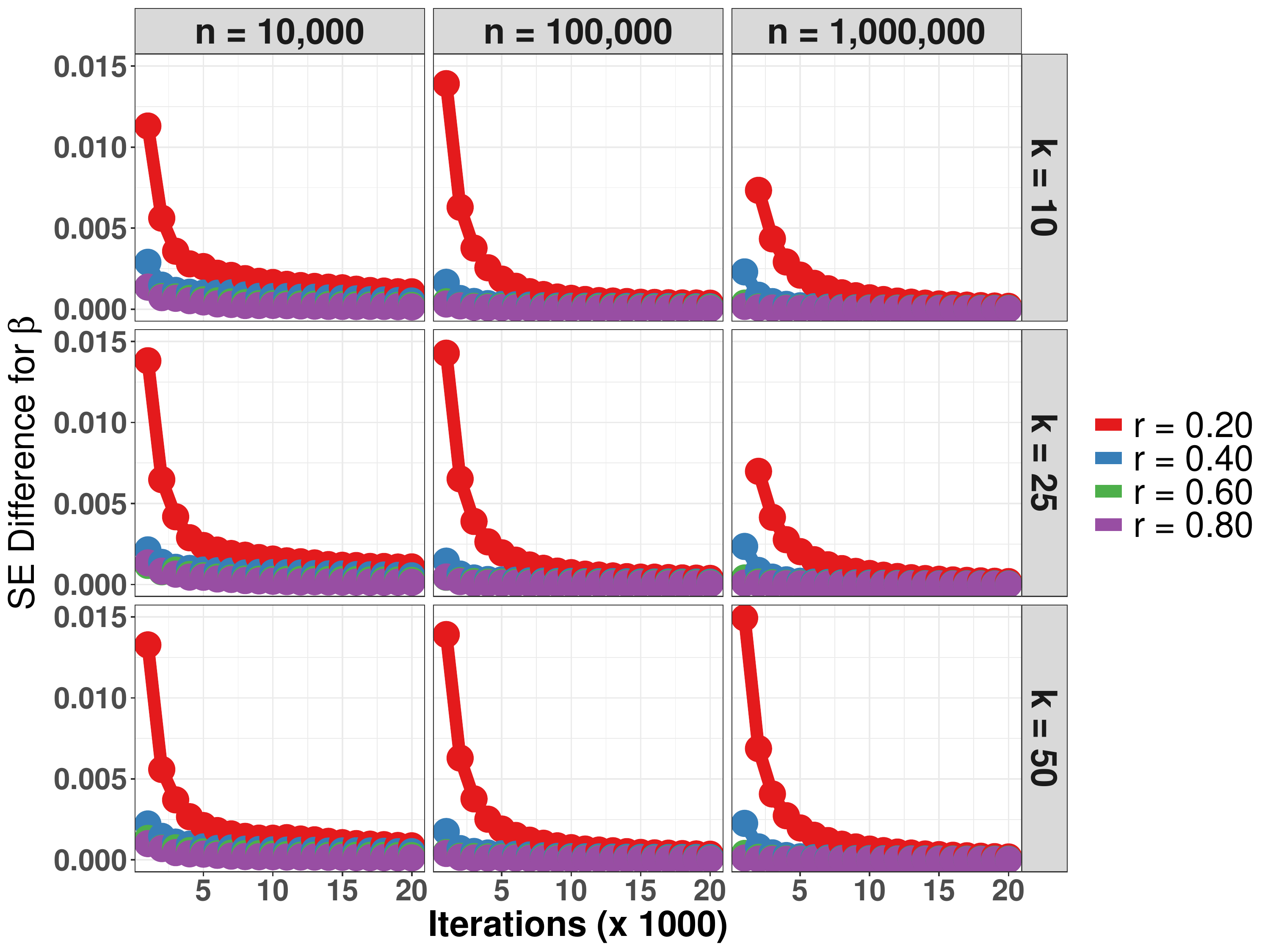}
  \caption{$\beta$}
  \label{fig:mcse_log1}
\end{subfigure}%
\begin{subfigure}{.5\textwidth}
  \centering
  \includegraphics[scale=0.27]{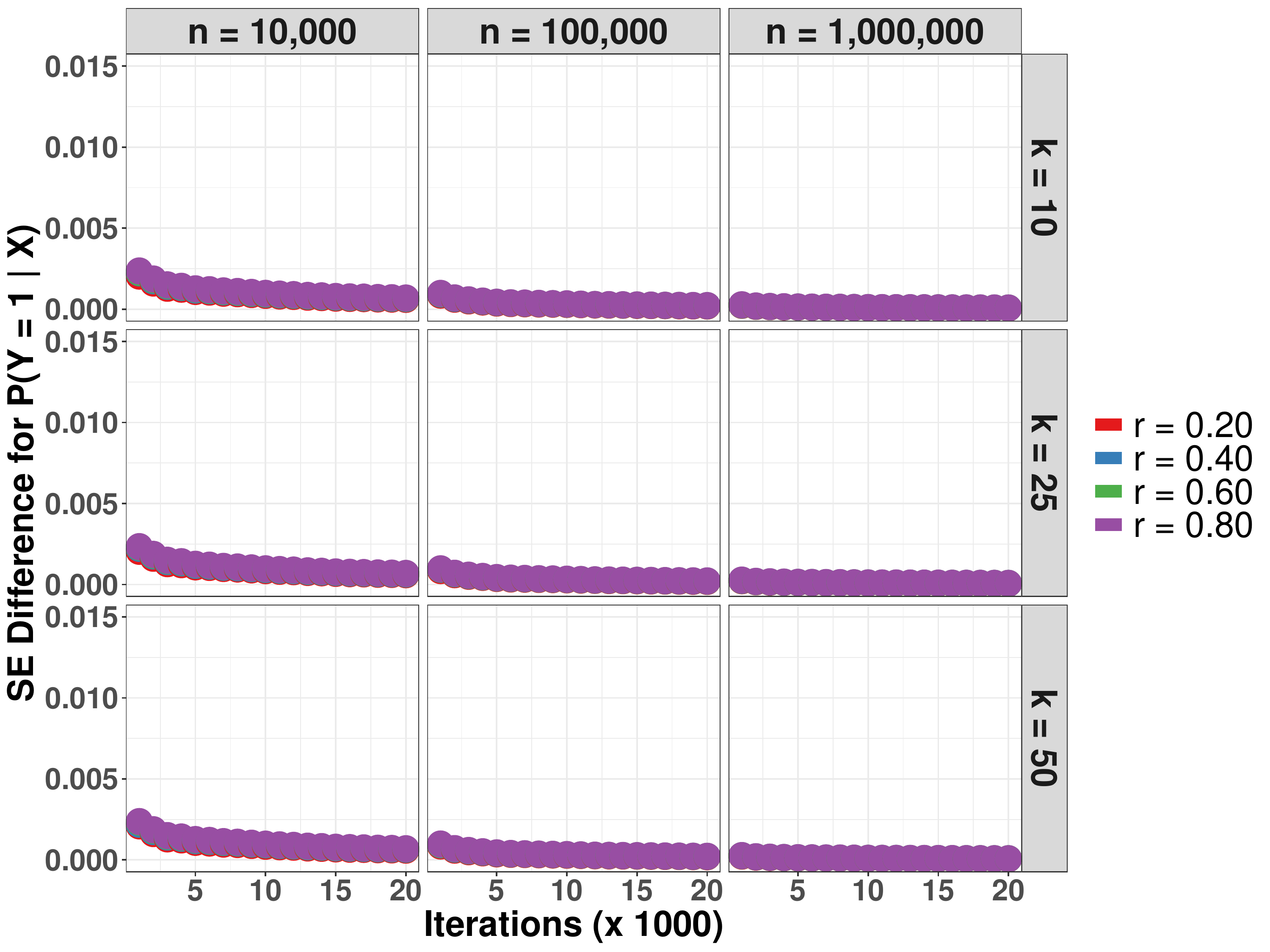}
  \caption{ $\text{P}(Y = 1 \mid x_1 = 1, \ldots, x_{10}=1)$}
  \label{fig:mcse_log2}
\end{subfigure}
\caption{$\text{SE}_{rk}(t)$ metric inference on $\beta$ and $\text{P}(Y = 1 \mid x_1 = 1, \ldots, x_{10}=1)$ using the ADDA Algorithm \ref{ADPG} based on P\'olya-Gamma DA in logistic regression with $r = 0.20, 0.40, 0.60, 0.80$ and $k=10, 25, 50$.}
\label{fig:mcse_log}
\end{figure}

The $\text{SE}_{rk}(t)$ metric for $\beta$ and $\text{P}(Y = 1 \mid x_1 = 1, \ldots, x_p = 1)$, respectively, decays to 0 as $t$ increases (Figure \ref{fig:mcse_log}). For every $n, k,$ and $t$, the $\text{SE}_{rk}(t)$ values for $\beta$  are the largest when $r=0.2$ but quickly decay to 0 as $t$ increases. The $\text{SE}_{rk}(t)$ values for $\text{P}(Y = 1 \mid x_1 = 1, \ldots, x_p = 1)$ quickly decay to 0 as $t$ increases and are comparatively insensitive of the choices of $(n, r, k)$. This shows that $\text{SE}_{rk}(t)$ values for both parameters are close to 0 for every $n, k$, and $ r$ if $t$ is sufficiently large and that the Monte Carlo standard errors of parameter estimates obtained using ADDA Algorithm \ref{ADPG} and its parent are comparable in massive data settings.

Note that Theorem \ref{pg_thm} says that the ADDA Markov chain converges to the same stationary distribution as the 
parent DA chain for any $(r, k)$. Since the $\text{Acc}_{rk}(t)$ values for $\text{P}(Y = 1 \mid x_1 = 1, \ldots, x_p = 1)$ and 
$\beta$ converge to 1 for a sufficiently large $t$ irrespective of the choices of $r$ and $k$, our empirical results support the 
assertion in Theorem \ref{pg_thm}. 

\subsubsection{High Dimensional Variable Selection}
\label{sec:sim-adda-Lasso}

We evaluate the empirical performance of the ADDA Algorithm \ref{ADBL} for Bayesian variable selection. We vary $n$ and $p$ such that $n = p$ and $n \in \{50, 500, 5000\}$. The first $0.90 p$ entries of $\beta$ in \eqref{eq:v1} are set to 0 and the next $0.10 p$ entries are set to $-2$ and $2$ alternatively starting from $-2$. The entries of $X$ are independent standard normal random variables. We use them to simulate $y_1, \ldots, y_n$ using the model in \eqref{eq:v1} with $\sigma^2 = 0.01$. The columns of $X$ are randomly split into $k$ subsets that are stored on the worker machines. The ADDA algorithm and its parent DA are run for 20000 iterations and $\beta$ draws are collected at the end of every iteration.

\begin{figure}[H]
\begin{subfigure}{.5\textwidth}
  \centering
  \includegraphics[scale=0.27]{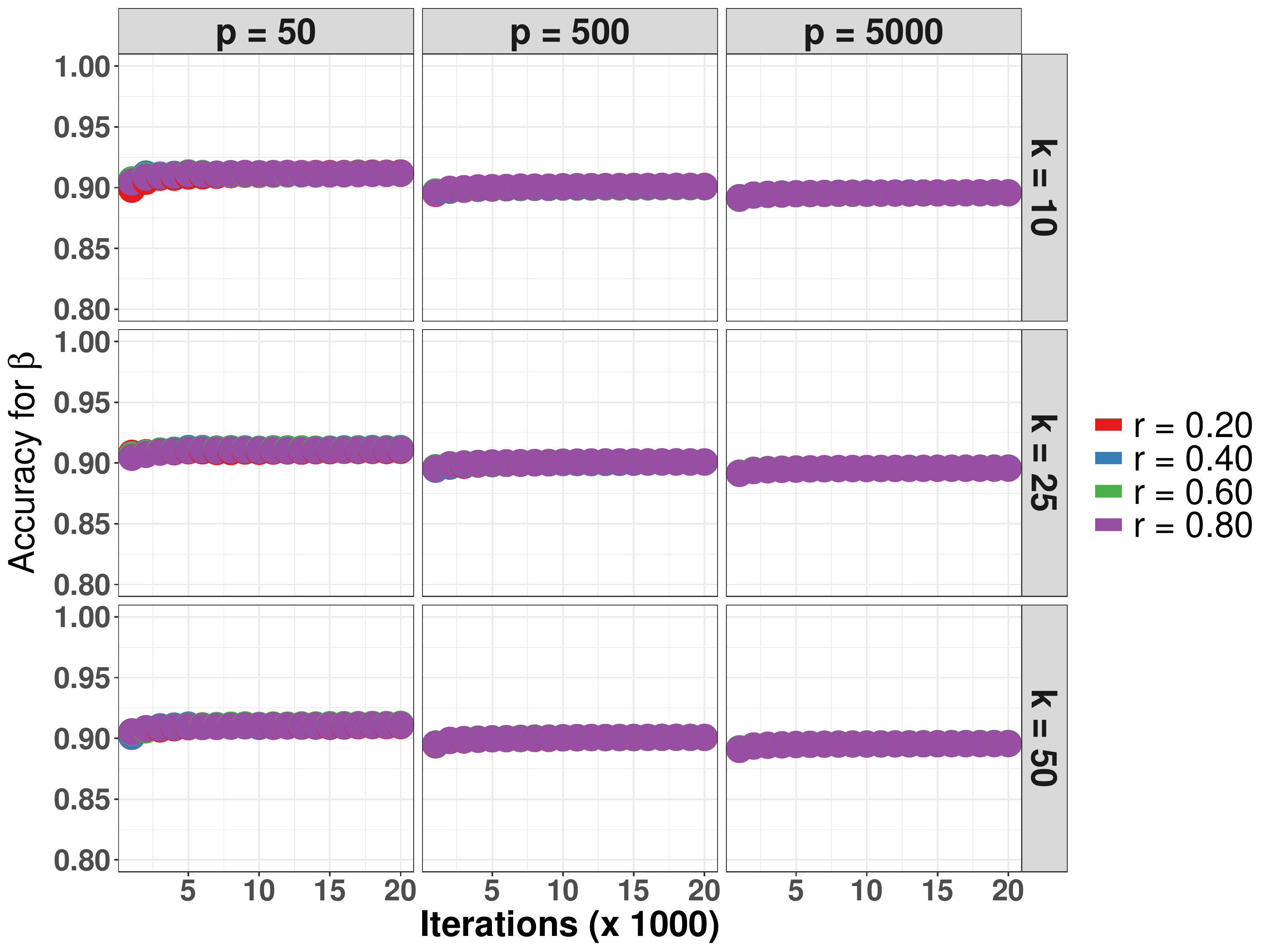}
  \caption{$\text{Acc}_{rk}(t)$}
  \label{fig:acc_bvs}
\end{subfigure}%
\begin{subfigure}{.5\textwidth}
  \centering
  \includegraphics[scale=0.27]{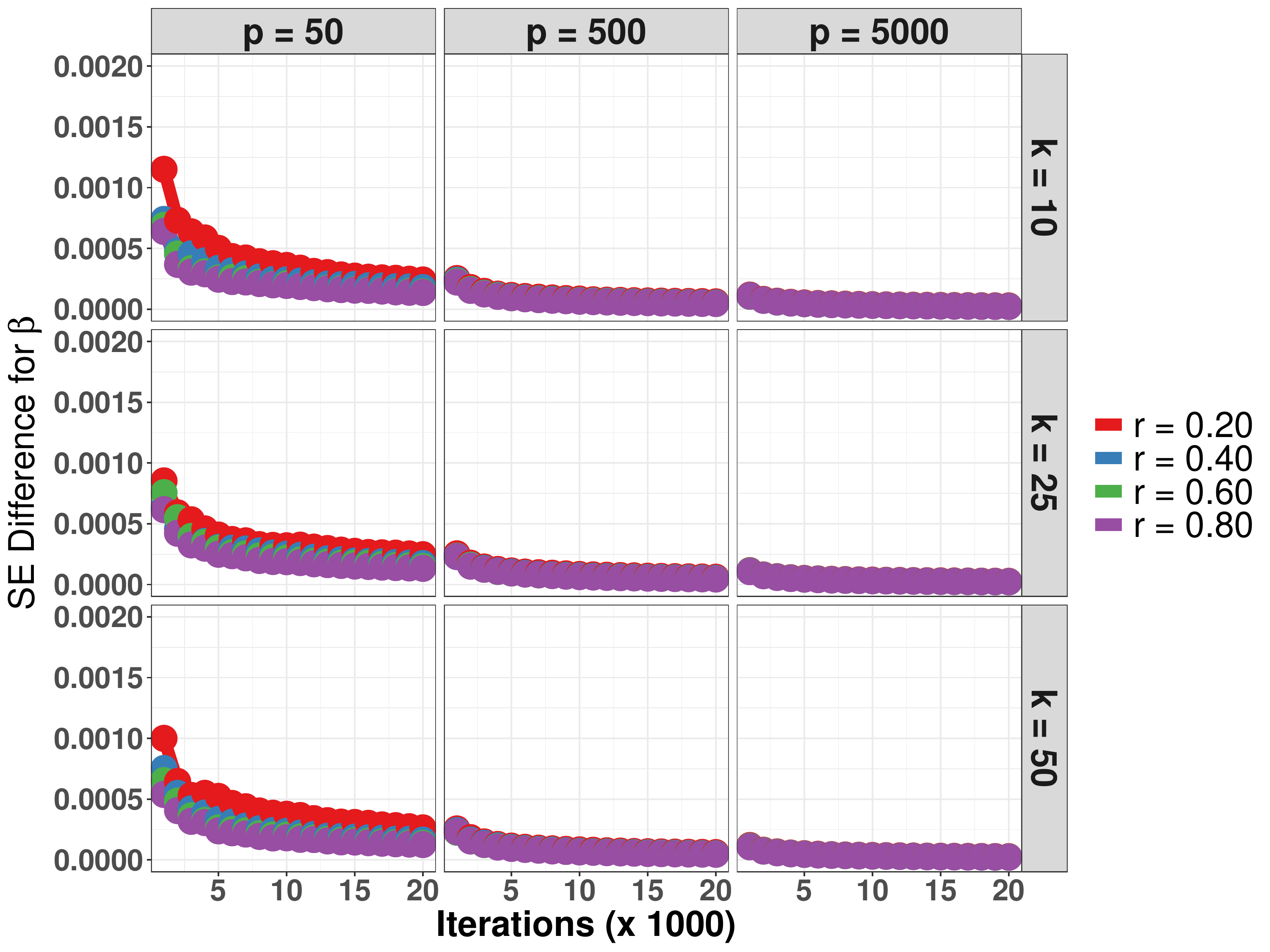}
  \caption{$\text{SE}_{rk}(t)$}
  \label{fig:mcse_bvs}
\end{subfigure}
\caption{$\text{Acc}_{rk}(t)$ and $\text{SE}_{rk}(t)$ metrics for inference on $\beta$ in \eqref{eq:v1} using the ADDA Algorithm \ref{ADBL} based on the Bayesian lasso prior in high-dimensional variable selection with $r = 0.20, 0.40, 0.60, 0.80$ and $k=10, 25, 50$.}
\label{fig:bvs}
\end{figure}

For $\eta = \beta$ and every $p, r, k, t$, the $\text{Acc}_{rk}(t)$ value is high and $\text{SE}_{rk}(t)$ value is low (Figure \ref{fig:bvs}). The $\text{Acc}_{rk}(t)$ values are fairly high even for small $t$, increase slightly with $t$, and are insensitive to the choice of $k$ and $r$. The high-dimensionality of the variable selection problem increase with $p$, so $\text{Acc}_{rk}(t)$ values decrease slightly with $p$ for a given $k$. Agreeing with this observation, $\text{SE}_{rk}(t)$ values also decay to 0 as $t$ increases and the decay is insensitive to the choices of $p,k$, and $r$. We conclude that ADDA Algorithm \ref{ADBL} is an accurate asynchronous and distributed generalization of the Bayesian Lasso DA algorithm. 

Theorem \ref{hdim_varsel} implies that the Bayesian Lasso ADDA chain converges to the same stationary distribution as 
the parent DA chain for any $(r, k)$ in the high-dimensional regime. Our empirical results support this assertion irrespective 
of the choice of $(p, r, k)$, even when $\epsilon = 0$ and $t$ is small. 

\subsubsection{Linear Mixed-Effects Modeling}
\label{sec:sim-adda-Mixed-Effects}

We evaluate the empirical performance of the ADDA Algorithm \ref{ADGLM} for linear mixed-effects modeling with $m \in \{100, 1000, 10,000\}$, $p=4$, $q=3$, and $n_i= n/m$ for every $i$. Following \citet{Lietal16}, the entries of $\beta = (-2, 2, -2, 2)^\T$, the entries of $X, Z$ are set to $1$ or $-1$ with equal probability satisfying Assumption \ref{GLM_As1}, $\sigma^2 =1$, and $\Sigma_{ii} = i$ ($i=1, 2, 3$), $\Sigma_{12} = -0.56$, $\Sigma_{31} = 0.52$, $\Sigma_{23} = 0.0025$. We use them to simulate $y_1, \ldots, y_m$ using the hierarchical model in \eqref{eq:mix1}. The rows of $X, Z$, and $y$ are split into $k$ disjoint subsets that are stored on the worker machines. The ADDA algorithm and its parent are run for 20000 iterations and $\beta, \Sigma$, and $\sigma^2$ draws are collected at the end of every iteration. 

\begin{figure}[H]
\begin{subfigure}{.5\textwidth}
  \centering
  \includegraphics[scale=0.27]{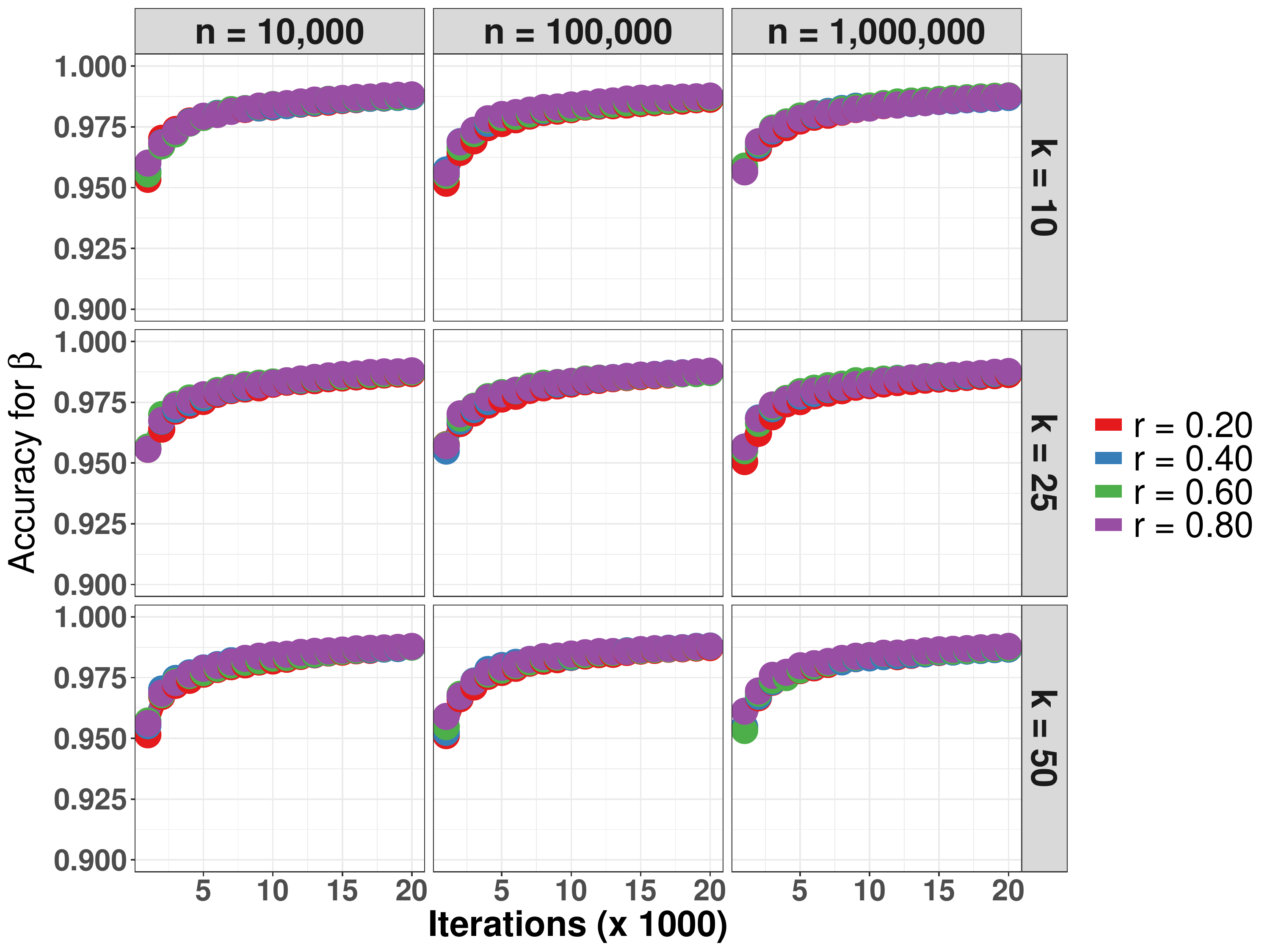}
  \caption{$\beta$}
  \label{fig:lme_beta}
\end{subfigure}%
\begin{subfigure}{.5\textwidth}
  \centering
  \includegraphics[scale=0.27]{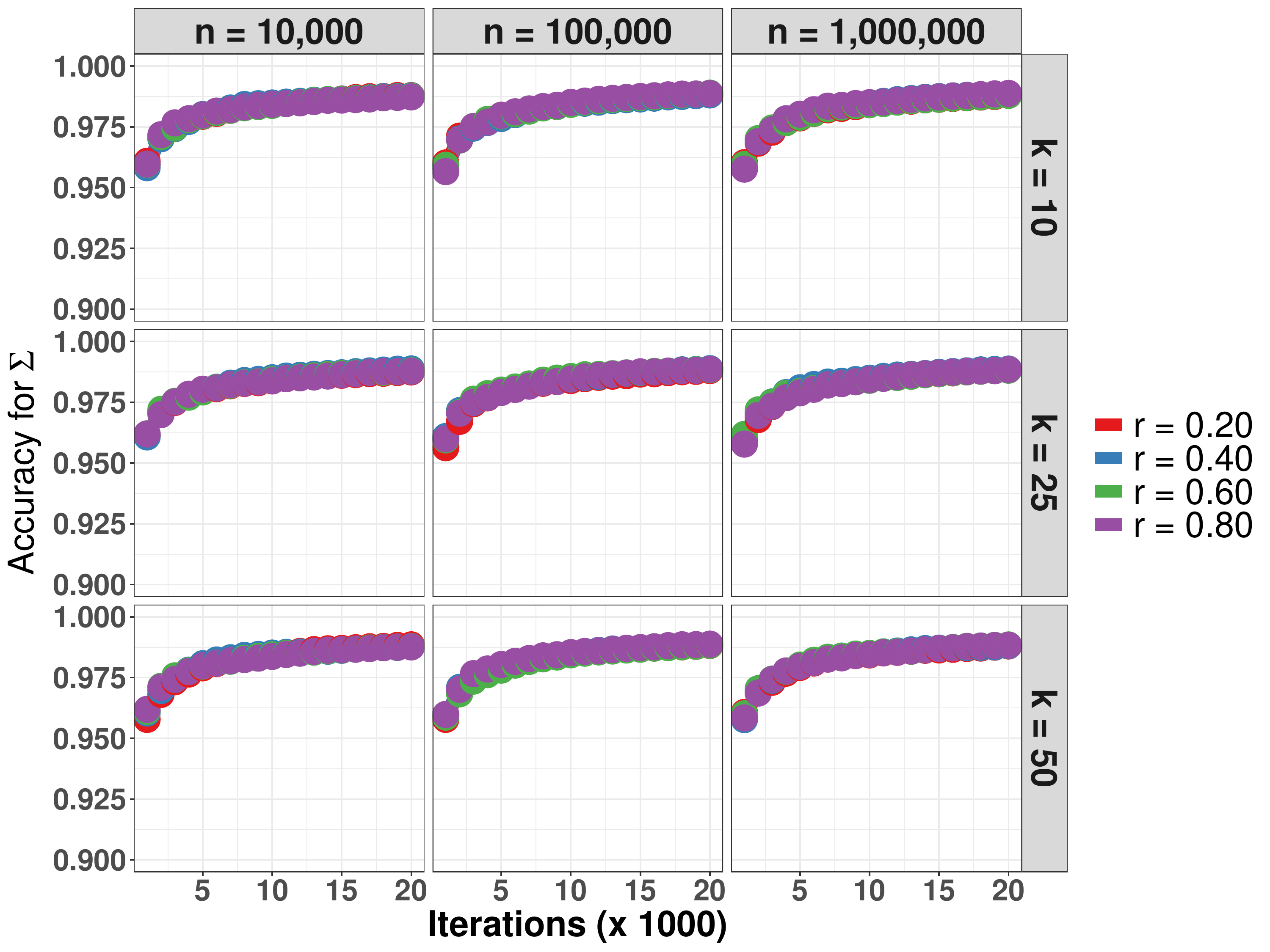}
  \caption{ $\Sigma$}
  \label{fig:lme_dmat}
\end{subfigure}
\caption{$\text{Acc}_{rk}(t)$ metric for inference on $\beta$ and $\Sigma$ in \eqref{eq:mix1} using the ADDA Algorithm \ref{ADGLM} based on the marginal DA algorithm in linear mixed-effects modeling with $r = 0.20, 0.40, 0.60, 0.80$ and $k=10, 25, 50$.}
\label{fig:lme_beta_dmat}
\end{figure}

\begin{figure}[H]
\begin{subfigure}{.5\textwidth}
  \centering
  \includegraphics[scale=0.27]{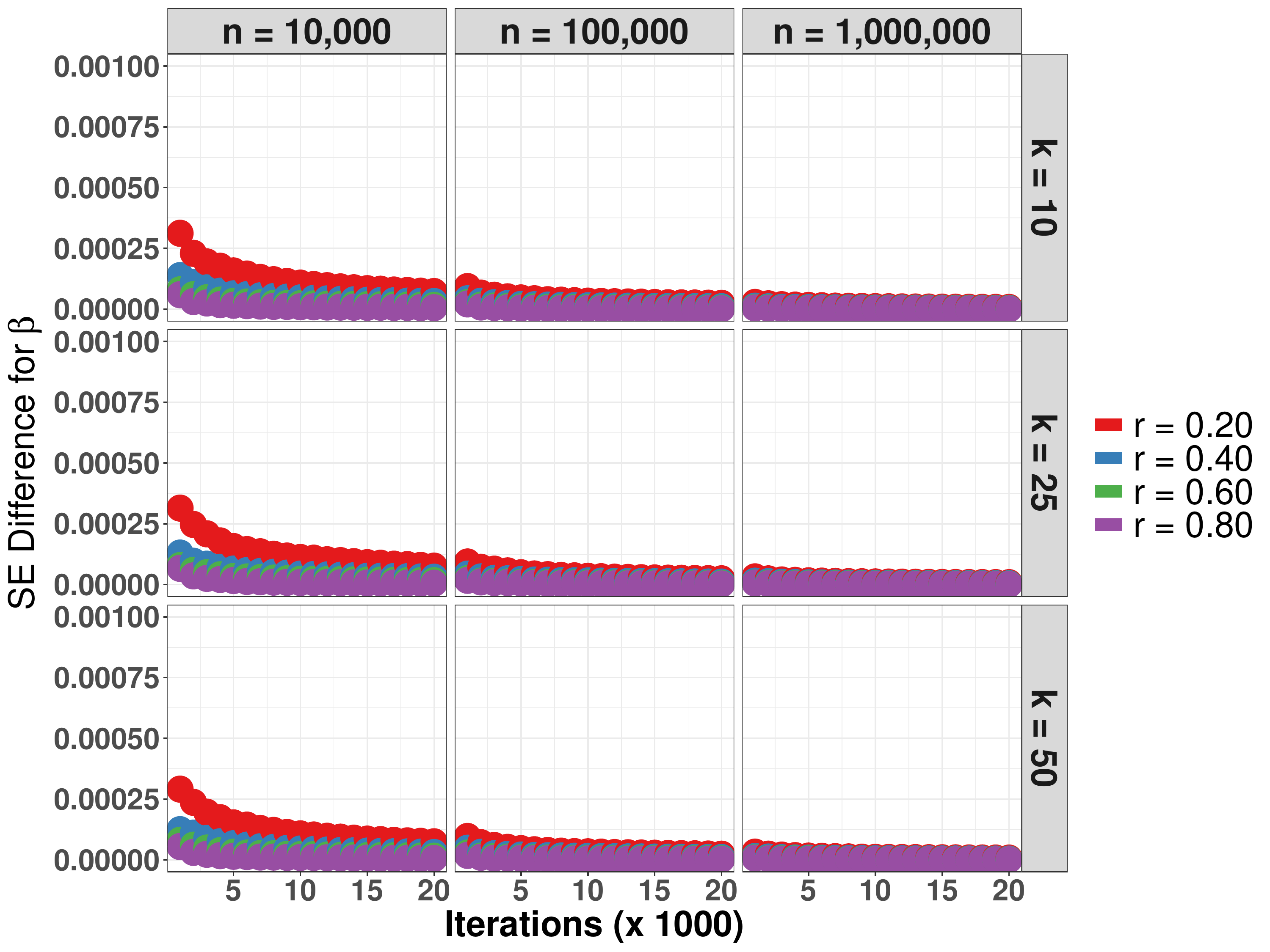}
  \caption{$\beta$}
  \label{fig:mcse_lme_beta}
\end{subfigure}%
\begin{subfigure}{.5\textwidth}
  \centering
  \includegraphics[scale=0.27]{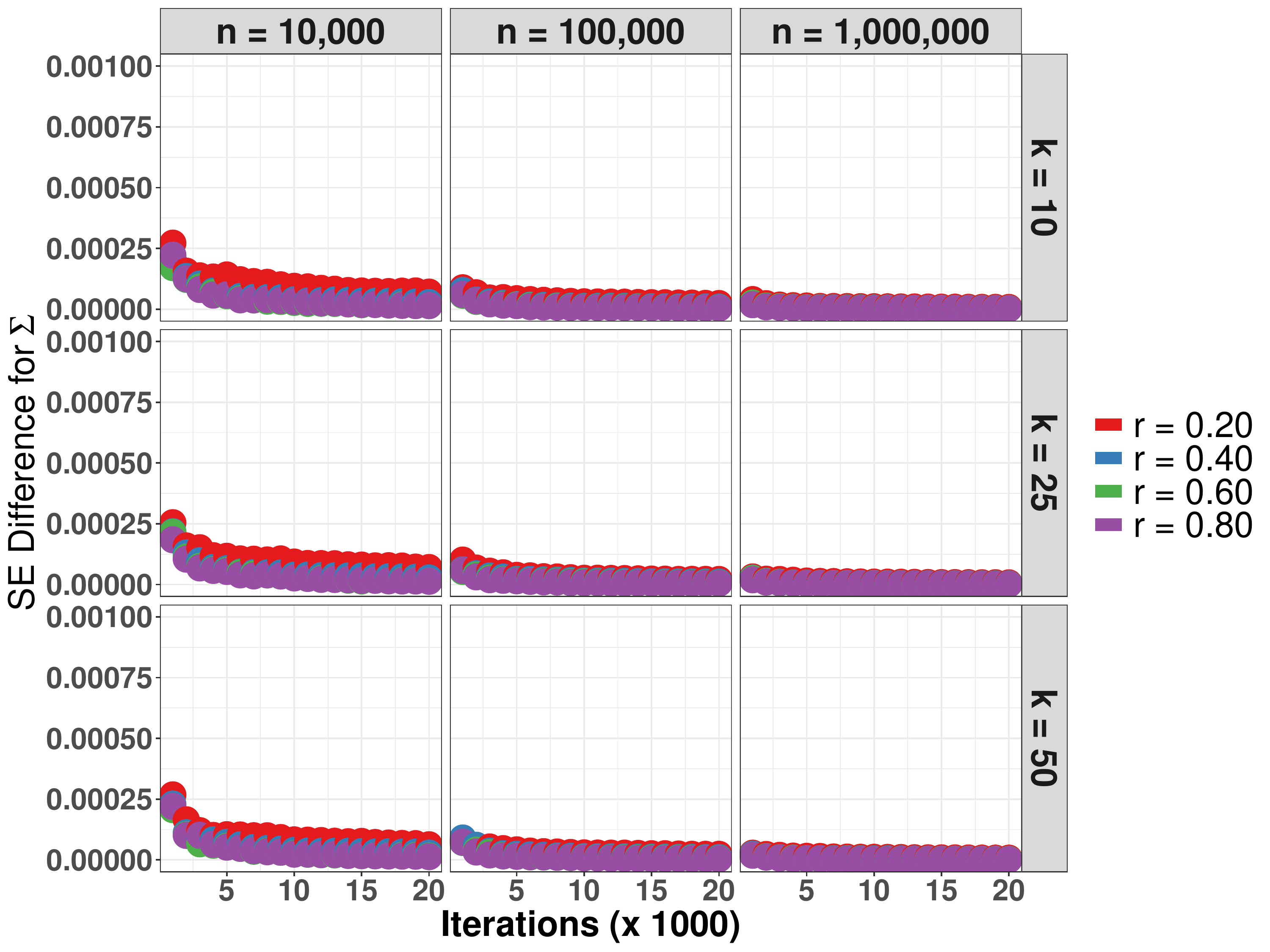}
  \caption{ $\Sigma$}
  \label{fig:mcse_lme_dmat}
\end{subfigure}
\caption{$\text{SE}_{rk}(t)$ metric for inference on $\beta$ and $\Sigma$ in \eqref{eq:mix1} using the ADDA Algorithm \ref{ADGLM}  based on the marginal DA algorithm in linear mixed-effects modeling with $r = 0.20, 0.40, 0.60, 0.80$ and $k=10, 25, 50$.}
\label{fig:mcse_lme_beta_dmat}
\end{figure}

\begin{figure}[H]
\begin{subfigure}{.5\textwidth}
  \centering
  \includegraphics[scale=0.27]{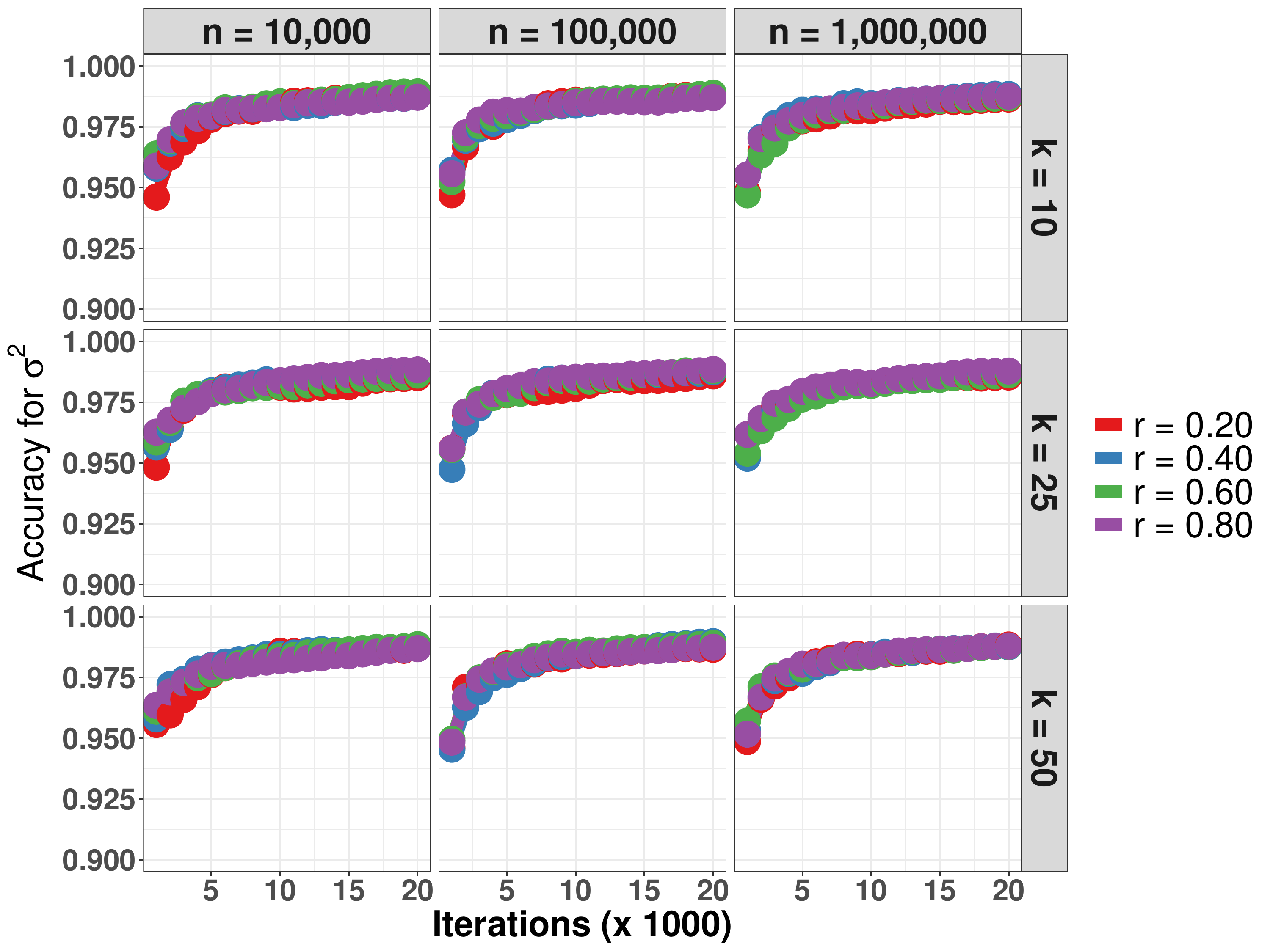}
  \caption{$\text{Acc}_{rk}(t)$}
  \label{fig:acc_lme_sig}
\end{subfigure}%
\begin{subfigure}{.5\textwidth}
  \centering
  \includegraphics[scale=0.27]{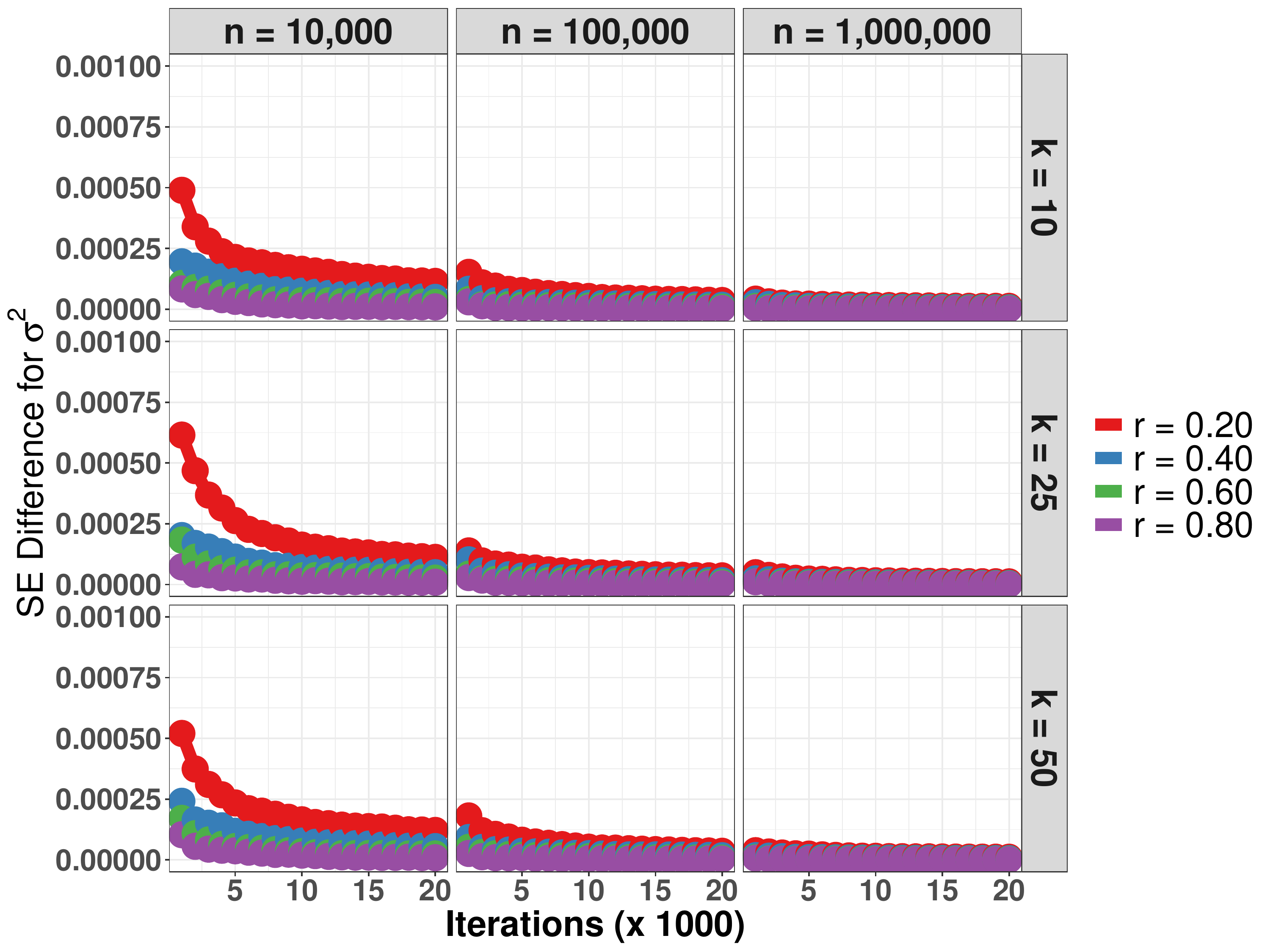}
  \caption{$\text{SE}_{rk}(t)$}
  \label{fig:mcse_lme_sig}
\end{subfigure}
\caption{$\text{Acc}_{rk}(t)$ and $\text{SE}_{rk}(t)$ metrics for inference on $\sigma^2$ in \eqref{eq:mix1} using the ADDA Algorithm \ref{ADGLM}  based on the marginal DA algorithm in linear mixed-effects modeling with $r = 0.20, 0.40, 0.60, 0.80$ and $k=10, 25, 50$.}
\label{fig:acc_mcse_lme_sig}
\end{figure}

We evaluate the $\text{Acc}_{rk}(t)$ metric for $\eta$ equalling $\beta$,  $\Sigma$, and $\sigma^2$, respectively (Figures \ref{fig:lme_beta_dmat} -- \ref{fig:acc_mcse_lme_sig}). The $\text{Acc}_{rk}(t)$ values for $\beta$, $\Sigma$, and $\sigma^2$ show similar patterns for every $n, r, k$ values as $t$ increases from 1 to 20000. The accuracies for all the three parameters are very high and increase with $t$. The $\text{Acc}_{rk}(t)$ values are insensitive to the choices of $r$, $k$, and $n$. We conclude that ADDA estimates the posterior distributions for these parameters with the same accuracy as its parent DA.

The SE$_{rk}(t)$ metric  for  $\beta$, $\Sigma$, and $\sigma^2$, respectively, decays to 0 as $t$ increases  (Figures \ref{fig:mcse_lme_beta_dmat} and \ref{fig:acc_mcse_lme_sig}). The decay patterns of SE$_{rk}(t)$ values for $\beta, \Sigma$, and $\sigma^2$ are similar in that they are the highest when $r=0.20$ for every $n, k, t$ and quickly decay to 0 as $t$ increases. For all the three parameters and every $r$, we observe that as $n$ increases, SE$_{rk}(t)$ values decays to 0 more quickly. This shows that parameter estimates obtained using ADDA algorithm and its parent have similar standard errors in massive data settings.

The empirical results show that the $\text{Acc}_{rk}(t)$ values are fairly similar for different choices of $r, k$, and $n$. The standard error of the Monte Carlo estimates obtained using  Algorithm \ref{ADGLM} and its parent DA also agree for a sufficiently large $t$ and for every $r, k$, and $n$. This supports the conclusions of Theorem \ref{thm_GLM}. 

\subsection{Real Data Analysis}
\label{sec:movi-data-analys}

\subsubsection{MovieLens Data}

We use the MovieLens ratings data (\texttt{http://grouplens.org}) that contain more than 10 million movie ratings from about 72 thousand users, where each rating ranges from 0.5 to 5 in increments of 0.5. Starting with \citet{Per17}, a modified form of this data has been used for illustrating scalable inference in linear mixed-effects models \citep{Srietal18,Srietal19,XuSri21}. Specifically, a movie's category is defined using its genre: \emph{action} category includes action, adventure, fantasy, horror, sci-fi, or thriller genres; \emph{children} category includes animation or children; \emph{drama} category includes crime, documentary, drama, film-noir, musical, mystery, romance, war, or western; and \emph{comedy} category includes comedy genre only. 

Three predictors are defined using this data. The first predictor encodes a movie's category using three dummy variables with \emph{action} category as the baseline. If a movie belongs to $C$ categories, then its category predictor is $1/C$ for each of the $C$ categories. The second and third predictors capture a movie's popularity and a user's mood. The popularity predictor of a movie equals logit$\{(l + 0.5) / (r + 1)\}$, where  $r$ users rated the movie and $l$ of them gave a rating of 4 or higher. The user's mood predictor equals 1 if the user gave the previously 30 rated movies a rating of 4 or higher and 0 otherwise. The second and third predictors are treated as numeric variables.

We model this data using logistic regression and linear mixed-effects model in Sections \ref{mov-log} and \ref{mov-lme} 
below. Before fitting the logistic regression model, the response in MovieLens data is modified for defining the $0/1$-valued 
response. If the user's rating in a given sample is greater than 3, then the response is 1; otherwise, the response is 0. No 
modification is required for the application of linear mixed-effects model. The original PG DA and marginal DA algorithms for 
logistic regression and linear mixed-effects modeling, respectively, are slow if we use with the full MovieLens data; therefore, 
we analyze randomly selected subsets of the MovieLens data to facilitate posterior computations and estimation of 
Acc$_{rk}(t)$  and SE$_{rk}(t)$ metrics.

\subsubsection{Logistic Regression}
\label{mov-log}

We randomly select two subsets of users in the MovieLens data such that the total number of ratings in them are about $10^6$ and $10^7$, respectively. The MovieLens data with modified $0/1$-valued responses are analyzed using logistic regression model in \eqref{eq:l1} with six predictors, including an intercept. There are three dummy variables for the movie categories and one each for a movie's popularity and a user's mood. The $\beta$ vector is $6$-dimensional. The responses and predictors in the modified data are collected into the response vector $y$ and design matrix $X$. We apply the ADDA Algorithm \ref{ADPG} used in Section \ref{sec:sim-adda-logist-reg} with two modifications: the number of iterations is 10,000 instead of 20,000 and $\epsilon = 0, 0.01, 0.10$. This setup is replicated ten times by randomly choosing the users in each replication.

\begin{figure}[H]
\begin{subfigure}{.5\textwidth}
  \centering
  \includegraphics[scale=0.23]{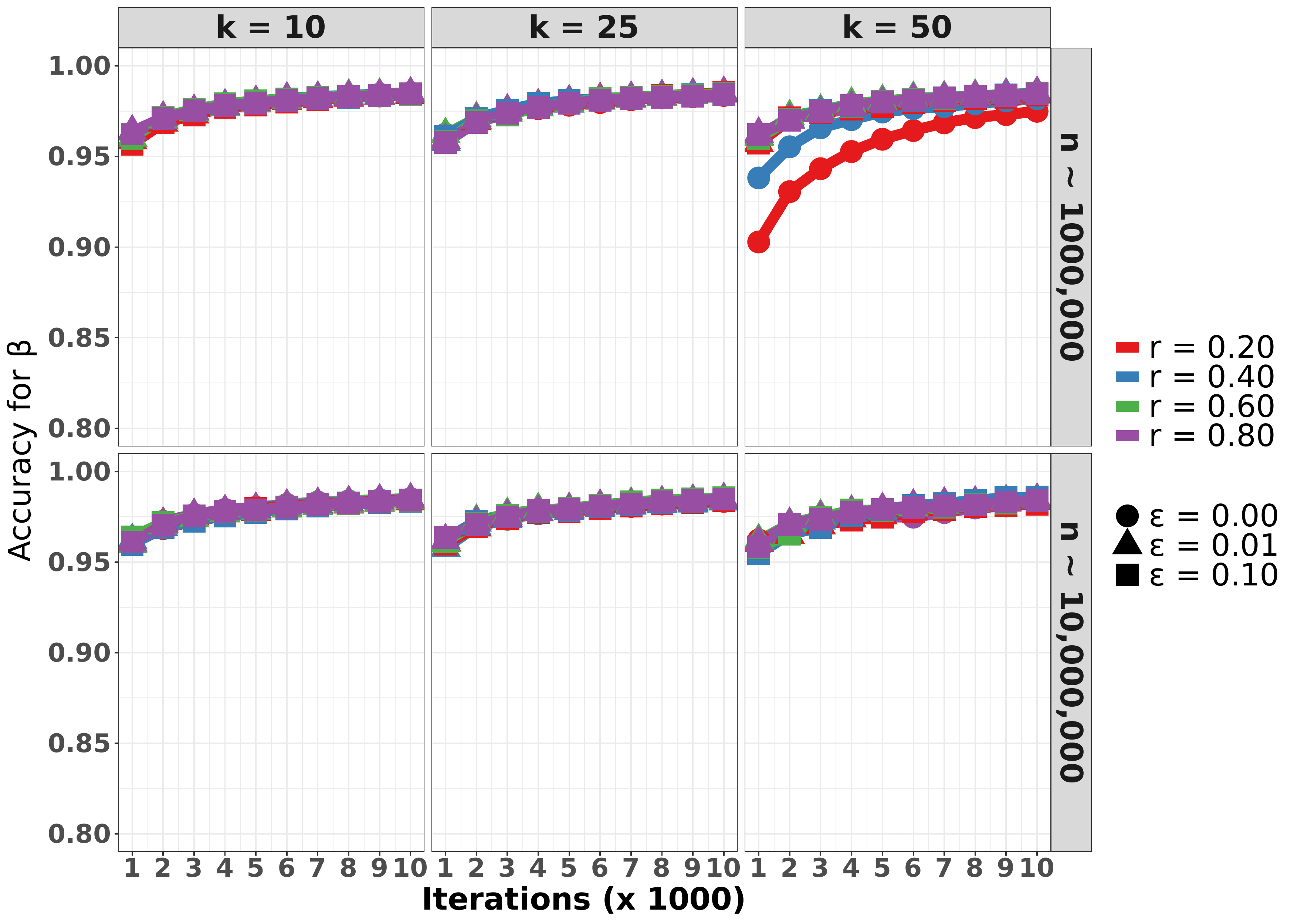}
  \caption{$\beta$}
  \label{fig:acc_log_mov_beta}
\end{subfigure}%
\begin{subfigure}{.5\textwidth}
  \centering
  \includegraphics[scale=0.23]{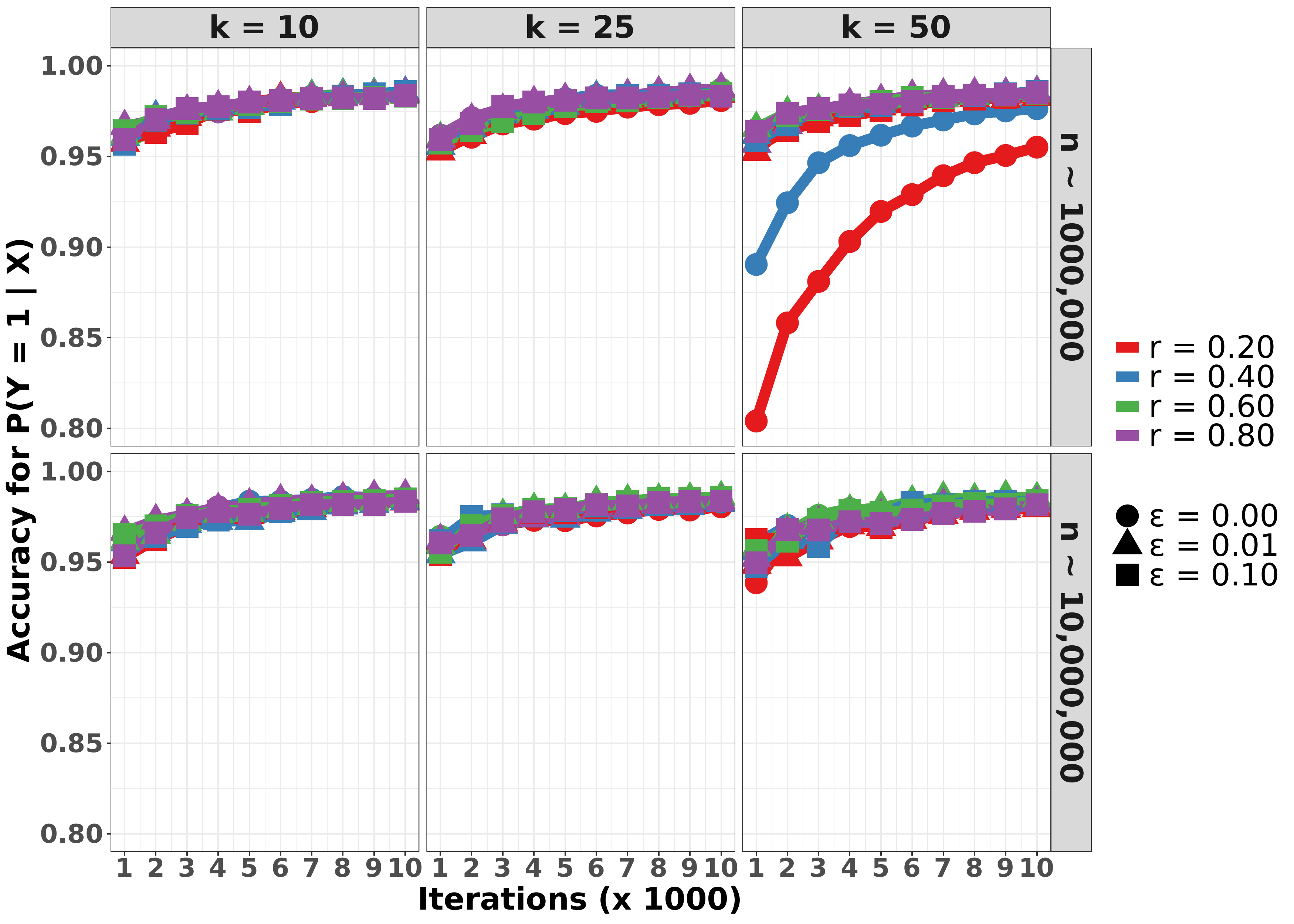}
  \caption{$\text{P}(Y = 1 \mid x_1 = 0, \ldots, x_{4}=0, x_5=1, x_6=0)$}
  \label{fig:acc_log_mov_prob}
\end{subfigure}
\caption{$\text{Acc}_{rk}(t)$ metric for inference on $\beta$ and $\text{P}(Y = 1 \mid x_1 = 0, \ldots, x_{4}=0, x_5=1, x_6=0)$ in the logistic regression model for MovieLens data with $r = 0.20, 0.40, 0.60, 0.80$ and $k=10, 25, 50$.}
\label{fig:acc_log_mov}
\end{figure}

\begin{figure}[H]
\begin{subfigure}{.5\textwidth}
  \centering
  \includegraphics[scale=0.23]{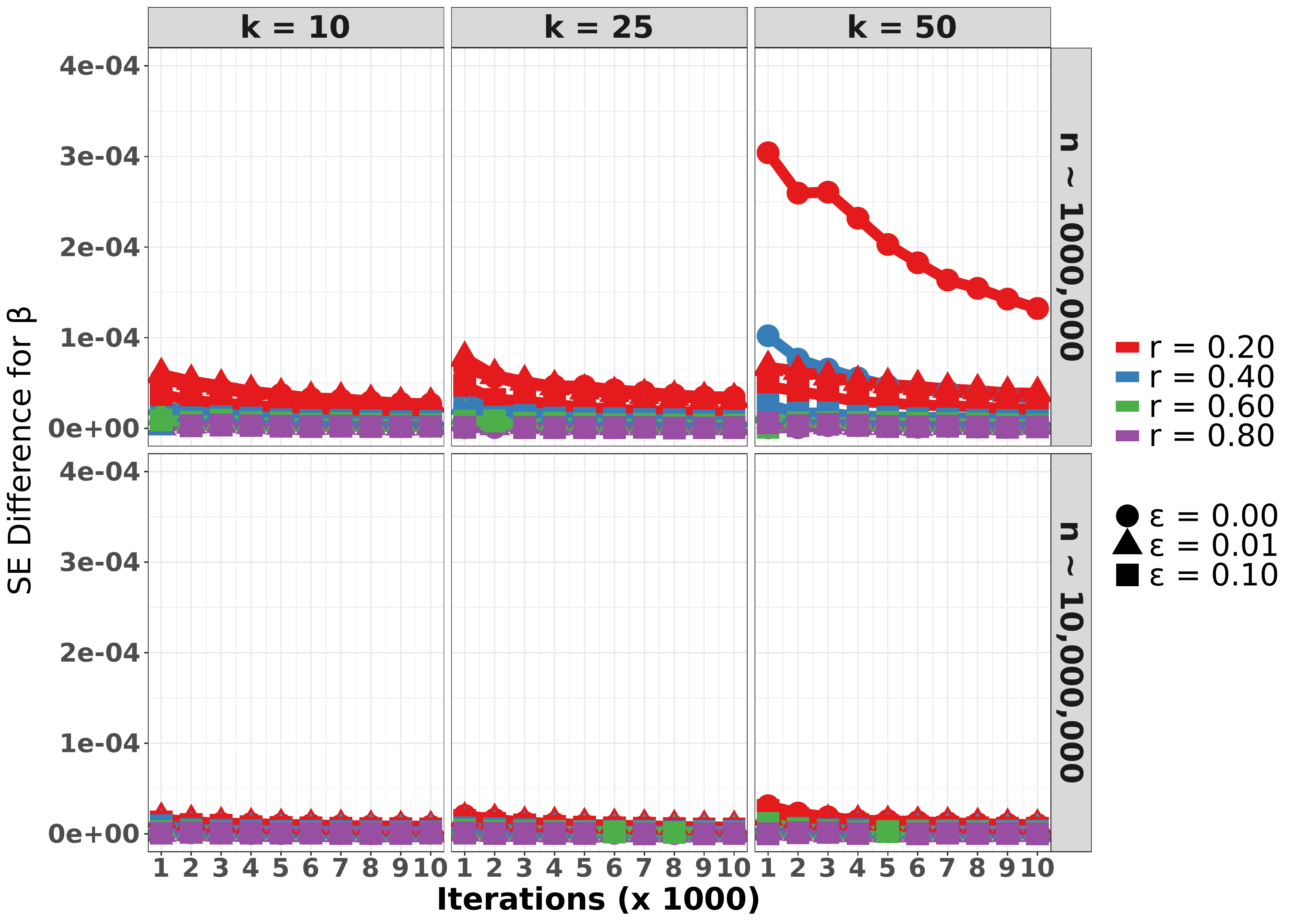}
  \caption{$\beta$}
  \label{fig:mcse_log_mov_beta}
\end{subfigure}%
\begin{subfigure}{.5\textwidth}
  \centering
  \includegraphics[scale=0.23]{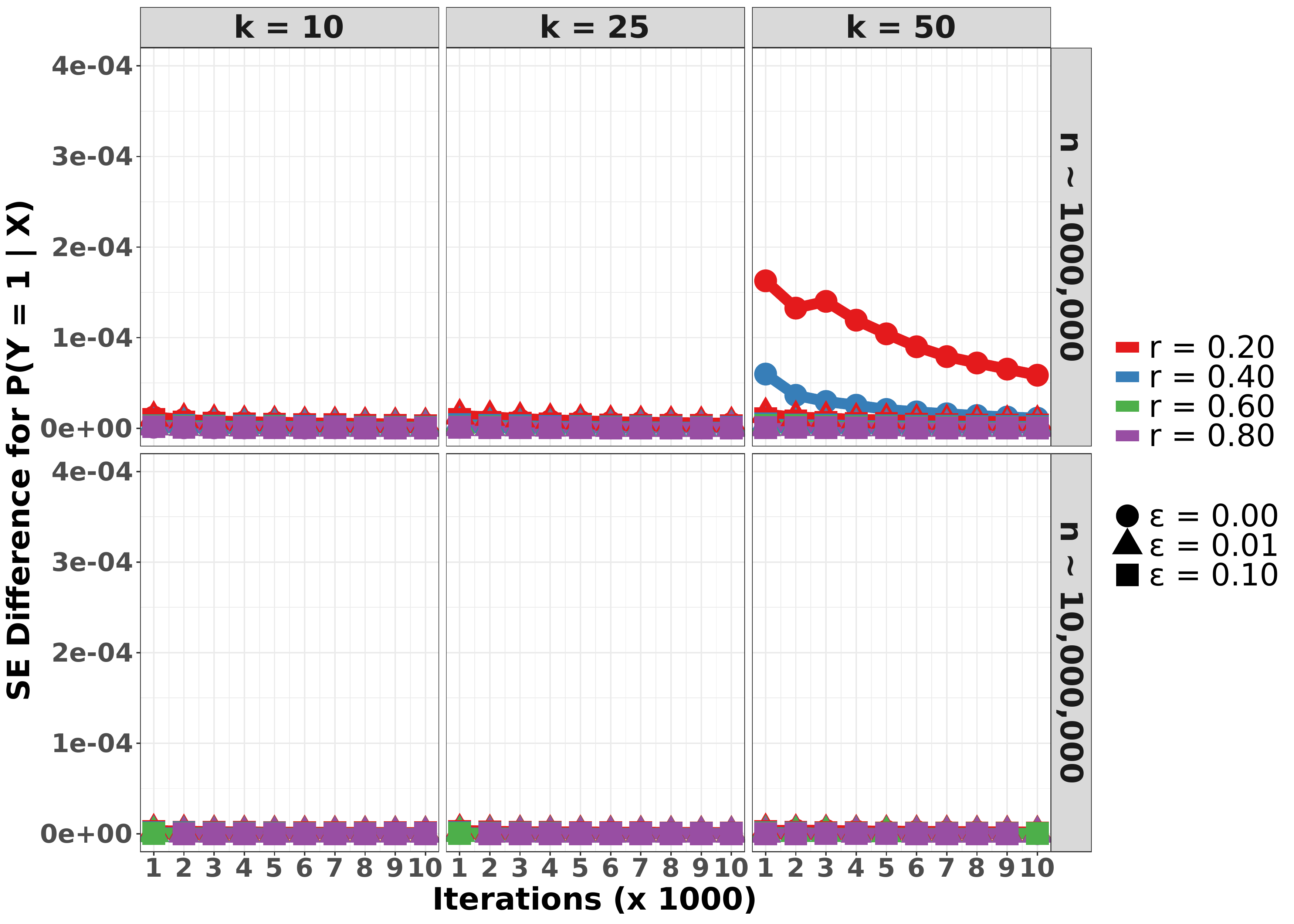}
  \caption{$\text{P}(Y = 1 \mid x_1 = 0, \ldots, x_{4}=0, x_5=1, x_6=0)$}
  \label{fig:mcse_log_mov_prob}
\end{subfigure}
\caption{$\text{SE}_{rk}(t)$ metric for inference on $\beta$ and $\text{P}(Y = 1 \mid x_1 = 0, \ldots, x_{4}=0, x_5=1, x_6=0)$ in the logistic regression model for MovieLens data with $r = 0.20, 0.40, 0.60, 0.80$ and $k=10, 25, 50$.}
\label{fig:mcse_log_mov}
\end{figure}

\textul{\emph{Accuracy comparisons.}} The $\text{Acc}_{rk}(t)$ and $\text{SE}_{rk}(t)$ metrics are evaluated for $\beta$ and $\text{P}(Y = 1 \mid x)$, where $x = (0, 0, 0, 0, 1, 0)$ is the predictor for a popular movie. 
When $\epsilon \neq 0$, the $\text{Acc}_{rk}(t)$ values for $\beta$ and $\text{P}(Y = 1 \mid x)$ are fairly similar and very high for every $r, k, t$ (Figures \ref{fig:acc_log_mov} and \ref{fig:mcse_log_mov}). If $\epsilon=0$, then  the setup of Theorem \ref{pg_thm} is violated. In this setting we observe slower convergence of the $\text{Acc}_{rk}(t)$ values to 1 when $k$ and $r$ are relatively large and small, respectively ($k=50$; $r=0.20, 0.40$). 

\textul{\emph{Monte Carlo standard error comparisons.}} Likewise, when $\epsilon \neq 0$, the $\text{SE}_{rk}(t)$ values are insensitive to the choice of $n, r, k$ and decay to 0 quickly after $t = 2000$. For the same reasons as discussed in the previous paragraph,  the convergence of $\text{SE}_{rk}(t)$ to 0 is much slower when $\epsilon=0$, especially when $k$ and $r$ are relatively large and small, respectively.

\textul{\emph{Run-time comparisons.}} 
The ADDA Algorithm \ref{ADPG} is three to five times faster than its parent DA (Figure \ref{fig:log_mov_time}). The gain in run-times are insensitive to the choices of $n, r,k,$ or $\epsilon$. The minor variations in run-time gains are present due to the varying loads on the cluster when the computations are performed. For example, when $k=50$ and $n=10^7$, the gain in run-times are close to three for $r = 0.60, 0.80$ and $\epsilon=0$. On the other hand, the gains are close to five for the same values of $n, k, r$ but $\epsilon$ equalling $0.01$ or $0.10$. A distributed (or parallel) implementation of the parent DA corresponds to an ADDA with $r=1.00$ (or $\epsilon = 1.00$). For every choice of $n, k$, and $\epsilon \in \{0.00, 0.01, 0.10\}$, the asynchronous implementations of ADDA with $r \in \{0.20, 0.40, 0.60, 0.80\}$ are two to three times faster than the distributed implementation of parent DA. An interesting observation is that the gain in runtime going from $r=1$ to $r=0.8$ is more significant than  going from $r=0.8$ to $r=0.2$. One possible reason is that there is a small minority of workers in each iteration who take significantly more time than others. Introducing asynchronous computation can allow us (with probability $1-\epsilon$) to bypass these workers for that particular iteration and speed up the computation. Overall, the results strongly demonstrate the utility of the proposed ADDA approach, as summarized below. 

\textul{\emph{Conclusions.}} 
We conclude that for a sufficiently large $t$ and every $n, r, k$, the $\text{Acc}_{rk}(t)$ and $\text{SE}_{rk}(t)$ values for the ADDA chain corresponding to Algorithm \ref{ADPG} are close to 1 and 0, respectively, as long as $\epsilon$ is positive. The  ADDA chain is also three to five times faster than its parent DA depending on $(k,r)$. The asynchronous computations are advantageous in that ADDA with $r < 1$ are two to three times faster than the distributed implementation of parent DA. This shows that the ADDA Algorithm \ref{ADPG} is a promising alternative to PG DA for accurate and efficient Bayesian inference in logistic regression models for massive data. 

\subsubsection{Mixed-Effects Modeling}
\label{mov-lme}

We randomly select two subsets of users in the MovieLens data such that the total number of ratings in them are approximately $10^5$ and $10^6$, respectively. We apply Algorithm \ref{ADGLM} for fitting linear mixed-effects models to the two subsets. There are six fixed and random effects predictors, including an intercept, three dummy variables for the movie categories, and the remaining two for a movie's popularity and a user's mood. The dimensions of $\beta$ and $\Sigma$ are $6 \times 1$ and $6 \times 6$, respectively. We fit the linear mixed effects model in \eqref{eq:mix1} following the previous section, except that we run the DA algorithm for 10,000 instead of 20,000 iterations and use three choices of $\epsilon = 0, 0.01, 0.10$. This setup is replicated ten times by randomly choosing the users in each replication. 

\begin{figure}[H]
\begin{subfigure}{.5\textwidth}
  \centering
  \includegraphics[scale=0.21]{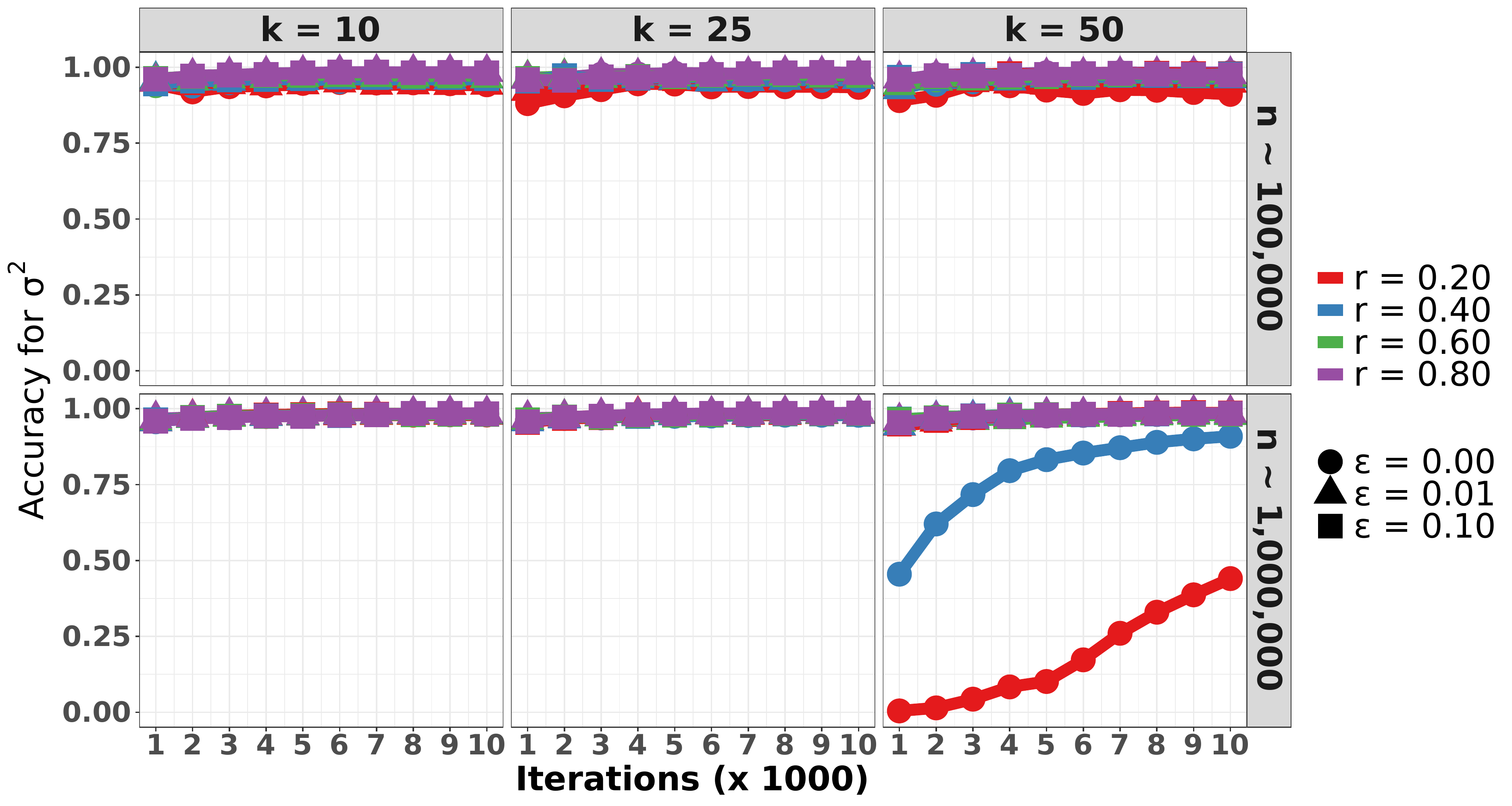}
  \caption{$\text{Acc}_{rk}(t)$}
  \label{fig:acc_mov_sig}
\end{subfigure}%
\begin{subfigure}{.5\textwidth}
  \centering
  \includegraphics[scale=0.21]{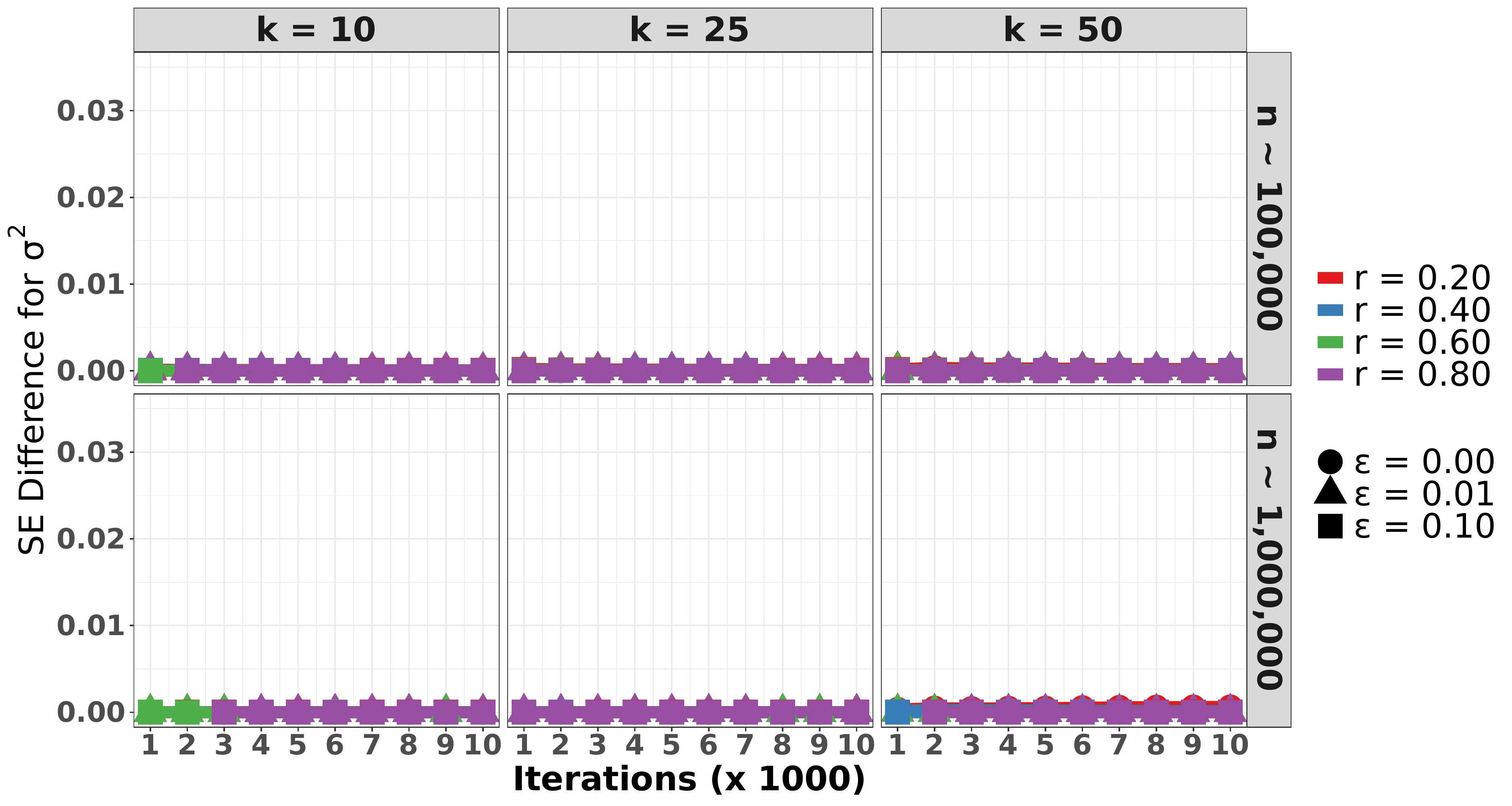}
  \caption{$\text{SE}_{rk}(t)$}
  \label{fig:mcse_mov_sig}
\end{subfigure}
\caption{$\text{Acc}_{rk}(t)$ and $\text{SE}_{rk}(t)$ metrics for inference on $\sigma^2$ in the mixed-effects model for MovieLens data with $r = 0.20, 0.40, 0.60, 0.80$ and $k=10, 25, 50$.}
\label{fig:acc_mcse_mov_sig}
\end{figure}

\textul{\emph{Accuracy comparisons.}} 
We evaluate $\text{Acc}_{rk}(t)$ and $\text{SE}_{rk}(t)$ metrics for $\beta$, $\Sigma$, and $\sigma^2$, respectively.
When $\epsilon \neq 0$ and $r \neq 0.2$, the $\text{Acc}_{rk}(t)$ values for $\sigma^2$, $\beta$, and $\Sigma$ converge to 1 quickly for every $n, k, t$ (Figures \ref{fig:mov_beta_dmat} and \ref{fig:acc_mcse_mov_sig}). For smaller $t$ values, $\text{Acc}_{rk}(t)$ is relatively low for $n=10^6$ and $r=0.2$; however, the differences between  $\text{Acc}_{rk}(t)$ values disappear for a sufficiently large $t$ given $\epsilon \neq 0$. 

On the other hand, when $\epsilon = 0$ the convergence of $\text{Acc}_{rk}(t)$ to 1 is very slow compared to the previous cases. For example, $\text{Acc}_{rk}(t)$ values for $\sigma^2$ are noticeably small when $r \in \{0.20, 0.40\}$ and $k=50$. This is a consequence of violating the assumptions of Theorem \ref{thm_GLM}; therefore, we recommend choosing a small but positive $\epsilon$ in practice.

\textul{\emph{Monte Carlo standard error comparisons.}} Theorem \ref{thm_GLM} also impacts the decay of $\text{SE}_{rk}(t)$ values to 0 when $\epsilon = 0$. Specifically, if $\epsilon \neq 0$, then $\text{SE}_{rk}(t)$ values for $\sigma^2$, $\beta$, and $\Sigma$ converge to 0 quickly for every $n, k, t$ (Figures \ref{fig:mcse_mov_beta_dmat} and \ref{fig:acc_mcse_mov_sig}). This pattern is maintained for $\sigma^2$ even when $\epsilon = 0$. The $\text{SE}_{rk}(t)$ values for $\beta$ and $\Sigma$ decay to 0 relatively slowly for every $k$ when $n$ is large, $r=0.20$, and $\epsilon=0$; however, if $t$ is sufficiently large, then the $\text{SE}_{rk}(t)$ value is small for every choice of $n, r, k$, and $\epsilon$. 

\textul{\emph{Run-time comparisons.}} 
The  ADDA Algorithm \ref{ADGLM} is three to five times faster than its parent DA for all the choices of $n, r,k,$ and $\epsilon$ (Figure \ref{fig:mov_time}). For a given $k$, the gain in run-time increases with $n$ for every $\epsilon$ and $r$.  On the other hand, for a given $n$, the gain in run-times decreases with increasing $k$ for every $\epsilon$ and $r$ due to the increased communication among manager and worker processes. Similar to our observations in   logistic regression, a distributed (or parallel) implementation of the parent DA  is two to three times slower than the asynchronous implementations of ADDA with $r < 1$ and $\epsilon < 1$ for every choice of $n, k$, and $\epsilon$. 
Unlike the case for logistic regression, varying loads on the cluster have a minimal impact on the run-times. 

\textul{\emph{Conclusions.}}
If $\epsilon$ is positive and $t$ is sufficiently large, then the $\text{Acc}_{rk}(t)$ and $\text{SE}_{rk}(t)$ values of Algorithm \ref{ADGLM} are close to 1 and 0, respectively, for every $n, r, k$. The ADDA chain corresponding to Algorithm \ref{ADGLM} is also three to five times faster than its parent DA depending on $(k,r)$. The asynchronous computations are more efficient than parallel computations in that ADDA with $r < 1$ is faster than the distributed implementation of parent DA.   
We conclude that ADDA Algorithm \ref{ADGLM} is a promising alternative to the marginal DA for accurate and efficient Bayesian inference in linear mixed-effects models for massive data.

\section{Discussion} \label{discussion}

It is interesting to explore extensions of our ADDA schemes and theoretical results. First, linear mixed-effects have been an important class of models for evaluating the performance of a new class of DA algorithms \citep{MenDyk99,DykMen01,DykPar08}. The excellent empirical performance of ADDA in Sections \ref{sec:sim-adda-Mixed-Effects} and \ref{mov-lme} suggests that it is possible to extend the ADDA scheme to other classes of hierarchical Bayesian models. For example, if the random effects in a linear mixed-effects model are assigned a multivariate $t$ distribution, then the I and P steps of the new model are similar to those in Section \ref{sec:adda-Mixed-Effects} \citep{Pieetal01}. This implies that the Algorithm \ref{ADGLM} is  easily extended to robust extensions of linear mixed-effects models.

\begin{figure}[H]
  \centering
  \includegraphics[scale=0.25]{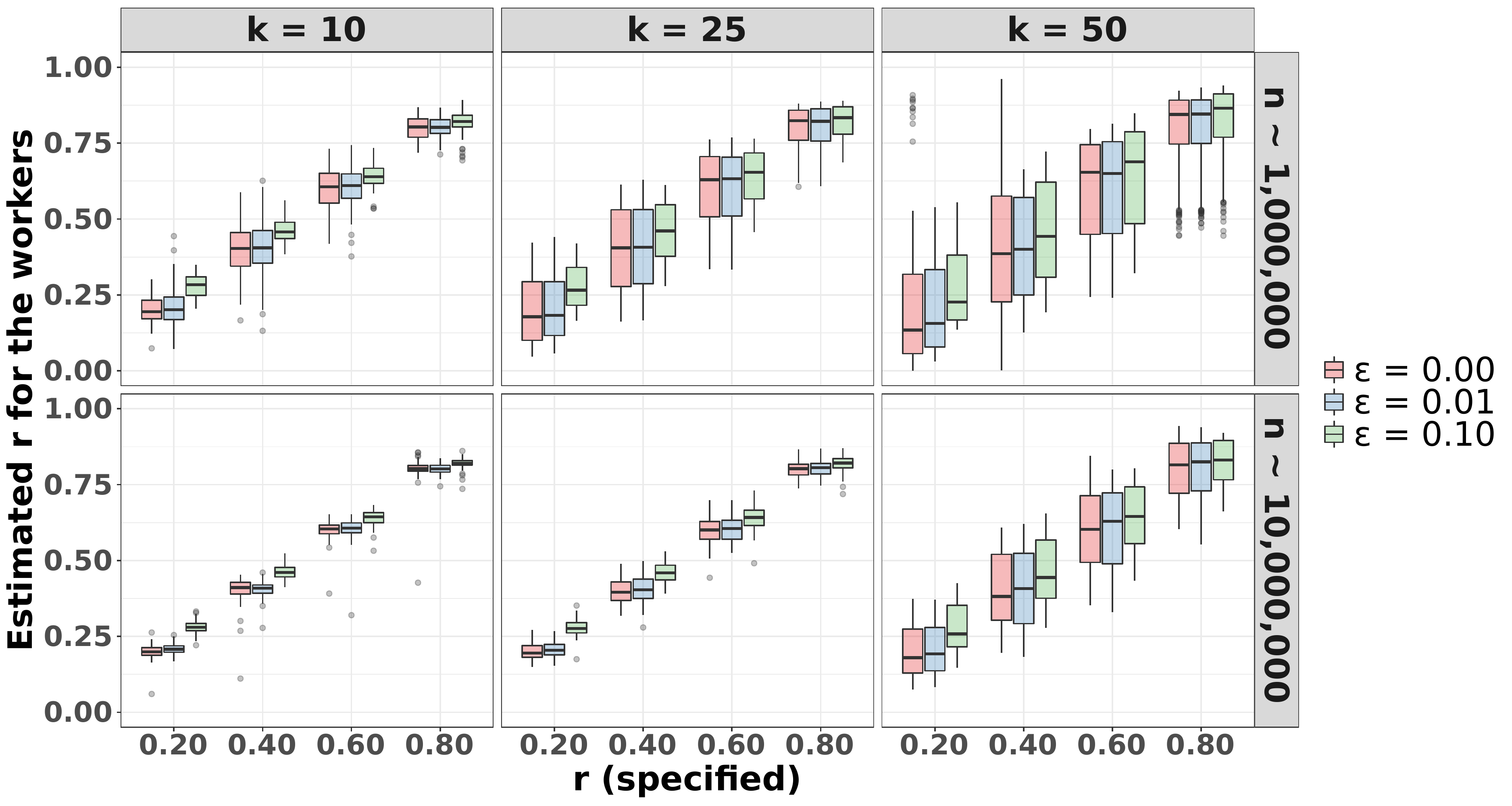}
  \caption{Estimated $r$ for every worker process vs the $r$ specified in the logistic regression model for MovieLens data. The estimated $r$ for a worker is defined as the fraction of the number of times the manager accepts its I step results over the total number of DA iterations.}
  \label{fig:est-r}
\end{figure}

Second, our real data analysis illustrates the dependence of accuracy and Monte Carlo standard errors on the choice of $(k, r, \epsilon)$. Due to varying loads on worker processes in a cluster, some workers return their I step results to the manager more often than others. Our empirical results suggest this variability decreases with $n$. For example, if we define the empirical estimate of $r$ for a worker as the fraction of the number of times the manager accepts its I step results over the total number of DA iterations, then empirical estimates of $r$ are closer to the true $r$ as $n$ increases for every $k$ and worker process; see Figure \ref{fig:est-r}. Furthermore, the Bernstein-von Mises theorem implies that the posterior distribution becomes increasingly Gaussian as $n$ increases. We are exploring techniques for balancing the Monte Carlo error (depending on $t, k, r, \epsilon$) and the statistical error (depending on $n$) for providing some guidance on the choice of $k$ and total number of MCMC iterations so that a desired accuracy is achieved by the ADDA algorithm.

Our DA extensions based on ADDA exploit two features of the augmented data model. First, the $k$ augmented data subsets are mutually independent given the original parameter and the observed data. Second, the conditional posterior distribution of the missing data on any subset depends only on the local copy of the observed data. Both assumptions are violated in models for time series data, including hidden Markov models (HMMs) and Gaussian state-space models. DA has been widely used for posterior inference and predictions in these models, but it suffers from similar computational bottlenecks in massive data settings that have been highlighted in this paper. Recently, this problem has been addressed for HMMs using online EM \citep{Cap11,LeFor13} and the divide-and-conquer technique \cite{WanSri21}. As part of future research, we will focus on extending the original DA algorithms using ADDA for scalable posterior inference and predictions in HMMs and related state-space models.


\section*{Acknowledgements}
Sanvesh Srivastava is partially supported by grants from the Office of Naval Research (ONR-BAA N000141812741) and the National Science Foundation (DMS-1854667/1854662). The code used in the experiments is available at \url{https://github.com/blayes/ADDA}. 

\bibliographystyle{Chicago}
\bibliography{papers}

\newpage

\pagenumbering{gobble}
\pagenumbering{arabic}
\renewcommand*{\thesection}{\Alph{section}}
\setcounter{section}{0}

\section{Supplemental Document for ``Asychronous and Distributed Data Augmentation for Massive Data Settings"}

\subsection{Proof of Theorem \ref{Harris}} \label{prove_harris}
Since both the AD-I and AD-P step leave the joint posterior distribution $\Pi (\cdot \mid \Dcal_{\text{obs}})$ invariant, it is 
enough to establish that the ADDA chain is irreducible, aperiodic and Harris recurrent. Note that the ADDA chain is a mixture 
of the parent DA chain(with prob $\epsilon$), and a pure asynchronous chain where only an $r$-fraction update regime is 
chosen for the workers (with prob $1-\epsilon$). Since the original DA chain is irreducible and aperiodic, it follows that the 
ADDA chain is also irreducible and aperiodic. { Lastly, let $K_{ADDA}$ and $K_{AD}$ denote the transition kernels 
for the ADDA chain and the purely asynchronous chain mentioned above. Then, for any $A \in\mathcal{B}(\mathbb{M}
\times\Theta)$ with $\Pi(A \mid \Dcal_{obs})=0$, $K_{ADDA} ((\Dcal, \theta), A) = \epsilon K_{DA}((\Dcal, \theta), A) + 
(1-\epsilon)K_{AD}((\Dcal, \theta), A))\leq 1-\epsilon$, where the inequality follows from the assumption that $K_{DA}$ is 
absolute continuous wrt. $\Pi$. Let $\left\{ (\Dcal^{(n)}, \theta^{(n)}) \right\}_{n=0}^\infty$ denote a realization of the ADDA 
Markov chain. Then for any $(\Dcal^{(0)}, \theta^{(0)})\in \MM \otimes  \Theta$, we have 
\begin{eqnarray*}
& & Pr((\Dcal^{(n)}, \theta^{(n)}) \in A \text{ for } 1 \leq n \leq N \vert (\Dcal^{(0)}, \theta^{(0)}))\\
&=& E \left[ \prod_{n=1}^{N-1}I_{A}((\Dcal^{(n)}, \theta^{(n)}) \in A) K_{ADDA} ((\Dcal^{(N-1)}, \theta^{(N-1)}), A) \vert 
(\Dcal^{(0)}, \theta^{(0)}) \right]\\ 
&\leq& (1-\epsilon) Pr((\Dcal^{(n)}, \theta^{(n)}) \in A \text{ for } 1 \leq n \leq N-1 \vert (\Dcal^{(0)}, \theta^{(0)})). 
\end{eqnarray*}

\noindent
Here $I_{A}(\cdot)$ is the indicator function. By repeating the above argument, we get that 
$$
Pr((\Dcal^{(n)}, \theta^{(n)}) \in A \text{ for } 1 \leq n \leq N \vert (\Dcal^{(0)}, \theta^{(0)})) \leq (1-\epsilon)^N. 
$$

\noindent
With $\epsilon>0$, by letting $N$ go to infinity, it follows that $Pr((D^{(n)}, \theta^{(n)})\in \mathcal{A} \text{ for all } n 
\vert(D^{(0)}, \theta^{(0)})) =0$. Then by Theorem 6 of \cite{roberts2006harris}, the ADDA chain is Harris recurrent.}

\subsection{The switched ADDA chain and geometric ergodicity}

\noindent
{ Let $K_{ADDA}$ denote the transition kernel of the ADDA chain described in Section \ref{sec:adda:-gener-fram}, 
and $\tilde{K}_{ADDA}$ denote the transition kernel of the switched ADDA chain, where the AD-P step is performed before 
the AD-I step. Hence, the one-step dynamics of the switched ADDA chain from $(\theta, \Dcal)$ to $(\theta', \Dcal')$ is 
described as follows. The manager draws $\theta'$ from $f( \cdot \mid \Dcal, \Dcal_{\mbox{obs}})$ and distributes it to 
the $K$ workers. The $i^{th}$ worker generates a draw $\Dcal'_i$ from $f( \cdot \mid \theta', \Dcal_{\mbox{obs}})$ and 
sends to manager (the draw is truncated if the next $\theta$ value is received from the manager before the draw is 
finished). As soon as the manager receives the new $\Dcal'_i$s from the required fraction of workers ($r$ with probability $1-
\epsilon$, and $1$ with probability $\epsilon$), it assigns $\Dcal'_i = \Dcal_i$ for the remaining fraction, and proceeds with 
the next $\theta$ draw. For both the ADDA and the switched ADDA chain, the marginal $\theta$-process is not in general 
Markov. However, it can be easily seen that the marginal $\Dcal$ process for both chains is Markov. This follows from the 
fact that for both chains, the distribution of the latest $\theta$ value given the entire past depends only on the immediately 
previous $\Dcal$ value, and the distribution of the latest $\Dcal$ value given the entire past depends only on the immediately  
previous $\theta$ and $\Dcal$ values. 

Let $\left\{ (\tilde{\theta}^{(t)}, \tilde{\Dcal}^{(t)}) \right\}_{t \geq 0}$ denote a realization of the switched ADDA chain. Since 
\begin{eqnarray*}
P(\tilde{\Dcal}^{(t)} \in A \mid \tilde{\Dcal}^{(0)}) 
&=& P(\tilde{\Dcal}^{(t)} \in A \mid \tilde{\Dcal}^{(0)}, \tilde{\theta}^{(0)})\\
&=& P((\tilde{\theta}^{(t)}, \tilde{\Dcal}^{(t)}) \in \Theta \times A \mid \tilde{\Dcal}^{(0)}, \tilde{\theta}^{(0)}), 
\end{eqnarray*}

\noindent
for any Borel subset $A$ of $\mathbb{M}$, it follows that if the switched ADDA chain $\left\{ (\tilde{\theta}^{(t)}, 
\tilde{\Dcal}^{(t)}) \right\}_{t \geq 0}$ is geometrically ergodic for any choice of initial distribution, then so is its marginal chain 
$\left\{ \tilde{\Dcal}^{(t)}) \right\}_{t \geq 0}$. 

Now, let $\left\{ (\theta^{(t)}, \Dcal^{(t)}) \right\}_{t \geq 0}$ denote a realization of the ADDA chain. Note that $\Dcal^{(t)}$ 
can be denoted as $h(\theta^{(t)}, \Dcal^{(t)})$ for an appropriate function $h$. Also 
$$
(\theta^{(t)}, \Dcal^{(t)}) \mid \Dcal^{(t)}, (\theta^{(0)}, \Dcal^{(0)}) \stackrel{d}{=} (\theta^{(t)}, \Dcal^{(t)}) \mid \Dcal^{(t)}. 
$$

\noindent
It follows that the marginal chain $\left\{ \Dcal^{(t)} \right\}_{t \geq 0}$ is homogeneously functionally de-initializing for 
$\left\{ (\theta^{(t)}, \Dcal^{(t)}) \right\}_{t \geq 0}$. It follows by Corollary 2 of \cite{roberts2001markov}  that if 
$\left\{ \Dcal^{(t)} \right\}_{t \geq 0}$ is geometrically ergodic for any choice of initial distirbution, then so is $\left\{ (\theta^{(t)}, 
\Dcal^{(t)}) \right\}_{t \geq 0}$. Finally, by noting that the time homogeneous Markov chains $\left\{ \tilde{\Dcal}^{(t)} \right\}_{t 
\geq 0}$ and $\left\{ \Dcal^{(t)} \right\}_{t \geq 1}$ have the same transition mechanism gives the following lemma. 
\begin{lemma} \label{Switched:ADDA:Chain}
If the switched ADDA chain with transition kernel $\tilde{K}_{ADDA}$ is geometrically ergodic for any choice of initial 
distribution, then so is the ADDA chain with transition kernel $K_{ADDA}$. 
\end{lemma}}

\subsection{Proof of Theorem \ref{pg_thm}} \label{prove_pg} \label{sec:proof-theor-refpg_th}

	{ We will prove geometric ergodicity of the switched ADDA chain, which will be enough to establish geometric 
	ergodicity of the ADDA chain based on Lemma \ref{Switched:ADDA:Chain}.} 
	
	In the following calculation of conditional expectations, if not otherwise specified, we assume conditioning on the data. 
	Define the drift function ${\bf V}(\omega,\beta)=\beta^{T}\beta + \sum_{i=1}^{K} {\bf V}_{i}(\omega_{(i)})$ where 
	${\bf V}_{i}(\omega_{(i)})=\sum_{j=1}^{n_{i}}((\frac{1}{w_{j(i)}})^c+w_{j(i)})$ where $c>0$ is a fixed positive number 
	that will be defined later in the proof. Notice the support for $(\omega,\beta)$ is $S = \mathbb{R}_{+}^{n}\times 
	\mathbb{R}$, and for any $x>0$, $\left\{ (\omega,\beta) : V(\omega,\beta)\leq x \right\}$ is compact in $S$. Thus the 
	function introduced is unbounded off compact sets in $S$. It remains to show that a geometric drift condition holds, i.e. 
	to show that there exist a $\rho \in [0,1)$ and $L>0$, such that
	\begin{eqnarray} \label{drift_cond}
	E({\bf V}(\omega,\beta)|(\omega^{*},\beta^{*}))\leq \rho {\bf V}(\omega^{*},\beta^{*}) + L
	\end{eqnarray}
	where the expectation is taken with respect to the transition kernel of $\Phi_{ADPG}$. 
	
	Firstly notice that
	\begin{eqnarray}\label{cond_ex}
	E({\bf V}(\omega,\beta)|\omega^{*},\beta^{*})= E(\beta^{T}\beta|\omega^{*})+\sum_{i=1}^{K}E(E({\bf V}_{i}(\omega_{(i)})|\beta,\omega^{*})|
	\omega^{*}).
	\end{eqnarray}
	
	Here 
	\begin{eqnarray} \label{bdd1}
	E(\beta^{T}\beta|\omega^{*}) = && tr((X^{T}\Omega X+\Sigma_{\beta}^{-1})^{-1}) +d^{T}V_{\omega^{*}}^{2}d \nonumber
	\\ && \leq L_{1}
	\end{eqnarray}
	where $d = (X^T \kappa + \Sigma_{\beta}^{-1} \mu_{\beta})$ and  $L_{1}=tr(\Sigma_{\beta})+d^{T}\Sigma_{\beta}^{2}d$.
	
	Now consider the inner part of the double expectation in \eqref{cond_ex}.  In the AD-I step, there is at least a probability 
	 $p_m=\epsilon$ that worker $i$ gets to send an updated value to the manager. It follows that:
	\begin{eqnarray}\label{eq_inner}
	& & E({\bf V}_{i}({\omega_{(i)}})|\beta,\omega^{*}) \nonumber\\
	&\leq& \left( 1 - p_{m} \right) \sum_{i=1}^{K}{\bf V}_{i}({\omega_{(i)}^{*}}) + 
	\sum_{i=1}^{K} E({\bf V}_{i}(\omega_{(i)})|\beta,\text{worker i successfully sends update to manager}). \nonumber\\
	\end{eqnarray}
	
	\noindent
	We now focus on the second term in \eqref{eq_inner}. 
	\begin{eqnarray}
	\sum_{i=1}^{K} E({\bf V}_{i}({\bf w})|\beta,\text{worker i successfully sends update to manager}) 
	\\= \sum_{i=1}^{K}\sum_{j=1}^{n_{i}}\left[E(\frac{1}{\omega_{j(i)}^c}|\omega_{j(i)}\sim PG(s_{j(i)}, |x_{j(i)}^{T}\beta|)) 
	+ E(\omega_{j(i)}|\omega_{j(i)}\sim PG(s_{j(i)}, |x_{j(i)}^{T}\beta|))\right]
	\end{eqnarray}
	
	By \cite{Poletal13}, 
	\begin{eqnarray}
	E(\omega_{j(i)}|\omega_{j(i)}\sim PG(s_{j(i)}, |x_{j(i)}^{T}\beta|))= \frac{s_{j(i)}}{2}(\frac{e^{|x_{j(i)}^{T}\beta|}-1}{e^{|x_{j(i)}^{T}\beta|}+1}) \nonumber
	\end{eqnarray}
	where by Lemma 1 of \cite{WanRoy18}, $\frac{1}{2}(\frac{e^{|x_{j(i)}^{T}\beta|}-1}{e^{|x_{j(i)}^{T}\beta|}+1})\leq\frac{1}{4}$. Thus  
	\begin{eqnarray}
	E(\omega_{j(i)}|\omega_{j(i)}\sim PG(s_{j(i)}, |x_{j(i)}^{T}\beta|))\leq \frac{s_{j(i)}}{4}.\nonumber
	\end{eqnarray}	
	
	Then to deal with $E(\frac{1}{\omega_{j(i)}^c}|\omega_{j(i)}\sim PG(s_{j(i)}, |x_{j(i)}^{T}\beta|)) $, we establish the 
	following Lemma:
	\begin{lem}\label{pg_lm1}
	For $w\sim PG(a, b)$, and any integer $c\geq a$, there exist a positive number $M_{(a,c)}>0$, which is a function of 
	$a$ and $c$ only,  such that $E(\frac{1}{w^{c}}) \leq e^{\frac{ab}{2}}M_{(a,c)}$
	\end{lem}
	
	The proof of this Lemma is given in Supplemental Section \ref{prove_lem1}. By this Lemma and setting $c = 
	max_{ij}\{s_{j(i)}\}$, we have
	\begin{eqnarray} \label{}
	E(\frac{1}{\omega_{j(i)}^c}|\omega_{j(i)}\sim PG(s_{j(i)}, |x_{j(i)}^{T}\beta|)) \leq 
	e^{\frac{s_{j(i)}|x_{j(i)}^{T}\beta|}{2}}M(s_{j(i)}, c).
	\end{eqnarray}
	Thus for the outer expectation of \eqref{cond_ex}, we need to consider 
	$M(s_{j(i)}, c)E(e^{\frac{s_{j(i)}|x_{j(i)}^{T}\beta|}{2}}|\omega^{*},\beta^{*})$. 
	
	Recall that $\beta|\omega^{*} \sim N(m_{\omega^{*}}, V_{\omega^{*}})$, thus $\frac{1}{2} x_{j(i)}^{T}\beta|\omega^{*}
	\sim N(\mu_{j(i)}, \sigma^2_{j(i)})$ where 
	\begin{eqnarray}
	\sigma^{2}_{j(i)} =\frac{1}{4}x_{j(i)}^{T}(X^{T}\Omega X+\Sigma_{\beta}^{-1})^{-1}x_{j(i)} 
	\leq 
	\frac{1}{4}x_{j(i)}^{T}\Sigma_{\beta} x_{j(i)},
	\end{eqnarray}
	and
	\begin{eqnarray}
	|\mu_{j(i)}| =\frac{1}{2} |x_{j(i)}^T V_{\omega^{*}} d|\leq \frac{1}{2}\sqrt{x_{j(i)}^T V_{\omega^{*}} x_{j(i)}} \sqrt{d^T 
	V_{\omega^{*}} d} \leq \frac{1}{2} \sqrt{x_{j(i)}^T \Sigma_{\beta} x_{j(i)}} \sqrt{d^T \Sigma_{\beta} d}.
	\end{eqnarray}
	
	Then by the moment generating function of folded normal random variable, and combining the above equations, we 
	have
	\begin{eqnarray}
	E(e^{\frac{s_{j(i)}|x_{j(i)}^{T}\beta|}{2}}|\omega^{*},\beta^{*})  \leq 
	\exp \{s_{j(i)}\mu_{j(i)} + \frac{s_{j(i)}^{2}\sigma^2_{j(i)}}{2}\} + \exp \{-s_{j(i)}\mu_{j(i)} + \frac{s_{j(i)}^{2}\sigma^2_{j(i)}}{2}\}\leq L_{ij}
	\end{eqnarray}
	where $L_{ij}=2 \exp \{\frac{1}{2}s_{j(i)}\sqrt{x_{j(i)}^T \Sigma_{\beta} x_{j(i)}} \sqrt{d^T \Sigma_{\beta} d} + \frac{s_{j(i)}^{2}x_{j(i)}^{T}\Sigma_{\beta} x_{j(i)}}{8}\}$, which is a function of known constants only. 
	
	Thus 
	\begin{eqnarray}
	\sum_{i=1}^{K}E(E({\bf V}_{i}(\omega_{(i)})|\beta)|\omega^{*},\beta^{*}) \leq \left( 1 - p_{m} \right) \sum_{i=1}^{K}{\bf V}_{i}(\omega_{i}^{*}) + \sum_{i=1}^{K}\sum_{j=1}^{n_{i}}(L_{ij} +\frac{s_{j(i)}}{4}).
	\end{eqnarray}
	Then by letting $\rho  = 1-p_m$ and $L = L_1 + \sum_{i=1}^{K}\sum_{j=1}^{n_{i}}(L_{ij} +\frac{s_{j(i)}}{4})$, we establish \eqref{drift_cond} as desired.

\subsection{Proof of Lemma \ref{pg_lm1}} \label{prove_lem1}
	
	The density of PG$(a, b)$ is written as 
	\begin{eqnarray}
	f(x|a,b) = \left\{\cosh^{a} \left(\frac{b}{2} \right) \right\}\frac{2^{a-1}}{\Gamma(a)}\sum(-1)^{n}\frac{\Gamma(n+a)}{\Gamma(n+1)}\frac{(2n+a)}{\sqrt{2\pi}}x^{-\frac{3}{2}}\exp\{-\frac{(2n+a)^{2}}{8x}-\frac{b^{2}}{2}x\}
	\end{eqnarray}
	\subsubsection*{Case 1: b = 0}
	\begin{eqnarray}
	E(\frac{1}{x^{c}})	&& =\frac{2^{a-1}}{\Gamma(a)}\sum_{n=0}^{\infty}(-1)^{n}\frac{\Gamma(n+a)}{\Gamma(n+1)}\frac{(2n+a)}{\sqrt{2\pi}}\int_{0}^{\infty}x^{-\frac{3}{2}-c}\exp\{-\frac{(2n+a)^{2}}{8x}\}dx \\ \nonumber
	&& =\frac{2^{a-1}}{\Gamma(a)}\sum_{n=0}^{\infty}(-1)^{n}\frac{\Gamma(n+a)}{\Gamma(n+1)}\frac{1}{\sqrt{2\pi}}\frac{\Gamma(c+\frac{1}{2})4^{2c+1}}{(2n+a)^{2c}} \\
	&& \leq \frac{2^{a-1}}{\Gamma(a)}\sum_{n=0}^{\infty}\frac{\Gamma(n+a)}{\Gamma(n+1)}\frac{1}{\sqrt{2\pi}}\frac{\Gamma(c+\frac{1}{2})4^{2c+1}}{(2n+a)^{2c}} 
	\end{eqnarray}
	For $c\geq a$, $\frac{\Gamma(n+a)}{\Gamma(n+1)(2n+a)^{2c}}\leq\frac{1}{(n+a)^{2}}$, thus
	\begin{eqnarray}
	E(\frac{1}{x^{c}})	\leq\frac{2^{a-1}\Gamma(c+\frac{1}{2})4^{2c+1}}{\Gamma(a)\sqrt{2\pi}}\sum_{n=0}^{\infty}\frac{1}{(n+a)^{2}}
	\end{eqnarray}
	where $\sum_{n=0}^{\infty}\frac{1}{(n+a)^{2}}$ converges. Thus by letting $M_{(a,c)}=\frac{2^{a-1}\Gamma(c+\frac{1}{2})4^{2c+1}}{\Gamma(a)\sqrt{2\pi}}\sum_{n=0}^{\infty}\frac{1}{(n+a)^{2}}$ the proof is complete.

	\subsubsection*{Case 2: b $>$ 0}
	When $b>0$ and $c\geq a$, 
	\begin{eqnarray}
	E(\frac{1}{x^{c}})	&&=\cosh^{a}(\frac{b}{2})\frac{2^{a-1}}{\Gamma(a)}\sum_{n=0}^{\infty}(-1)^{n}  \frac{\Gamma(n+a)}{\Gamma(n+1)}\frac{(2n+a)}{\sqrt{2\pi}} \int_{0}^{\infty} x^{-\frac{3}{2}-c}\exp\{-\frac{(2n+a)^{2}}{8x}-\frac{b^{2}}{2}x\}dx
	\nonumber \\
	&&\leq \cosh^{a}(\frac{b}{2})\frac{2^{a-1}}{\Gamma(a)}\sum_{n=0}^{\infty}\frac{\Gamma(n+a)}{\Gamma(n+1)}\frac{(2n+a)}{\sqrt{2\pi}} \int_{0}^{\infty} x^{-\frac{3}{2}-c}\exp\{-\frac{(2n+a)^{2}}{8x}\}dx
	\nonumber \\
	&&\leq \cosh^{a}(\frac{b}{2})M_{(a,c)}=[\frac{e^{\frac{b}{2}}+e^{-\frac{b}{2}}}{2}]^{a}M_{(a,c)}
	\nonumber \\
	&&=e^{\frac{ab}{2}}[\frac{1+e^{-b}}{2}]^{a}M_{(a,c)}\leq e^{\frac{ab}{2}}M_{(a,c)}.
	\end{eqnarray}
	where the first and last inequality follows from $e^{-x}\in(0,1]$ for any $x\geq 0$, and the second inequality follows from results in Case 1. 

\subsection{Proof of Theorem \ref{hdim_varsel}} \label{prove_BL}
\label{sec:proof-theor-refhd}

	{ We will prove geometric ergodicity of the switched ADDA chain, which will be enough to establish geometric 
	ergodicity of the ADDA chain based on Lemma \ref{Switched:ADDA:Chain}.} 
	
	Define the following drift function:
	\begin{eqnarray}
	{\bf V}(\beta,\sigma^2,\tau) &&=\beta^{T}D_{\tau}^{-1}\beta+\sum_{j=1}^{p}\tau_{j}+w(y-X\beta)^{T}(y-X\beta)\nonumber 
	\\&&+ \sum_{j=1}^{p}\frac{1}{(\tau_{j})^{s/2}}
	+\sigma^2 + \frac{1}{\sigma^2}
	\end{eqnarray}
	where $w>0$ and $0<s<\frac{1}{2}$ are positive numbers which are to be defined later. It can be easily seend that 
	the above drift function is unbounded off compact sets for the parameters $(\beta,\sigma^2,\tau)$. Thus, we can 
	proceed to show that by choosing proper $w$ and $s$, there exist some constants $0\leq\rho<1$ and $\psi>0$ such 
	that 
	\begin{eqnarray} \label{drift_BL}
	E({\bf V}(\beta,\sigma^{2},\tau)|\beta^{*},\sigma^{2,*},\tau^{*})\leq \rho {\bf V}(\beta^{*},\sigma^{2,*},\tau^{*})+\psi
	\end{eqnarray}
	Here this expectation is with respect to the transition kernel of $\Phi_{ADBL}$, and can be written as 
	\begin{eqnarray}\label{bl_drift}
	E({\bf V}(\beta,\sigma^{2},\tau)|\beta^{*},\sigma^{2,*},\tau^{*})=E(E(E({\bf V}|\beta,\sigma^{2},\tau^{*})|\sigma^2,\tau^{*})|\tau^{*})
	\end{eqnarray}
	
	\noindent
	Under the assumption $\epsilon>0$, for a fixed acceptance rate $r$, there is a probability of at least $p_m=\epsilon$ 
	that $\tau_j$ is updated at a given iteration. Hence, 
	\begin{eqnarray}\label{etau}
	E(\tau_{j}|\beta,\sigma^2,\tau^{*})&&\leq (\sqrt{\frac{\beta_{j}^{2}}{\lambda^{2}\sigma^{2}}}+\frac{1}{\lambda^{2}}) + 
	(1-p_{m})(\tau_{j}^{*}) \nonumber \\
	&& =(\sqrt{\frac{A\tau_{j}^{*}}{\lambda^{2}\sigma^2}\frac{\beta_{j}^{2}}{A\tau_j^{*}}}+\frac{1}{\lambda^{2}})+(1-p_{m})
	(\tau_{j}^{*}) \nonumber \\
	&& \leq (\frac{A\tau_j^{*}/(2\sigma^2)+1}{\lambda^{2}}+\frac{\beta_{j}^{2}}{2A\tau_j^{*}})+(1-p_m)(\tau_{j}^{*}), 
	\end{eqnarray}
	where $A$ is an arbitrary positive number, and the last inequality holds because $\sqrt{xy}\leq \frac{x+y}{2}$ for any positive number $x$ and $y$.  Similarly we can establish
	\begin{eqnarray}\label{etauinv}
	E(1/\tau_{j}|\beta,\sigma^2,\tau^{*}) && \leq (\sqrt{\frac{\lambda^{2}\sigma^{2}}{\beta_{j}^{2}}})+(1-p_{m})(1/\tau_{j}^{*})\nonumber \\
	&&\leq (\frac{\sigma^{2}}{2\beta_{j}^{2}}B\tau_{j}^{*}+\frac{\lambda^{2}}{2B\tau_{j}^{*}})+(1-p_m)(1/\tau_{j}^{*})
	\end{eqnarray}
	where $B$ is an arbitrary positive number.
	
	Then if we define $w=\left(1-p_{m}+\frac{\lambda^{2}}{2B}+\frac{1}{2A}\right)$, by using results of \eqref{etau} and \eqref{etauinv} we have
	\begin{eqnarray}\label{ble1}
	E({\bf V}(\beta,\sigma^2,\tau)|\beta,\sigma^{2},\tau^{*}) &&\leq \left(1-p_m+\frac{\sigma^{2}B}{2}+\frac{A}{2\lambda^{2}\sigma^{2}}\right)\sum_{j=1}^{p}(\tau_{j}^{*}) 
	+\sum_{j=1}^{p}\beta_{j}^{2}\frac{1}{\tau_{j}^{*}}\left(1-p_m+\frac{\lambda^{2}}{2B}+\frac{1}{2A}\right) \nonumber \\
	&& +\frac{p}{\lambda^{2}} +w(Y-X\beta)^{T}(Y-x\beta) +\sum_{j=1}^{p}E[\frac{1}{\tau_{j}^{1/4}}|\beta,\sigma^{2}] \nonumber \\ 
	&&= c_{\sigma^{2}}\sum_{j=1}^{p}(\tau_{j}^{*})+w\left[(Y-X\beta)^{T}(Y-x\beta)+\beta^{T}D_{\tau^{*}}^{-1}\beta\right]\nonumber \\
	&& +\sum_{j=1}^{p}E[\frac{1}{\tau_{j}^{s/2}}|\beta,\sigma^{2},\tau^{*}]+\frac{p}{\lambda^{2}}+\sigma^2 + \frac{1}{\sigma^2}
	\end{eqnarray}
	where $c_{\sigma^{2}}=\left(1-p_m+\frac{\sigma^{2}B}{2}+\frac{A}{2\lambda^{2}\sigma^{2}}\right)$. 
	
	To calculate $E[\frac{1}{\tau_{j}^{s/2}}|\beta,\sigma^{2},\tau^{*}]$, note that $\tau_{j}$, given $\beta, \sigma^{2}$ and that 
	the corresponding worker successfully sends update to the manager, is Generalized Inverse Gaussian(GIG) distributed 
	with parameters $(\frac{1}{2}, \lambda^2, \frac{\beta_j^{2}}{\sigma^{2}})$. More specifically, if the corresponding 
	worker successfully sends update to the manager, then 
	\begin{eqnarray}
	\pi(\tau_{j}|\beta,\sigma^{2}) 
	&=&\frac{\lambda}{\sqrt{2\pi}}\tau_{j}^{-1/2}\exp\left\{-\frac{\lambda^{2}(1-\tau_{j}\sqrt{\frac{\lambda^{2}\sigma^{2}}{\beta_{j}^{2}}})^{2}}{2\tau_{j}\times\frac{\lambda^{2}\sigma^{2}}{\beta_{j}^{2}}}\right\} \nonumber \\
	&& \propto\tau_{j}^{-1/2}\exp\{-\frac{\lambda^{2}\tau_{j}}{2}-\frac{\beta_{j}^{2}}{2\sigma^{2}}\frac{1}{\tau_{j}}\}
	\end{eqnarray}
	By \cite[eq, (3.15)]{pal2014geometric},  for $0<s<\frac{1}{2}$, 
	\begin{eqnarray}
	E(\frac{1}{\tau_{j}^{s/2}}|\beta,\sigma^{2},\tau^{*})\leq(1-p_{m})(\frac{1}{\tau_{j}^{*}})^{s/2}+C_{2}(\frac{\sigma}{|\beta_{j}|})^{s}+C_{3}.
	\end{eqnarray}
	Here $C_{2}=(1+\varepsilon^{*})\frac{[\Gamma(\frac{1}{2}-\frac{s}{2})] 2^{s/2}}{\Gamma(\frac{1}{2}-s)}$, where 
	$\varepsilon^{*}$ is an arbitrary positive number. Here $C_{3}=\lambda^{s}[\frac{[2(1-s)]^{\frac{s}{2}}}{\varepsilon^{s}}+\frac{1}{\varepsilon^{\frac{s}{2}}}]$ where  $\varepsilon$ depends on $s$ and $\varepsilon^{*}$ only, thus $C_{3}$ is irrelevant of the parameters $(\beta,\sigma^{2},\tau^{*})$.
	
	Then we proceed to calculate expectation of \eqref{ble1} conditional on $\tau^{*}, \sigma^2$. Notice
	\begin{eqnarray}
	&&E(\beta^{T}D_{\tau^{*}}^{-1} \beta+(Y-X\beta)^{T}(Y-X\beta)|\tau^{*},\sigma^{2})\nonumber\\
	=&&E(\beta^{T}(X^{T}X+D_{\tau^{*}}^{-1})\beta|\tau^{*},\sigma)-2Y^{T}X(X^{T}X+D_{\tau^{*}})^{-1}X^{T}Y+Y^{T}Y\nonumber\\
	=&&Y^{T}Y-Y^{T}X(X^{T}X+D_{\tau^{*}})^{-1}X^{T}Y+p\sigma^{2}\nonumber\\
	\leq&&Y^{T}Y+p\sigma^{2}\nonumber
	\end{eqnarray}
	Further more By Proposition A1 in \cite{pal2014geometric}
	\begin{eqnarray}
	E[(\frac{1}{|\beta_{j}|})^{s}|\tau^{*},\sigma^{2}] 
	\leq \frac{\Gamma(\frac{1-s}{2})2^{\frac{1-s}{2}}}{\sqrt{2\pi}[\sigma^{2}A_{\tau^{*}}^{-1}]_{ii}^{\frac{s}{2}}}
	\leq
	\frac{\Gamma(\frac{1-s}{2})2^{\frac{1-s}{2}}}{\sqrt{2\pi}\sigma^{s}}[\lambda_{X}^{s/2}+(\frac{1}{\tau_{i}^{*}})^{s/2}]
	\end{eqnarray}
	where $[\cdot]_{ii}$ is the $(i,i)$'th entry of a matrix and $\lambda_{X}$ is the maximum eigenvalue of $X^{T}X$. Then it 
	follows that 
	\begin{eqnarray}
	& & E(E({\bf V}(\beta, \sigma^2, \tau)|\beta,\sigma^2, \tau^{*})|\sigma^2, \tau^{*})\\
	&\leq& c_{\sigma^{2}}\sum(\tau_{j}^{*})+w(Y^{T}Y+p\sigma^{2})+
	[(1-p_{m}+C_{2}\frac{\Gamma(\frac{1-s}{2})2^{\frac{1-s}{2}}}{\sqrt{2\pi}})\sum_{j=1}^{p}(\frac{1}{\tau_{j}^{*}})^{s/2}] \nonumber
	\\
	&&+p\lambda_{X}^{s/2}+pC_{3}+\frac{ p}{\lambda^{2}} \nonumber\\
	&=& c_{\sigma^{2}}\sum(\tau_{j}^{*})+wp\sigma^{2}
	+(1-p_{m}+C_{2}\frac{\Gamma(\frac{1-s}{2})2^{\frac{1-s}{2}}}{\sqrt{2\pi}})\sum_{j=1}^{p}(\frac{1}{\tau_{i}^{*}})^{s/2}+c_{1}
	\end{eqnarray}
	where $c_{1}=wY^{T}Y+p\lambda_{X}^{s/2}+pC_{3}+\frac{ p}{\lambda^{2}}$. Notice $C_{2}\frac{\Gamma(\frac{1-s}{2})2^{\frac{1-s}{2}}}{\sqrt{2\pi}}$ goes to zero when $s\rightarrow \frac{1}{2}$, by choosing $s$ close enough to $\frac{1}{2}$, we can have $C_{2}\frac{\Gamma(\frac{1-s}{2})2^{\frac{1-s}{2}}}{\sqrt{2\pi}}\leq\frac {}{2}$, which leads to 
	\begin{eqnarray}
	& & E(E({\bf V}(\beta, \sigma^2, \tau)|\beta,\sigma^2, \tau^{*})|\sigma^2, \tau^{*})\\
	&\leq& c_{\sigma^{2}}\sum(\tau_{j}^{*})+wp\sigma^{2}+(1-\frac{1}{2}p_{m})\sum_{j=1}^{p} 
	(\frac{1}{\tau_{i}^{*}})^{s/2}+c_{1}
	\end{eqnarray}  
	
	Then for the outer expectation of \eqref{bl_drift}, we first establish the following inequalities:
	\begin{eqnarray}\label{ineq_sigma}
	E(\sigma^{2}|\tau^{*})	=\frac{Y^{T}(I-XA_{\tau^{*}}^{-1}X^{T})Y+2b}{n+2\alpha-2}\leq\frac{Y^{T}Y+2b}{n+2\alpha-2} \nonumber\\
	E(\frac{1}{\sigma^{2}}|\tau^{*})	=\frac{n+2\alpha}{Y^{T}(I-XA_{\tau^{*}}^{-1}X^{T})Y+2b}\leq\frac{n+2\alpha}{2b}
	\end{eqnarray} 
	where $n\geq 3$ guarantees that the first bound is positive. It follows that 
	\begin{eqnarray}
	& & E(E(E({\bf V}(\beta, \sigma^2, \tau)|\beta,\sigma^2, \tau^{*})|\sigma^2, \tau^{*})|\tau^{*})\\ 
	&\leq& (1-p_{m}+\frac{p_{M} B}{2}\times(\frac{Y^{T}Y+2b}{n+2\alpha-2})+\frac{p_{M} A}{2\lambda^{2}} 
	\times (\frac{n+2\alpha}{2b}))\sum(\tau_{j}^{*})\nonumber\\
	& & +(1-\frac{1}{2}p_{m})\sum_{j=1}^{p}(\frac{1}{\tau_{i}^{*}})^{s/2}+\psi\nonumber
	\end{eqnarray}
	where $\psi = c_{1}+wp(\frac{Y^{T}Y+2b}{n+2\alpha-2})$. 
	
	Finally by setting $A=\frac{\lambda^{2}p_{m}}{2p_{M}}\times(\frac{n+2\alpha}{2b})^{-1}$,  $B=\frac{p_{m}}{2p_{M}}(\frac{Y^{T}Y+2b}{n+2\alpha-2})^{-1}$ we can establish the drift condition as:
	\begin{eqnarray}\label{drift_bl}
	E({\bf V}(\beta,\sigma,\tau)|\beta^{*},\sigma^{*},\tau^{*})
	&&
	\leq
	(1-\frac{1}{2}p_{m})\sum(\tau_{j}^{*})+(1-\frac{1}{2}p_{m})\sum_{j=1}^{p}(\frac{1}{\tau_{i}^{*}})^{s/2}+\psi \nonumber \\
	&&\leq(1-\frac{1}{2}p_{m}){\bf V}(\beta^{*},\sigma^{*},\tau^{*})+\psi
	\end{eqnarray}  
	Thus the proof of \eqref{drift_BL} is completed by letting $\rho=1-\frac{1}{2}p_{m}$, and $\psi$ as defined above.

\subsection{Proof of Theorem \ref{thm_GLM}} \label{prove_LME}

	
	{ We will prove geometric ergodicity of the switched ADDA chain, which will be enough to establish geometric 
	ergodicity of the ADDA chain based on Lemma \ref{Switched:ADDA:Chain}.} 

	By Assumption \ref{GLM_As1} holding, there exists a symmetric positive definite matrix $Q$ such that $Z_i^T 
	Z_i\succ Q$.  Also since we are fixing $\Gamma$ as identity matrix, $b$ is the same as $d$, $\Sigma$ is the same as 
	$\tilde{\Sigma}$, and $\alpha$ will simply be $\beta$ (since the $\gamma$ entries are fixed). So in the following we will 
	only use $b$, $\Sigma$ and $\beta$. Hereby we can define some matrices that will be used later in the proof. Let 
	$$Z=\left[\begin{array}{ccccc}
	Z_{1}\\
	& Z_{2}\\
	&  & \ddots\\
	&  &  & Z_{m-1}\\
	&  &  &  & Z_{m}
	\end{array}\right],$$
	$b=(b_{1}^{T},\dots,b_{m}^{T})^{T}$, $H=X(V_{\beta}+X^{T}X)^{-1}X^{T}$, $\mu_{i}=(\frac{1}{\sigma^{2}}Z_{i}^{T}Z_{i}+\Sigma^{-1})^{-1}\frac{1}{\sigma^{2}}Z_{i}^{T}(y_{i}-X_{i}\beta)$ and $V_{i}=(\frac{1}{\sigma^{2}}Z_{i}^{T}Z_{i}+\Sigma^{-1})^{-1}$. When  $\Gamma$ is fixed as identity, the conjugate priors are in the forms as $\sigma^{2}\sim\frac{M}{\chi_{a}^{2}}$, $\beta|\sigma^{2}\sim N(0,\sigma^{2}V_{\beta}^{-1})$ and $\Sigma\sim IW(W,s)$, and the corresponding posterior conditional distributions will be the following:
	\begin{eqnarray}
	b_{i}|(\sigma^{2},\Sigma,\beta, \text{ worker i successfully sends update to the manager}) \stackrel{ind}{\sim} N(\mu_{i},V_{i}) \nonumber \\
	\Sigma^{-1}|b \sim Wishart((\sum b_{i}b_{i}^{T}+W)^{-1},m+s) \nonumber \\
	\sigma^{2}|b \sim \frac{(y-Zb)^{T}(I-H)(y-Zb)+M}{\chi_{n+a-p}^{2}} \nonumber \\
	\beta|\sigma^{2},b \sim N((V_{\beta}+X^{T}X)^{-1}X^{T}(y-Zb),\sigma^{2}(V_{\beta}+X^{T}X)^{-1})
	\end{eqnarray}
	
	\noindent
	We proceed by defining the following drift function:
	\begin{eqnarray}\label{drift_glm}
	{\bf V}(b,\beta,\sigma^{2},{\Sigma}) &=& \|y-Zb\|_{2}^{2} + \frac{1}{\sigma ^2 }+ tr(\Sigma^{-1})+ c_{1}\sigma^{2} 
	\nonumber\\
	&& +c_{2}\frac{1}{\sigma^{2}}\|y-X\beta\|_{2}^{2}+c_{3}\sum_{i}b_{i}^{T}b_{i}+c_{4}[tr(\Sigma)]. 
	\end{eqnarray}

        \noindent
	Here $c_1,c_2, c_3, c_4$ are fixed positive constants to be defined later. This drift function is unbounded off compact 
	sets, and it remains to show that there exist some constants $0\leq\rho<1$ and $\psi>0$ such that 
	\begin{eqnarray} \label{dc_glm}
	E({\bf V}(b,\beta,\sigma^{2},{\Sigma})|b^{*},\beta^{*},(\sigma^{2})^{*},{\Sigma}^{*})\leq \rho {\bf V}(b^{*},\beta^{*},
	(\sigma^{2})^{*},{\Sigma}^{*})+\psi.
	\end{eqnarray}
	
	\noindent
	where this expectation is with respect to the transition kernel of $\Phi_{ADLME}$ (adjusted for fixed $\Gamma = I_q$). 
	By the definition of $\Phi_{ADLME}$, the expectation can be written as
	$$
	E({\bf V}(b,\beta,\sigma^{2},{\Sigma})|b^{*},\beta^{*},(\sigma^{2})^{*},{\Sigma}^{*}) = E(E(E({\bf V}|\sigma^{2},\Sigma,\beta,b^{*})|\sigma^{2},b^{*})|b^{*}). 
	$$
	
	\noindent
	Using linearity, we proceed to simplify each term in \eqref{drift_glm}.  
	
	{Let $p_{i,r} (\sigma^2, \Sigma, \beta)$ denote the probability that $b_{i}$ is updated and drawn from the conditional 
	distribution given $\sigma^2, \Sigma, \beta$ (and not set to $b_i^*$). Note that $p_{i,r}$ is uniformly bounded below by 
	$\epsilon$ and uniformly bounded above by $1$.} Then
	\begin{eqnarray}\label{p1}
	&&E(\|y-Zb\|_{2}^{2}|\sigma^{2},\Sigma,\beta,b^{*})
	\nonumber\\=&&y^{T}y+\sum_{i=1}^{m}p_{i,r}\times tr \left( \left( \frac{1}{\sigma^{2}}Z_{i}^{T}Z_{i}+\Sigma^{-1} 
	\right)^{-1}Z_{i}^{T}Z_{i} \right)+\nonumber\\
	&&\sum_{i=1}^{m}p_{i,r}(y_{i}-X_{i}\beta)^{T}\frac{1}{\sigma^{2}}Z_{i}(\frac{1}{\sigma^{2}}Z_{i}^{T}Z_{i}+\Sigma^{-1})^{-1}Z_{i}^{T}Z_{i}(\frac{1}{\sigma^{2}}Z_{i}^{T}Z_{i}+\Sigma^{-1})^{-1}\frac{1}{\sigma^{2}}Z_{i}^{T}(y_{i}-X_{i}\beta)
	\nonumber\\
	&&-2\sum_{i=1}^{m}p_{i,r}y_{i}^{T}Z_{i}(\frac{1}{\sigma^{2}}Z_{i}^{T}Z_{i}+\Sigma^{-1})^{-1}\frac{1}{\sigma^{2}}Z_{i}^{T}(y_{i}-X_{i}\beta)\nonumber\\
	&&+\sum_{i=1}^{m}(1-p_{i,r})b_{i}^{*,T}Z_{i}^{T}Z_{i}b_{i}^{*}-2\sum_{i=1}^{m}(1-p_{i,r})y_{i}^{T}Z_{i}b_{i}^{*}\nonumber\\
	\leq&&y^{T}y+\sum_{i}p_{i,r}q\sigma^{2}\nonumber\\
	&&+\sum_{i=1}^{m}p_{i,r}(y_{i}-X_{i}\beta)^{T}\frac{1}{\sigma}Z_{i}(\frac{1}{\sigma^{2}}Z_{i}^{T}Z_{i}+\Sigma^{-1})^{-1}\frac{1}{\sigma}Z_{i}^{T}(y_{i}-X_{i}\beta)\nonumber\\
	&&-2\sum_{i=1}^{m}p_{i,r}y_{i}^{T}Z_{i}(\frac{1}{\sigma^{2}}Z_{i}^{T}Z_{i}+\Sigma^{-1})^{-1}\frac{1}{\sigma^{2}}Z_{i}^{T}(y_{i}-X_{i}\beta)\nonumber\\
	&&+\sum_{i=1}^{m}(1-p_{i,r})b_{i}^{*,T}Z_{i}^{T}Z_{i}b_{i}^{*}-2\sum_{i=1}^{m}(1-p_{i,r})y_{i}^{T}Z_{i}b_{i}^{*}
	\end{eqnarray}
	where 
	\begin{eqnarray}
	&&\sum_{i=1}^{m}p_{i,r}(y_{i}-X_{i}\beta)^{T}\frac{1}{\sigma}Z_{i}(\frac{1}{\sigma^{2}}Z_{i}^{T}Z_{i}+\Sigma^{-1})^{-1}\frac{1}{\sigma}Z_{i}^{T}(y_{i}-X_{i}\beta) \nonumber
	\\&&-2\sum_{i=1}^{m}p_{i,r}y_{i}^{T}Z_{i}(\frac{1}{\sigma^{2}}Z_{i}^{T}Z_{i}+\Sigma^{-1})^{-1}\frac{1}{\sigma^{2}}Z_{i}^{T}(y_{i}-X_{i}\beta) \nonumber
	\\=&&\sum_{i=1}^{m}p_{i,r}\beta^{T}X_{i}^{T}\frac{1}{\sigma}Z_{i}(\frac{1}{\sigma^{2}}Z_{i}^{T}Z_{i}+\Sigma^{-1})^{-1}\frac{1}{\sigma}Z_{i}^{T}X_{i}\beta \nonumber
	\\&&-\sum_{i=1}^{m}p_{i,r}y_{i}^{T}\frac{1}{\sigma}Z_{i}(\frac{1}{\sigma^{2}}Z_{i}^{T}Z_{i}+\Sigma^{-1})^{-1}\frac{1}{\sigma}Z_{i}^{T}y \nonumber
	\\\leq&&\sum_{i=1}^{m}p_{i,r}\beta^{T}X_{i}^{T}\frac{1}{\sigma}Z_{i}(\frac{1}{\sigma^{2}}Z_{i}^{T}Z_{i}+\Sigma^{-1})^{-1}\frac{1}{\sigma}Z_{i}^{T}X_{i}\beta \nonumber
	\\ \leq&&\beta^{T}(\sum_{i=1}^{m}p_{i,r}X_{i}^{T}X_{i})\beta \leq \beta^{T}X^{T}X\beta, 
	\end{eqnarray}
	where the second last inequality holds because$ \frac{1}{\sigma}Z_{i}(\frac{1}{\sigma^{2}}Z_{i}^{T}Z_{i}+\Sigma^{-1})^{-1}\frac{1}{\sigma}Z_{i}^{T}\prec I$. It follows that 
	\begin{eqnarray}\label{p2}
	&&E(\|y-Zb\|_{2}^{2}|\sigma^{2},\Sigma,\beta,b^{*}) \leq  \sum_{i=1}^{m}p_{i,r}y_{i}^{T}y_{i}\nonumber \\ &&+\sum_{i=1}^{m}(1-p_{i,r})(y_{i}-Z_{i}b_{i}^{*})^{T}(y_{i}-Z_{i}b_{i}^{*})+p_{M}\beta^{T}X^{T}X\beta+\sum_{i}p_{i,r}q\sigma^{2}. 
	\end{eqnarray}
	In addition, 
	\begin{eqnarray}\label{p3}
	&&E(\beta^{T}X^{T}X\beta|\sigma^{2},b^{*})\nonumber
	\\
	=&&(y-Zb^{*})^T X(V_{\beta}+X^{T}X)^{-1}X^{T}X(V_{\beta}+X^{T}X)^{-1}X^{T}(y-Zb^{*}) \nonumber
	\\&&+\sigma^{2}tr(X^{T}X(V_{\beta}+X^{T}X)^{-1})\nonumber
	\\\leq&&(y-Zb^{*})^{T}H(y-Zb^{*})+\sigma^{2}p
	\end{eqnarray}
	Combining \eqref{p2} and \eqref{p3}, we get 
	\begin{eqnarray}
	&&E(E(E(\|y-Zb\|_{2}^{2}+c_{1}\sigma^{2}|\sigma^{2},\Sigma,\beta,b^{*})|\sigma^{2},b^{*})|b^{*})\nonumber
	\\=&&\sum_{i}p_{i,r}y_{i}^{T}y_{i}+\sum_{i}(1-p_{i,r})(y_{i}-Z_{i}b_{i}^{*})^{T}(y_{i}-Z_{i}b_{i}^{*})\nonumber
	\\&&+(y-Zb^{*})^{T}H(y-Zb^{*}) \nonumber
	\\&&+\frac{p+\sum_{i}p_{i,r}q+c_{1}}{n+a-p-2}(y-Zb^{*})^{T}(I-H)(y-Zb^{*})+\frac{(p+\sum_{i}p_{i,r}q+c_{1})M}{n+a-q-2}\nonumber
	\\\leq&&(y-Zb^{*})^{T}(I_p+(H-\epsilon I_p)+(\frac{(p+mq)+c_{1}}{n+a-p-2})(I-H))(y-Zb^{*})\nonumber\\ &&+\frac{((p+mq)+c_{1})M}{n+a-p-2} + \sum_{i}y_{i}^{T}y_{i}.
	\end{eqnarray}
	By Assumption \ref{GLM_As1}, there exist a small enough $c_1>0$, such that $\frac{(p+mq)+c_{1}}{n+a-p-2}(I_{p}-H)\prec \epsilon I_{p}-H$. It guarantees that $$\rho_1=\lambda_{max}\{I_p+(H-\epsilon I_p)+(\frac{(p+mq)+c_{1}}{n+a-p-2})(I-H)\}<1$$
	where $\lambda_{max}\{A\}$ represent the maximum eignenvalue of $A$. Then it follows that
	\begin{eqnarray}\label{term12}
	E(E(E(\|y-Zb\|_{2}^{2}+c_{1}\sigma^{2}|\sigma^{2},\Sigma,\beta,b^{*})|\sigma^{2},b^{*})|b^{*})\leq \rho_1 \|y-Zb^{*}\|_{2}^{2}+\psi_{1}
	\end{eqnarray}
	where $\psi_{1}=\sum_{i}y_{i}^{T}y_{i}+\frac{((p+mq)+c_{1})M}{n+a-p-2}.$  
	
	For the later terms, notice that $E(\frac{1}{\sigma^{2}}|b^{*})\leq  \frac{n+a-p}{M}$.  And we also have 
	\begin{eqnarray}
	&&E(E(\frac{1}{\sigma^{2}}\|y-X\beta\|_{2}^{2}|\sigma^{2},b^{*})|b^{*}) \nonumber\\
	\leq&&E(\frac{1}{\sigma^{2}} 2y^{T}y)|b^{*})+E(E(\frac{2}{\sigma^{2}}\beta^{T}X^{T}X\beta|\sigma^{2},b^{*})|b^{*})\nonumber\\\leq&&\frac{n+a-p}{M}2y^{T}y+2p +E(\frac{2}{\sigma^{2}}(y-Zb^{*})^{T}H(y-Zb^{*})|b^{*}) \nonumber
	\\=&&\frac{n+a-p}{M}2y^{T}y+2p \nonumber
	\\&&+\frac{2(n+a-p)(y-Zb^{*})^{T}H(y-Zb^{*})}{(y-Zb^{*})^{T}(I-H)(y-Zb^{*})+M}
	\end{eqnarray}
	Thus when $ c_{2}=\frac{\lambda_{min}(I-H)}{\lambda_{max}(H)2(n+a-p)}$ we have 
	\begin{eqnarray} \label{term3}
	E(\frac{1}{\sigma^2}+\frac{c_{2}}{\sigma^{2}}\|y-X\beta\|_{2}^{2})\leq\psi_{2}
	\end{eqnarray}
	where $\psi_{2}=\frac{n+a-p}{M}+c_{2}[\frac{n+a-p}{M}2y^{T}y+2p]+1$.  
	
	Furthermore, 
	\begin{eqnarray} \label{ee1}
	&&E(c_{3}\sum_{i}b_{i}^{T}b_{i}\nonumber|\beta,\sigma^2,b^{*})
	\\=&&c_{3}\sum_{i} p_{i,r}(y_{i}-X_{i}\beta)^{T}\frac{1}{\sigma^{2}}Z_{i}(\frac{1}{\sigma^{2}}Z_{i}^{T}Z_{i}+\Sigma^{-1})^{-1}(\frac{1}{\sigma^{2}}Z_{i}^{T}Z_{i}+\Sigma^{-1})^{-1}\frac{1}{\sigma^{2}}Z_{i}^T(y_{i}-X_{i}\beta)\nonumber
	\\&&+c_{3}\sum_{i}p_{i,r}tr((\frac{1}{\sigma^{2}}Z_{i}^{T}Z_{i}+\Sigma^{-1})^{-1})+ c_{3}\sum_{i}(1-p_{i,r})b_{i}^{T,*}b_i^{*}
	\nonumber
	\\\leq&&c_{3}\max\{\lambda_{max}\left( (Z_{i}^{T}Z_{i})^{-1} \right)\}\sum_{i}(y_{i}-X_{i}\beta)^{T}Z_{i}(Z_{i}^{T}Z_{i}+
	\sigma^2 \Sigma^{-1})^{-1} Z_{i}(y_{i}-X_{i}\beta)\nonumber 
	\\&&+c_{3}mtr(\Sigma) +c_{3}(1-\epsilon)\sum_{i}b_{i}^{T,*}b_i^{*}
	\nonumber
	\\\leq&&c_{3}\delta\|y-X\beta\|_{2}^{2}+mtr(\Sigma)+c_{3}(1-\epsilon)\sum_{i}b_{i}^{T,*}b_i^{*}
	\end{eqnarray}
	where $\delta=\max_{i}\{\lambda_{max}\left( (Z_{i}^{T}Z_{i})^{-1} \right)\}$.  
	
	For the first term in \eqref{ee1}, notice that
	$E(\|y-X\beta\|_{2}^{2}|\sigma^2)\leq 2y^{T}y+2(y-Zb^{*})^{T}H(y-Zb^{*})+2\sigma^{2}p$. It follows that 
	\begin{eqnarray} \label{term4}
	E(E(c_{3}\delta\|y-X\beta\|_{2}^{2}|\sigma^{2},b^{*})|b^{*})\leq l\|y-Zb^{*}\|_{2}^{2}+\psi_{3}
	\end{eqnarray}
	where $l=c_{3}\delta\lambda_{max}\{(\frac{2p}{n+a-p})(I_p-H)+2H)\}$, $\psi_{3} = 4pc_{3}\delta M/(n+a-p)+2c_{3}\delta 
	y^{T}y$. We can choose $c_3>0$ small enough such that $l<\frac{1-\rho_1}{2}$.  
	
	For the second term in \eqref{ee1}, combined together with the last term in \eqref{drift_glm}, we have
	\begin{eqnarray}
	&&E(c_{3}mtr(\Sigma))+c_{4}[tr(\Sigma)]|b^{*}) \nonumber
	\\=&&\frac{c_{3}m+c_{4}}{m+s-q-1}[\sum_{i}b_{i}^{*,T}b_{i}^{*}+tr(W)]
	\end{eqnarray}
	where $c_4$ can be chosen small enough such that $\frac{m+c_{4}/c_{3}}{m+s-q-1}<\epsilon$. This choice is 
	attainable because of Part (b) of Assumption \ref{GLM_As1} which states that $s-q-1>\frac{(1-\epsilon)m}{\epsilon}$. 
	Then it follows that 
	\begin{eqnarray}\label{term5}
	&&E(c_{3}mtr(\Sigma))+c_{4}[tr(\Sigma)] +c_{3}(1-\epsilon)\sum_{i}b_{i}^{T,*}b_i^{*}|b^{*}) \nonumber
	\\\leq&&\rho_2 c_3\sum_{i}b_{i}^{*,T}b_{i}^{*}+ \psi_4
	\end{eqnarray}
	where $\rho_2 = \frac{m+c_{4}/c_{3}}{m+s-q-1} + 1 - \epsilon <1$ and $\psi_4 = \frac{c_{3}m+c_{4}}{m+s-q-1} tr(W)$.  
	
	Notice that $E[tr(\Sigma^{-1})|b^{*}]\leq (m+s)tr(W^{-1})$. Combining this result with \eqref{term12}, \eqref{term3}, 
	\eqref{term4} and \eqref{term5}, and letting $\psi_{5} = (m+s)tr(W^{-1})$, we get 
	\begin{eqnarray}
	E({\bf V}(b,\beta,\sigma^{2},{\Sigma})|b^{*},\beta^{*},(\sigma^{2})^{*},{\Sigma}^{*})\leq \frac{1+\rho_1}{2} \|y-Zb^{*}\|_{2}^{2}+\rho_2  c_3\sum_{i}b_{i}^{*, T}b_{i}^{*} \nonumber\\+ \psi_{1}+\psi_{2}+\psi_{3}+\psi_4 + \psi_5
	\end{eqnarray}
	By letting $\rho = max\{\frac{1+\rho_1}{2}, \rho_2 \}$ and $\psi = \psi_{1}+\psi_{2}+\psi_{3}+\psi_4+\psi_5$, we establish \eqref{dc_glm} as desired. 

\end{document}